\documentclass[twocolumn]{aastex631}

\usepackage{xcolor}
\usepackage{multirow}
\usepackage{comment}
\usepackage{array}
\usepackage{float}
\usepackage{enumitem}

\DeclareUnicodeCharacter{0301}{\'{e}}
\begin{document}

\title{ALMA-IMF XX: Core fragmentation in the W51 high-mass star-forming region}

\author[0000-0003-2968-5333]{T. Yoo}
\affiliation{Department of Astronomy, University of Florida, PO Box 112055, Florida, USA}

\author[0000-0001-6431-9633]{A. Ginsburg}
\affiliation{Department of Astronomy, University of Florida, PO Box 112055, Florida, USA}

\author{J. Braine}
\affiliation{Laboratoire d’Astrophysique de Bordeaux, Univ. Bordeaux, CNRS, B18N, allée Geoffroy Saint-Hilaire, 33615 Pessac, France}

\author[0000-0002-0533-8575]{N. Budaiev}
\affiliation{Department of Astronomy, University of Florida, PO Box 112055, Florida, USA}

\author{F. Louvet}
\affiliation{Univ. Grenoble Alpes, CNRS, IPAG, 38000 Grenoble, France}
\affiliation{DAS, Universidad de Chile, 1515 camino el observatorio, Las Condes, Santiago, Chile}

\author{F. Motte}
\affiliation{Univ. Grenoble Alpes, CNRS, IPAG, 38000 Grenoble, France}

\author[0000-0003-2300-8200]{A. M.\ Stutz}
\affiliation{Departamento de Astronom\'{i}a, Universidad de Concepci\'{o}n, Casilla 160-C, Concepci\'{o}n, Chile}
\author{B. Thomasson}
\affiliation{Univ. Grenoble Alpes, CNRS, IPAG, 38000 Grenoble, France}

\author{M. Armante}
\affiliation{Laboratoire de Physique de l’École Normale Supérieure, ENS, Université PSL, CNRS, Sorbonne Université, Université de Paris, Paris, France}
\affiliation{Observatoire de Paris, PSL University, Sorbonne Université,
LERMA, 75014 Paris, France}

\author{M. Bonfand}
\affiliation{Departments of Astronomy and Chemistry, University of Virginia, Charlottesville, VA 22904, USA
}

\author{S. Bontemps}
\affiliation{Laboratoire d’Astrophysique de Bordeaux, Univ. Bordeaux, CNRS, B18N, allée Geoffroy Saint-Hilaire, 33615 Pessac, France}

\author{L. Bronfman}
\affiliation{Departamento de Astronomía, Universidad de Chile, Casilla 36-D, Santiago, Chile}

\author[0000-0002-2189-6278]{G. Busquet}
\affiliation{Departament de F ́ısica Quàntica i Astrofísica (FQA), Universitat de Barcelona, Martí  i Franquès 1, E-08028 Barcelona, Catalonia, Spain}
\affiliation{Institut de Ciències del Cosmos (ICCUB), Universitat de Barcelona, Martí  i Franquès 1, E-08028 Barcelona, Catalonia, Spain}
\affiliation{Institut d'Estudis Espacials de Catalunya (IEEC), Esteve Terradas 1, edifici RDIT, Parc Mediterrani de la Tecnologia (PMT) Campus del Baix Llobregat - UPC 08860 Castelldefels (Barcelona), Catalonia, Spain}

\author{T. Csengeri}
\affiliation{Laboratoire d’Astrophysique de Bordeaux, Univ. Bordeaux, CNRS, B18N, allée Geoffroy Saint-Hilaire, 33615 Pessac, France}

\author{N. Cunningham}
\affiliation{SKA Observatory, Jodrell Bank, Lower Withington, Macclesfield,
SK11 9FT, United Kingdom}

\author{J. Di Francesco}
\affiliation{Herzberg Astronomy and Astrophysics Research Centre, National Research Council of Canada, 5071 West Saanich Road, Victoria, Canada}

\author[0000-0002-6325-8717]{D. J. D\'iaz-Gonz\'alez}
\affiliation{Instituto de Radioastronomía y Astrofísica, Universidad Nacional Autónoma de México, Morelia, Michoacán, 58089, Mexico} 

\author{M. Fernández-Lopez}
\affiliation{Instituto Argentino de Radioastronomía (CCT-La Plata, CONICET; CICPBA), C.C. No. 5, 1894, Villa Elisa, Buenos Aires, Argentina}

\author{R. Galván-Madrid}
\affiliation{Instituto de Radioastronomía y Astrofísica, Universidad Nacional Autónoma de México, Morelia, Michoacán, 58089, Mexico}
\author[0000-0002-2542-7743]{C. Goddi}
\affiliation{Instituto de Astronomia, Geofísica e Ciências Atmosféricas, Universidade de São Paulo, R. do Matão, 1226, São Paulo,
SP 05508-090, Brazil}
\affiliation{Dipartimento di Fisica, Universitá degli Studi di Cagliari, SP Monserrato-Sestu km 0.7, 09042 Monserrato (CA), Italy}
\affiliation{
INAF – Osservatorio Astronomico di Cagliari, Via della Scienza 5, 09047 Selargius (CA), Italy}
\author{A. Gusdorf}
\affiliation{Laboratoire de Physique de l’École Normale Supérieure, ENS, Université PSL, CNRS, Sorbonne Université, Université de Paris, Paris, France}
\affiliation{Observatoire de Paris, PSL University, Sorbonne Université,
LERMA, 75014 Paris, France}

\author{N. Kessler}
\affiliation{Laboratoire d’Astrophysique de Bordeaux, Univ. Bordeaux, CNRS, B18N, allée Geoffroy Saint-Hilaire, 33615 Pessac, France}
\author[0000-0003-2713-0211]{A. Koley}
\affiliation{Departamento de Astronom\'{i}a, Universidad de Concepci\'{o}n, Casilla 160-C, Concepci\'{o}n, Chile}

\author{H.-L Liu}
\affiliation{School of physics and astronomy, Yunnan University, Kunming, 650091, P. R. China}

\author{T. Nony}
\affiliation{Instituto de Radioastronomía y Astrofísica, Universidad Nacional Autónoma de México, Morelia, Michoacán, 58089, Mexico}
\affiliation{INAF - Osservatorio Astrofisico di Arcetri, Largo E. Fermi 5, 50125, Firenze, Italy}

\author{F. Olguin}
\affiliation{Institute of Astronomy, National Tsing Hua University, Hsinchu 30013, Taiwan}

\author[0000-0002-7125-7685]{P. Sanhueza}
\affiliation{Department of Astronomy, School of Science, The University of Tokyo, 7-3-1 Hongo, Bunkyo, Tokyo 113-0033, Japan}

\author{M. Valeille-Manet}
\affiliation{Laboratoire d’Astrophysique de Bordeaux, Univ. Bordeaux, CNRS, B18N, allée Geoffroy Saint-Hilaire, 33615 Pessac, France}
\author[0000-0003-2343-7937]{L. A. Zapata}
\affiliation{Instituto de Radioastronomía y Astrofísica, Universidad Nacional Autónoma de México, Morelia, Michoacán, 58089, Mexico}

\author{Q. Zhang}
\affiliation{Center for Astrophysics, Harvard \& Smithsonian, 60 Garden Street, Cambridge, MA 02138, USA}





\begin{abstract}

We present a study of core fragmentation in the W51-E and W51-IRS2 protoclusters in the W51 high-mass star-forming region. The identification of core fragmentation is achieved by the spatial correspondence of cores and compact sources which are detected in the short (low resolution) and the long baseline (high resolution) continuum images with the Atacama Large Millimeter/submillimeter Array (ALMA) in Bands 3 ($3\,{\rm mm}$) and 6 ($1.3\,{\rm mm}$), respectively. We characterize the compact sources found in the long baseline image, and conclude that the compact sources are pre/protostellar objects (PPOs) that are either prestellar dust cores or dust disks or envelopes around protostars. 
The observed trend of core fragmentation in W51 is that (i) massive cores host more PPOs, (ii) bright PPOs are preferentially formed in massive cores, (iii) equipartition of flux between PPOs is uncommon. Thermal Jeans masses of parent cores are insufficient to explain the masses of their fragments, and this trend is more prominent at high-mass cores. We also find that unfragmented cores are large, less massive, and less dense than fragmented cores. 

\end{abstract}



\section{Introduction} \label{sec:intro}

 Unveiling the origin of stellar masses is of great importance to many astrophysical fields. Not only is it an ultimate question in star formation theory, but also the stellar initial mass function (IMF) is a keystone to build models of galaxies, stellar clusters, and planet formation since the initial stellar mass is the most important factor in determining the stellar evolution. One of the remaining unresolved problems is whether the IMF is universal. While the observations for nearby clusters show almost invariant IMF (see the references in \citealt{bastian10}; \citealt{offner14}; \citealt{krumholz15}), extending the targets to diverse environments, e.g., the galactic center \citep[e.g.][]{lu13,hosek19}{}{}, other nearby galaxies \citep[e.g.][]{weisz15,wainer24}, and young massive clusters of the Galaxy \citep[e.g.][]{maia16}, provide some cases where the IMF slope has non-negligible scatter and deviation from the canonical slope ($dN/M\equiv\alpha=-2.35$; \citealt{salpeter55}).

Observations resolving the internal structure of molecular clouds has shown that they collapse under self-gravity into dense substructures, known as cores (which have $r\sim0.01-0.3$ pc), that are a first stage in the formation of new stars.
The mass function of these structures, the Core Mass Function (CMF), has often been found in nearby, low-mass protoclusters, to have a similar shape to the stellar initial mass function  \citep[IMF; e.g.,][]{motte98,alves07, simpson08, andre10}.
However, in high-mass star-forming protoclusters, the high-mass end of the CMF has been found to have a top-heavy (shallow) slope compared to the standard IMF \citep[e.g.][]{motte18, kong19, sanhueza19, lu20, pouteau22, nony23, louvet24}{}{}.


Overall, the primary parameter for setting the initial stellar mass is still an open question. The turbulent fragmentation model \citep[e.g.][]{padoan02} suggested that the power-law slope of the CMF derives from the density distribution of the supersonic turbulent ISM and demonstrated that the slope of the CMF is consistent with that of the Salpeter IMF \citep{salpeter55}. On the other hand, mass accretion onto cores has been suggested as another process affecting the initial stellar masses. For instance, the competitive accretion model \citep[e.g.][]{bonnell01,bonnell02} suggested the picture that star-forming cores grow by competing with each other for gas accretion, and the initial core mass in this model is less important in setting the initial stellar masses. A recent numerical study, \cite{padoan20}, proposed an initial inflow model where the gas inflow generated by supersonic turbulence has a more dominant role in setting stellar mass than the inertial core mass. 

 
 One important process affecting the transition between the CMF and the IMF is core fragmentation.
 \citet{offner14} suggested that several conditions, including that there is the same degree of core fragmentation over the full mass range, are required to maintain the same power-law slope of the CMF to the IMF. Over the decades, several analytic models of fragmentation have been developed. \cite{guszejnov15} established an analytic model of core fragmentation based on the turbulent fragmentation scheme, and found a slightly flatter CMF slope ($\alpha=-2.1$) compared to the IMF slope ($\alpha=-2.3$). On the other hand, \cite{vazquez-semadeni19} proposed the global hierarchical collapse model where gravitational infall rather than supersonic turbulence induces fragmentation (as initially suggested by \citealt{hoyle53}).
 
 Early observations of massive dense cores or clumps at a scale of $\sim0.1\,{\rm pc}$ have revealed that they host condensations or fragments in high spatial-resolution images \citep[e.g.][]{zhang09, bontemps10,zhang15}. With the advent of new telescopes such as the Submillimeter Array (SMA) and ALMA, the observation of smaller structures has been conducted for nearby, low-mass star-forming regions \citep[e.g.][]{pokhrel18} and the high-mass star-forming regions at a farther distance \citep[e.g.][]{palau21, tang22, budaiev24, li24}. A study of the Perseus star-forming region compiled various observational data with five different spatial scales--the entire cloud, clumps, cores, envelopes, and protostellar objects--showed that thermal Jeans mass is not high enough to explain the observed number of core fragments \citep{pokhrel18}. \cite{palau21} investigated the fragmentation of 18 massive dense cores and found a tentative correlation between the number of fragments and the mass-to-flux ratio, suggesting that magnetic field could have a role in determining core fragmentation. In contrast, \cite{tang22} argued that the number of core fragments in the central region of high-mass protocluster W51-IRS2 can be well explained by thermal Jeans mass. 

In this study, we utilize ALMA millimeter wavelength observational data of the W51 star-forming complex, which is the nearest ($5.1$--$5.41\,{\rm kpc}$; \citealt{xu09, sato10}) extreme high-mass ($M\gtrsim10^4\, M_\odot$; \citealt{kumar04}) star-forming region in the Milky Way (see the review in \citealt{ginsburg17b}). The main star-forming region, W51 A, has two sub-regions that are protoclusters, W51-E and W51-IRS2. This cloud is one of the best targets to study high-mass star formation given its very young age $(\lesssim1\,{\rm Myr}$; \citealt{okumura00}), such that most of the massive stars are (pre-)main-sequence; the region is at a very early stage of high-mass star formation \citep[e.g.][]{Zap2009,Zap2010,ginsburg15, ginsburg17b}{}{}. The cloud shows abundant evidence of high-mass star formation: hot cores and hyper-compact HII regions such as W51e2, W51e8 and W51 North \citep[e.g.][]{zhang97, goddi16, ginsburg16, ginsburg17}, X-ray stars \citep[e.g.][]{townsley14}, and masers from OH \citep{etoka12}, ${\rm H_2O}$ \citep[e.g.][]{genzel81, imai02, eisner02}, ${\rm CH_3OH}$ \citep[e.g.][]{phillips05, etoka12}{}{}, SiO \citep[e.g.][]{morita92, eisner02}, CS \citep{ginsburg19}, complex molecules \citep{Ron2016} and ${\rm NH_3}$ \citep[e.g.][]{brown91, gaume93, henkel13}. 
W51 therefore provides an optimal laboratory in which to study the formation of the upper end of the IMF.

Two sub-regions in the W51 cloud, W51-E and W51-IRS2, were targets of the ALMA-IMF large program \citep{motte22} surveying embedded massive protocluster with ALMA Band 3 (3 mm) and Band 6 (1.3 mm) continuum. The physical resolution of the ALMA-IMF image for W51 is $\sim2000\,{\rm au}$, at the size scale of dust cores. These two protoclusters were recently revisited with higher resolution ($\sim100$--$300\,{\rm au}$) ALMA observations \citep{goddi20} that detected a considerable number of compact sources. We will use ``high-resolution sources" (shortened to high-res sources) to refer to the compact, centrally-peaked sources identified in the high-resolution images throughout the paper. By spatially associating high-res sources with cores found in low-resolution ALMA-IMF images in W51, we will see how each ALMA-IMF core fragments in a high-mass star-forming region. 

The paper is organized as follows. In Section \ref{sec:obs}, we describe our observational data and the method to identify sources in the continuum images. In Section \ref{sec:YSO candidate_characterization}, we characterize the sources. In Section \ref{sec:fragmentation}, we investigate core fragmentation by linking cores and fragments, and we analyze the behavior of fragmentation as a function of core mass. In Section \ref{sec:discussion}, we discuss whether core fragmentation follows Jeans fragmentation and compare the physical properties of unfragmented/fragmented cores. In Section \ref{sec:conclusions}, we summarize our main results and discuss the implication of this study on high-mass star formation models and the relation between the CMF and the IMF.




\section{Observational data} \label{sec:obs}
In this section, we introduce the archival data we used for this study and the data reduction process. For the ALMA-IMF dataset, we illustrate the core catalog extracted from the continuum images.  

\subsection{ALMA-IMF continuum images and core catalog}
\label{subsec:obs_core}

We utilized the $3\,{\rm mm}$ (Band 3) and $1.3\,{\rm mm}$ (Band 6) continuum archival data of the W51 from the ALMA large program (2017.1.01355.L, PIs: Motte, Ginsburg, Louvet, Sanhueza), ALMA-IMF \citep[][]{motte22, ginsburg22}{}{}. The continuum images are obtained from 12 m array mosaic observations. As summarized in Table.~2 of \cite{motte22}, the $3\,{\rm mm}$ and $1.3\,{\rm mm}$ continuum images of W51-E have a field of view, $150\arcsec\times160\arcsec$ and $100\arcsec\times90\arcsec$, and angular resolution $0.28\arcsec$ and $0.30\arcsec$, respectively. For W51-IRS2, the fields of view of $3\,{\rm mm}$ and $1.3\,{\rm mm}$ are $160\arcsec\times150\arcsec$ and $92\arcsec\times98\arcsec$, and the angular resolutions are $0.27\arcsec$ and $0.48\arcsec$, respectively. We use \texttt{cleanest} continuum images, which use only the line-free channels, in this study. 
The detailed data processing of the continuum image is described in \cite{motte22} and \cite{ginsburg22}. A combined continuum image of W51-E and W51-IRS2 is shown in Fig.~\ref{fig:overview}.

The physical resolution of the ALMA-IMF continuum image is $\sim2000\,{\rm au}$, resolving the cores in protoclusters. \cite{louvet24} extracted the cores from the continuum image using the source finding algorithms, \texttt{getsf} \citep{menshchikov21} and \texttt{GExt2D} (Bontemps et al. in prep.). In this study, we used the core catalogs created by \texttt{getsf} from the $1.3\,{\rm mm}$ and $3\,{\rm mm}$ image smoothed to $2700\,{\rm AU}$ resolution as described in \cite{louvet24}.  

\begin{figure*}
    \centering
    \includegraphics[scale=0.5]{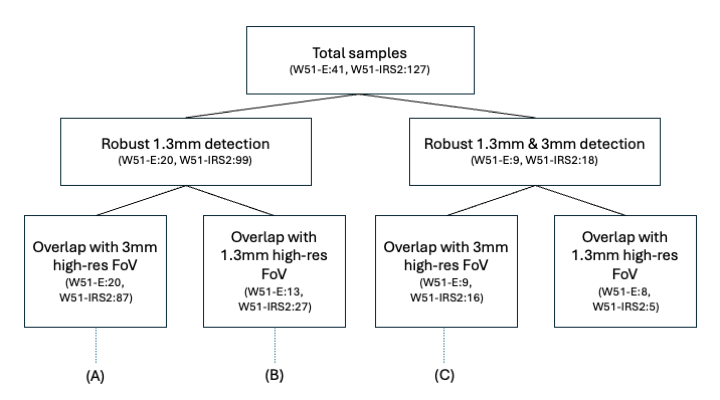}
    \caption{Core groups used in this study. Total ALMA-IMF cores detected in \texttt{getsf} are first divided into two groups by the criteria passing robust 1.3mm detection or robust 1.3mm \& 3mm detection of \texttt{getsf} core catalog. The core samples are further classified into two groups by examining the spatial overlap with the high-resolution continuum images. The number of cores in each group is listed in parentheses. Each group used in the analyses of this study are labeled.}
    \label{fig:core_subsample}
\end{figure*}

Among 41 and 127 \texttt{getsf} sources in W51-E and W51-IRS2, we applied selection criteria to remove unreliable flux and size measurements. We adopt the selection criteria described in \cite{menshchikov21} and \cite{pouteau22} to filter out sources with low signal-to-noise and high ellipticity in each band. Following the description in \cite{pouteau22}, we divided the core samples into the \texttt{getsf} sources with robust $1.3\,{\rm mm}$ measurement and sources with robust $1.3\,{\rm mm}$ and $3\,{\rm mm}$ measurement (Fig.~\ref{fig:core_subsample}) -- the robust $1.3\,{\rm mm}$ flux measurement criteria is used to ensure the presence of cores because $1.3\,{\rm mm}$ flux better traces dust emission, and the $3\,{\rm mm}$ criteria can be used if $3\,{\rm mm}$ flux is needed. For the analysis comparing core fluxes and high-res sources fluxes, we made groups depending on whether cores are found inside the field of view of either $1.3\,{\rm mm}$ or $3\,{\rm mm}$ high-resolution images. We label three of four groups as follows:
\begin{itemize}
    \item group A: robust $1.3\,{\rm mm}$ core flux measurement within the field of view of the $3\,{\rm mm}$ high-resolution image
    \item group B: robust $1.3\,{\rm mm}$ core flux measurement within the field of view of the $1.3\,{\rm mm}$ high-resolution image
    \item group C: robust $1.3\,{\rm mm}$ and $3\,{\rm mm}$ core flux measurement within the field of view of $3\,{\rm mm}$ high-resolution image
\end{itemize}
The fourth group in the Fig.~\ref{fig:core_subsample}, core samples with robust $1.3\,{\rm mm}$ and $3\,{\rm mm}$ core flux measurement within the field of view of the $1.3\,{\rm mm}$ image is not used in this study because no analyses require this sample.

Note that, the field of view of the $3\,{\rm mm}$ high-resolution image includes the whole area of the field of view of the $1.3\,{\rm mm}$ high-resolution image. Therefore, core group A provides the largest number of cores. This setup allows us to use different numbers of cores in different situations. For example, we have two options, group A and group B, in using $1.3\,{\rm mm}$ core flux. The former includes cores outside of $1.3\,{\rm mm}$ high-resolution image field of view; the $1.3\,{\rm mm}$ flux of high-res sources associated with these cores is not measurable. We use group A in the analysis in Sec.~\ref{subsubsec:nfrag_coremass}, where the number of fragments can be determined by single band detection. When we compare $1.3\,{\rm mm}$ flux of cores and high-res sources, we use core samples from group B instead. In all analyses using $3\,{\rm mm}$ flux of cores and high-res sources, we use group C.

In each group, we removed free-free contaminated sources with the spectral index $\alpha<2$ \citep{pouteau22, louvet24}. For those cores where $3\,{\rm mm}$ flux is not available, the pixel-by-pixel spectral index map was produced; if cores reside in the pixels with $\alpha<1$ then they are discarded \citep{louvet24}. We further discarded duplicate cores in the overlap between the two regions -- \#30, \#31, \#20 in W51-E and \#38, \#108, \#22 in W51-IRS2, respectively. We removed these three cores from W51-E. The number of cores in each group is listed in Fig.~\ref{fig:core_subsample}.

\subsection{High-resolution ALMA continuum image}

To find the signature of core fragmentation in the W51 region, we utilized archival high-resolution ALMA datasets from the ALMA cycle 3 program 2015.1.01596.S (PI: C. Goddi) at $1.3\,{\rm mm}$ and the ALMA cycle 5 program 2017.1.00293.S (PI: A. Ginsburg) at $3\,{\rm mm}$ observing W51-E and W51-IRS2. Both observations used 12 m array with a baseline ranging from 85 to 16196$\,{\rm m}$. The angular resolution of the continuum images is $27.3\,{\rm mas}\times20.0\,{\rm mas}$ in $1.3\,{\rm mm}$ and $66.1\,{\rm mas}\times41.8\,{\rm mas}$ in $3\,{\rm mm}$, corresponding to $\sim100$--$300\,{\rm au}$ in physical resolution. The central frequency of each image is approximately $86\,{\rm GHz}$ in Band 3 and $218\,{\rm GHz}$ in Band 6. The details about the observation in $1.3\,{\rm mm}$ can be found in \cite{goddi20}.

The data reduction and self-calibration were conducted by the Common Astronomy Software Applications (CASA) package (\cite{mcmullin07}; version 5.7.0). To produce the image files, we run Python scripts provided with ALMA raw data, \texttt{ScriptforPI.py} and \texttt{ScriptforImaging.py}. Then we self-calibrated the images with the script as follows. 
\begin{itemize}
    \item[1.] separate the science data from the whole dataset using CASA task \texttt{split}. 
    \item[2.] produce a dirty image of the long-baseline data with \texttt{tclean} task.
    \item[3.] regrid the startmodel using CASA task \texttt{imregrid} with the dirty image as a template to make \textit{startmodel} available for \texttt{tclean} process.
    \item[4.] start eight stages of iterative phase self-calibration process with \texttt{tclean}, \texttt{gaincal}, and \texttt{applycal}. 
\end{itemize}
During the \texttt{tclean} process, we used the ALMA-IMF continuum model image as \textit{startmodel} to complement the flux from the short baseline. We started our cleaning shallowly and gradually deepened the cleaning process during the iteration to include the diffuse emission into the continuum image.
For this reason, we used a gradually decreasing threshold of \texttt{tclean} for each step: \textit{threshold}=[0.2, 0.2, 0.15, 0.15, 0.15, 0.15, 0.1, 0.1]$\,{\rm mJy}$ for $1.3\,{\rm mm}$ and [0.1, 0.075, 0.075, 0.075, 0.075, 0.075, 0.075, 0.075]$\,{\rm mJy}$ for $3\,{\rm mm}$. We also applied two different types of \textit{cleanmask}, the shallower one in steps 1 to 3 and the deeper one in steps 4 to 8. Each mask covers any sources identified by visual inspection from the continuum image. For the weighting scheme, we employed Briggs weighting with robust=0. The \texttt{tclean} parameters were chosen as a result of a set of experiments with different parameters. We visually evaluated the quality of the final image produced by self-calibration and tweaked parameters to find the optimal set. The parts of the final continuum images in $1.3\,{\rm mm}$ are displayed in Fig.~\ref{fig:overview}.


\section{Characterization of high-res sources} \label{sec:YSO candidate_characterization}

In this section, we describe the identification process of high-res sources from high-resolution ALMA continuum images. We then present the physical properties of high-res sources such as spectral indices, sizes, masses, and separation. By doing so, we characterize and classify high-res sources to better understand the results of the core fragmentation process in high-mass star-forming regions. The characterized properties are summarized in Appendix.~\ref{appendix:ppo_tab}.

\subsection{High-res sources identification} 

\label{subsec:dendro}

\begin{figure*}
    \centering
    \includegraphics[scale=0.4]{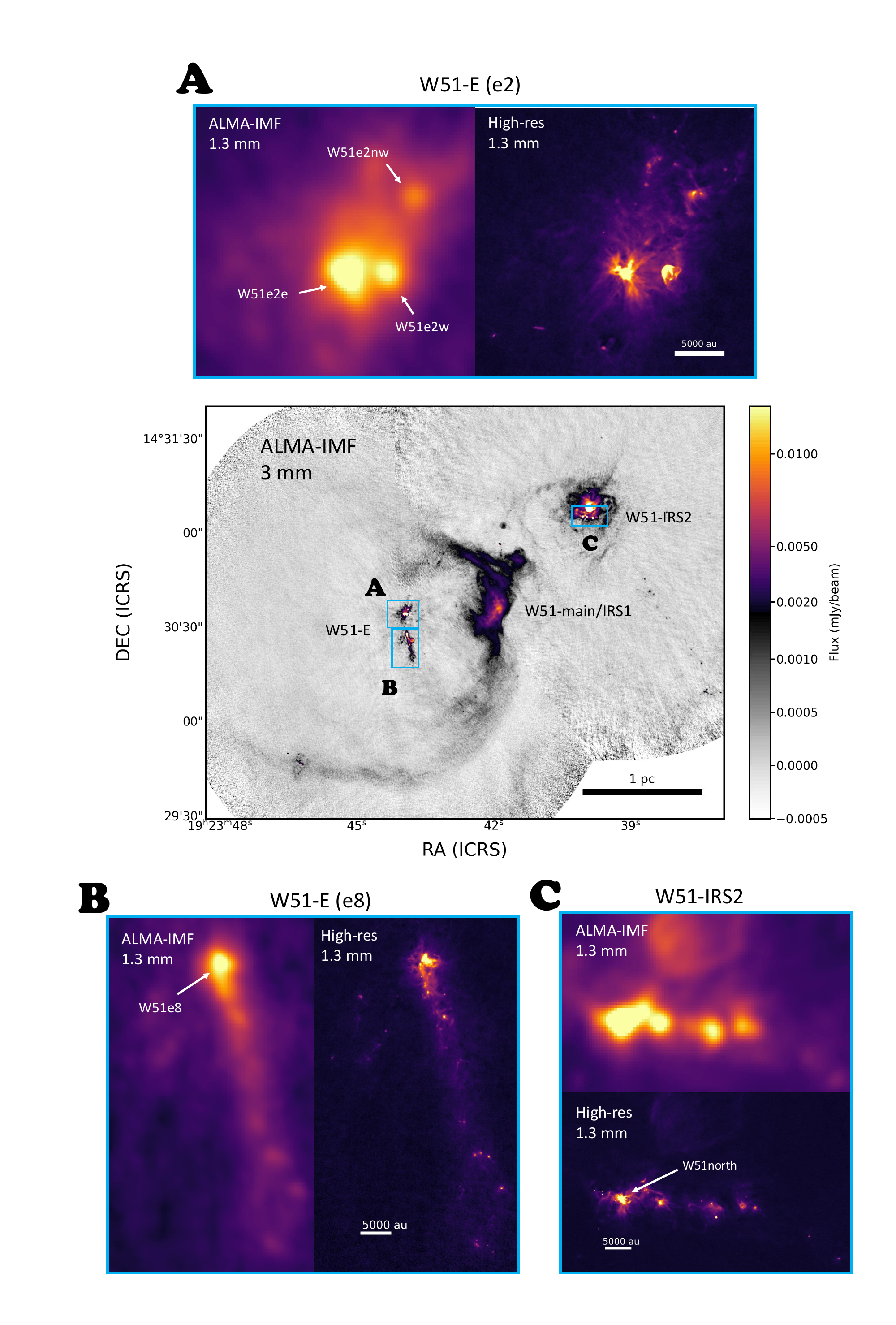}
    \caption{The overview of W51A region. The image at the center is the ALMA $3\,{\rm mm}$ continuum image of ALMA-IMF combining two sub-regions, W51-E and W51-IRS2. Each continuum image is truncated and merged in the middle to look seamless. Three regions marked with the blue boxes are magnified with $1.3\,{\rm mm}$ continuum image of ALMA-IMF data (left) and high-resolution data (right). We annotate massive YSOs studied in previous studies \citep[e.g.][]{goddi20}, W51e2 (\#39) and W51e8 (\#32) in W51-E and W51north (\#11) in W51-IRS2.}
    \label{fig:overview}
\end{figure*}

The high-resolution ALMA continuum images reveal many compact sources that were unresolved in the ALMA-IMF images (Fig.~\ref{fig:overview}). 
To make a catalog of high-res sources, we visually selected sources that are \textit{compact, centrally peaked, significant}, and \textit{independent} from background. Here, \textit{compact} source means that we did not include any extended structure, such as filamentary structure or diffuse HII regions (larger than several times of the image beam size). However, it is possible that compact sources surrounded by patchy extended emission exist, which still can be identified as high-res sources if the sources are \textit{centrally peaked}.

To ensure whether the selected sources are \textit{significant}, we used a watershed-based algorithm (\texttt{dendrogram}\footnote{ \texttt{astrodendro}\footnote{https://dendrograms.readthedocs.io/en/stable/} and \texttt{dendrocat}\footnote{https://dendrocat.readthedocs.io/en/latest/} Python packages.}; \citealt{rosolowsky08}). The \texttt{dendrogram} algorithm finds the local maxima and generates the hierarchical tree structure based on the parameters of the minimum value (\textit{min\_value}), the minimum significance (\textit{min\_delta}), and the minimum number of pixels (\textit{min\_npix}). We adopted the following set of parameters,  \textit{min\_value}=$3\sigma$, \textit{min\_delta}=$1.5\sigma$, \textit{min\_npix}=15 for our source finding scheme, where $\sigma$ is the rms noise. Any insignificant visually-selected sources that are not matched with \texttt{dendrogram}-selected sources are removed from the final catalog.

The high-res sources need to be \textit{independent} from the background, as the main goal of the source identification is to find fragments. The populated areas of the protocluster are often surrounded by extended emission, which possibly makes compact sources apparently connected to larger structures. Thus, \textit{independent} sources include any significant peaks that are partially detached from their parent structure. This unavoidably gives rise to some ambiguity in visual selection. To minimize the ambiguity, only the sources identified by more than two of the three coauthors are included into the final catalog. During the identification process, three independent coauthors changed the color scale stretches of the images. This helps human visual recognition to compact sources in a wide range of contrast. We have two groups of sources that will not be used in this study: i) \textit{low S/N} sources that were visually identified by more than two observers but not chosen by \texttt{dendrogram} and \textit{candidates} that were not selected by eye more than two times. We report these sources separately in Appendix~\ref {appendix:ambiguous} and will not use these sources in this paper.

We then cross-matched the sources in both bands when the sky coordinate offsets are smaller than the beam size at each wavelength. Among the matched sources, we found 13 common sources in the overlapping regions between the two regions, and removed those sources from the W51-E image. As a result, we identified 118 high-res sources in W51-E --73 matched sources, 23 sources seen in only $3\,{\rm mm}$, and 22 sources seen in only $1.3\,{\rm mm}$. For W51-IRS2, out of 93 high-res sources, we have 23 matched sources, 59 sources only seen in $3\,{\rm mm}$, and 11 sources only seen in $1.3\,{\rm mm}$. The number of detections is summarized in Table.~\ref{tab:ppo} and their snapshots and positions are displayed in Appendix.~\ref{appendix:ppo_map} and \ref{appendix:snapshots}.

\begin{deluxetable}{lcc} 
\label{tab:ppo}
\tablecaption{The number of high-res sources detected in W51-E and W51-IRS2.}
\tablehead{
    \colhead{} & \colhead{W51-E} & \colhead{W51-IRS2}
}
\startdata
Cross-matched (a+b+c) & 73 & 23 \\
\hspace{0.3cm} a) dust-dominated & 59 & 13\\
\hspace{0.3cm} b) optically thick dust or & \multirow{2}{*}{10} & \multirow{2}{*}{6} \\ free-free contaminated  \\
\hspace{0.3cm} c) free-free contaminated & 4 & 4\\
\hline
Single detection (d+e) & 45 & 70\\
\hspace{0.3cm} d) $1.3\,{\rm mm}$ & 22 & 11\\
\hspace{0.3cm} e) $3\,{\rm mm}$ & 23 & 59\\
\hline
Total (a+b+c+d+e) & 118 & 93\\
\enddata
\end{deluxetable}

\subsection{Spectral indices}
\label{subsec:spectral_indices}
\begin{figure*}
    \centering
    \includegraphics[scale=0.34]{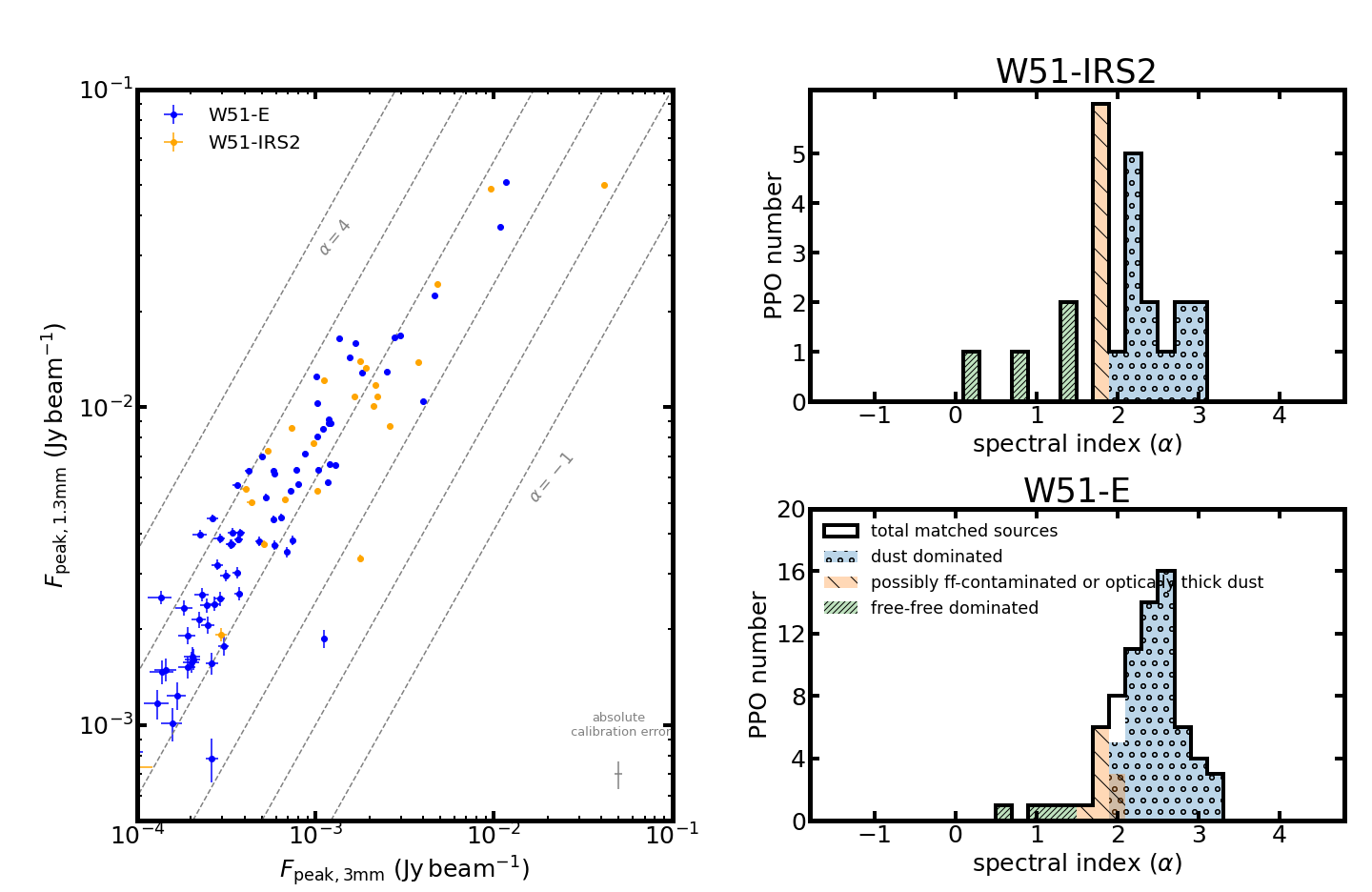}
    \caption{The spectral index distribution of high-res sources detected in both bands. (Left) $3\,{\rm mm}$ peak flux - $1.3\,{\rm mm}$ peak flux diagram. Here, $1.3\,{\rm mm}$ peak flux is measured from the convolved continuum image with the common beam. The error bar denotes the background noise level measured in each continuum image. Typical calibration errors, 5\% in Band 3 and 10\% in Band 6 are marked as gray crosses on the lower right corner. The gray dashed lines indicate the spectral index = [-1, 0, 1, 2, 3, 4] from bottom to top. (Middle, Right) Histograms of spectral index in W51-E and W51-IRS2. High-res sources in each region are grouped depending on whether the spectral index indicates certainly dusty sources or certainly HII regions or possibly free-free emission contamination.}
    \label{fig:alpha}
\end{figure*}
The spectral index, defined as the power-law slope of the SED at specific wavelength intervals, is a useful tool for characterizing the observation targets with photometry. Using the continuum flux of $3\,{\rm mm}$ and $1.3\,{\rm mm}$, we derive the spectral index as
\begin{equation}
    \label{eq:alpha}
    \alpha\equiv\log(F_{3\,{\rm mm}}/F_{1.3\,{\rm mm}})/\log(\nu_{3\,{\rm mm}}/\nu_{1.3\,{\rm mm}}),
\end{equation}
where $F_{3\,{\rm mm}}$ and  $F_{1.3\,{\rm mm}}$ are the flux density in $3\,{\rm mm}$ and $1.3\,{\rm mm}$ and $\nu_{3\,{\rm mm}}$ and $\nu_{1.3\,{\rm mm}}$ are central band frequency which is the intraband flux-weighted average frequency assuming power-law slope (i.e., $F_\nu\propto\nu^\alpha$). For example, the central frequency in $1.3\,{\rm mm}$ is calculated as 
\begin{equation}
    \nu_{1.3\,{\rm mm}}=\int_{1.3\,{\rm mm}} \nu^{\alpha+1} d\nu / \int_{1.3\,{\rm mm}} \nu^{\alpha} d\nu
\end{equation}
When we assume a power-law slope, $\alpha=2$, we obtain the central frequencies, $92.98\,{\rm GHz}$ and $226.69\,{\rm GHz}$ at $1.3\,{\rm mm}$ and $3\,{\rm mm}$, respectively.

The spectral index can be used to classify our samples to first order. When dust emission is optically thick, the SED follows the Rayleigh-Jeans law, $S_\nu\propto\nu^2$, leading to a spectral index $\alpha\sim2$ at mm wavelengths. 
On the other hand, when the dust emission is completely optically thin, the dust power-law index ($\propto\nu^\beta$; $\beta\sim1.75$ for typical ISM; \citealt[][]{ossenkopf94}{}{}) modifies the black body radiation to the spectral slope $\alpha\sim3.7$. Recently, \cite{budaiev24} suggested a model in which an optically thick part plus an optically thin envelope explains the spectral slope $3.7 > \alpha\gtrsim2$ measured for cores between ALMA $3\,{\rm mm}$ and $1.3\,{\rm mm}$ observations. When $T_{\rm dust} \sim 50\,{\rm K}$ is assumed, the spectral index ranges from 1.93 to 3.7 depending on the radius of an optically thick sphere and gas surface number density. Indices $\alpha\sim0$ are most likely to originate from optically-thin HII regions since the spectrum of free-free emission is almost flat over the turnover frequency \citep[e.g.][]{ghavamian98,wilson13}. 

We measured a spectral index using Eq.~\ref{eq:alpha} with the peak flux, which is the brightest pixel value, rather than the integrated flux. In most cases, the intensity from the brightest pixel includes the information along the line-of-sight of the densest part of the sources. For this reason, the peak intensity of each source was chosen and converted to the peak flux density for the spectral index measurement. Given that the beam size is not identical between the two bands, we conducted the convolution of the continuum image to the common beam of the two bands for fair comparison. Note that, the beam of $3\,{\rm mm}$ high-resolution image is always larger than that of $1.3\,{\rm mm}$ and therefore the common beam is identical to the $3\,{\rm mm}$ beam. This means that only the $1.3\,{\rm mm}$ continuum image needs to be convolved with the common beam. The convolution of the image was conducted with the \texttt{astropy} Python package and the common beam size is calculated by \texttt{radio\_beam}\footnote{https://radio-beam.readthedocs.io/en/latest/}. After the convolution, the image array is multiplied by the beam area ratio of Band 3 and $1.3\,{\rm mm}$ to conserve the peak flux value of the unresolved sources. 

The left panel of Fig.~\ref{fig:alpha} displays the peak fluxes extracted from the $3\,{\rm mm}$ and convolved $1.3\,{\rm mm}$ continuum image. We used 73 and 23 cross-matched sources in W51-E and W51-IRS2 for this analysis. The flux error is estimated from the background noise estimated over the area without any compact sources in each continuum image. The upper right and lower right panel of Fig.~\ref{fig:alpha} show the distribution of the spectral indices of fragments. Using spectral indices, we categorize our samples into three groups: dust-dominated sources with a spectral index greater than 2, considered to be dusty sources with $T\gtrsim20\,{\rm K}$ without free-free emission. The optically thick HII regions also can have $\alpha\sim2$ \citep{keto08}, and we will discuss this possibility in a later part of this section. 
 However, the spectral index $\alpha < 2$ can be produced even when the dust emission is optically thick. This occurs when free-free emission is mixed with dust emission or optically thick dust emission has temperature gradient along the line of sight \citep[e.g.][]{li17, galvan-madrid18}. In particular, \cite{galvan-madrid18} showed that the submm spectral index at the Rayleigh-Jeans tail can be as low as 1.5 without changing the dust opacity power-law exponent from the YSO models. Motivated by these studies, we created the second category with the spectral index $1.5 < \alpha < 2$, where the source can be optically thick or free-free contamination possibly exist.   
 The third group consists of sources with spectral indices lower than 1.5, indicating significant contamination from free-free emission. We summarize the number of each group in Table~\ref{tab:ppo}. We remind the reader that the spectral index measurement is carried out for cross-matched samples only. 

In both regions, more than half of the cross-matched samples (69 out of 73 in W51-E and 19 out of 23 in W51-IRS2) are either dust-dominated sources or possible free-free contaminated sources while fewer sources are free-free contaminated sources, indicating that we observe predominantly dusty sources. 
When the flux measurement uncertainties are expanded to include absolute calibration errors—assumed to be 10\% at $1.3\,{\rm mm}$ and 5\% at $3\,{\rm mm}$—the number of dust-dominated sources could vary between 54 and 62 in W51-E and between 10 and 15 in W51-IRS2. Similarly, the number of sources in the second group could range from 7 to 15 in W51-E and from 4 to 10 in W51-IRS2.
The peak of the spectral index distribution around $\alpha\sim2$ in our samples indicates that most of our sources have optically thick centers and are surrounded by optically thin dust parts as suggested in the study of Sagittarius B2 cloud in \cite{budaiev24}.  

On the other hand, we have only a few (4 in each region) matched sources in the free-free contaminated group in each region. The number of free-free contaminated sources does not change even though we consider the calibration error by error propagation--no change in W51-E and possibly $N=4$ to $N=3$ in W51-IRS2. Some of the sources are already known HII regions, including d2 (\#4; e.g. \citealt[][]{ginsburg16}) in W51-IRS2. The recent VLA observation in W51-IRS2 revealed two hyper-compact HII regions with sizes $33\,{\rm AU}$ and $170\,{\rm AU}$ (Deal et al. in prep), and they are also classified as free-free contaminated sources in our catalog ($\alpha=1.44$ and 0.22). Note that, we do not have any sources with $\alpha\lesssim-2$, which would indicate gyrosynchrotron emission from a magnetic field.

The second group includes 10 and 6 sources in W51-E and W51-IRS2, respectively. The three massive YSOs, W51e2e (\#39 in W51-E), W51e8 (\#32 in W51-E), and W51North (\#11 in W51-IRS2) studied in \cite{goddi20} are all include in this group. They all exhibit no sign of free-free emission at longer wavelengths \citep{ginsburg16}.  Therefore, these objects may be so optically thick at $1.3\,{\rm mm}$ that the index is inverted because of the temperature structure, i.e., the optically thick layer is colder at $1.3\,{\rm mm}$ than at $3\,{\rm mm}$ \citep{galvan-madrid18}.


Finally, we discuss the possibility of optically thick H\textsc{ii} regions. Optically thick free-free emission follows the Rayleigh-Jeans law as explained in Eq.~7 of \cite{keto08}, 
\begin{equation}
    S_\nu\approx2kT_e(\nu^2/c^2).
\end{equation}
In this case, the optically thick H\textsc{ii} regions can produce spectral index $\alpha=2$. 
Given a large fraction of unresolved high-res sources ($\sim60$--$80\%$; Sec.~\ref{subsec:sizes}), it is possible to have brightness temperature $T_B$ such that $T_B\eta=T_e$, where $\eta<1$ is a beam-filling factor. Assuming the upper limit radius of possible unresolved hyper-compact H\textsc{ii} (HCH\textsc{ii}) region $R\sim100\,{\rm AU}$ and Lyman continuum luminosity of OB stars $L_{LyC}\gtrsim5\times10^{47}\,s^{-1}$, we obtain electron density $n_e\gtrsim10^7\,{\rm cm^{-3}}$ inside the Str\"{o}mgen radius. At such high density, the confined HCH\textsc{ii} regions are mostly likely to be surrounded by dust; we should still observe dust emission rather than free-free emission even though high-res sources with $\alpha\sim2$ are hyper-compact H\textsc{ii} regions.


\subsection{Sizes}
\label{subsec:sizes}
\begin{figure}
    \centering
    \includegraphics[scale=0.36]{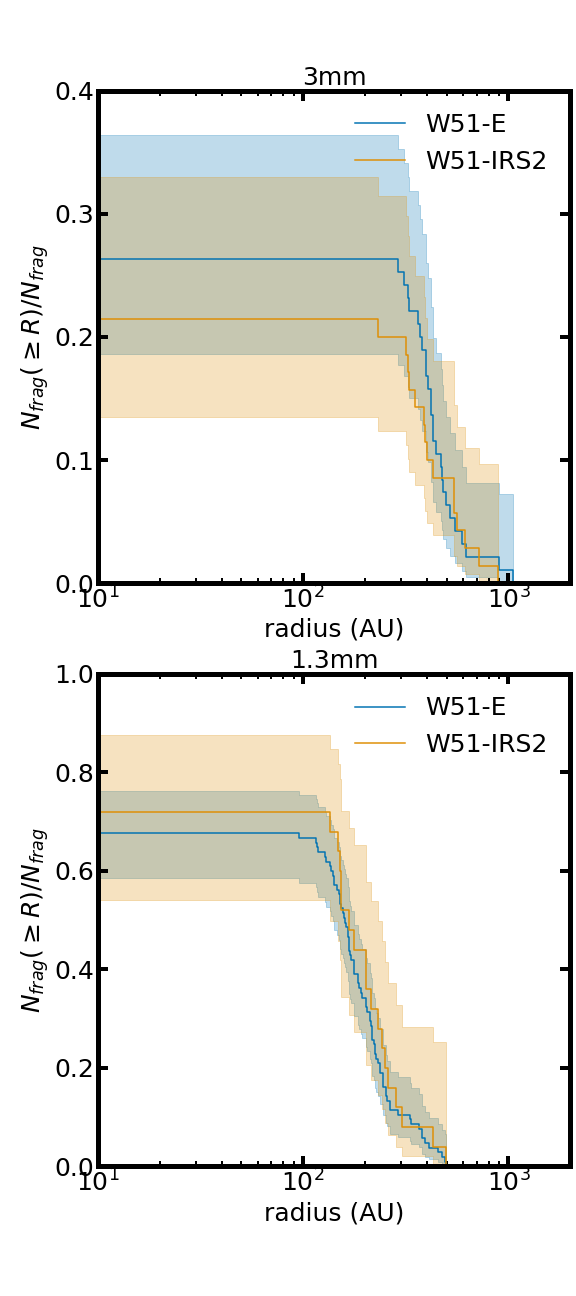}
    \caption{Cumulative distribution of radii of high-res sources. The cumulative function is generated by survival analysis with the Kaplan-Meier estimator provided by \textit{lifeline} package. The function is censored on the left side by excluding the unresolved or undetected sources in the analysis.}
    \label{fig:size}
\end{figure}

For the size of our sources, we utilize the definition of the recent ALMA survey of protoplanetary disks \citep{ansdell18} that the radius includes 95\% of the total flux. When it is adopted in the Gaussian model, the disk radius is equivalent to  $1/(\sqrt{2\ln(2)}) \times {\rm FWHM}$ of the Gaussian model \citep[][]{tobin20}. 
We use the FWHM of the major axis of the Gaussian model produced by our fitting method, \texttt{TGIF} (Appendix.~\ref{appendix:flux}). To estimate the physical sizes, we deconvolve the image beam ellipse from the FWHM ellipse of our 2D Gaussian model. We regarded the failure of deconvolution due to the observed size being smaller than the beam as unresolved status. 

Fig.~\ref{fig:size} shows the cumulative distribution of radii measured from the $1.3\,{\rm mm}$ and $3\,{\rm mm}$ high-resolution images. We perform survival analysis with the Kaplan-Meier estimator implemented in the Python package \texttt{lifelines} \citep[][]{davidson-pilon19}{}{}. Sources that are unresolved or undetected are censored on the left side of the cumulative function. Only $\sim70\%$ and $\sim20$--$30\%$ of sources in $1.3\,{\rm mm}$ and $3\,{\rm mm}$ images, respectively, are identified as resolved. The smaller fraction of resolved source in $3\,{\rm mm}$ images is mainly due to the larger beam size. The resolved sources have a range of size distribution, $\sim200$--$1000\,{\rm AU}$ in $3\,{\rm mm}$ and $\sim100$--$500\,{\rm AU}$ in $1.3\,{\rm mm}$. This tells us the upper limit of the sizes for unresolved sources, $100$ and $200\,{\rm AU}$ in $1.3\,{\rm mm}$ and $3\,{\rm mm}$.



\begin{figure}
    \centering
    \includegraphics[scale=0.39]{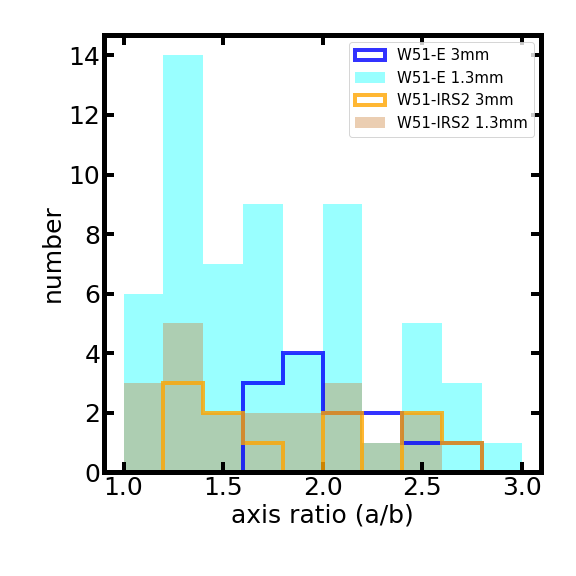}
    \caption{The axis ratio between the major and the minor axis of high-res sources.}
    \label{fig:inclination}
\end{figure}

We present the aspect ratio of high-res sources in Fig.~\ref{fig:inclination}. The distribution of axis ratios, defined as the ratio of the deconvolved major-axis size to the deconvolved minor-axis size, ranges from 1.0 to 2.8.

\subsection{Classification of high-res sources: Pre/Protostellar objects (PPOs)}
\label{subsec:classification}

A large number of sources are unresolved with the upper limit of the size ($\sim100$--$200\,{\rm AU}$) while the maximum sizes of our sources are estimated to be $\sim1000\,{\rm AU}$ in the previous section. Since our sources are smaller than the typical size of cores ($\sim0.05$--$0.1\,{\rm pc}$; \citealt{difrancesco07}; \citealt{motte22}), it is not appropriate to call these compact objects `cores'. 
In addition, the size range of our targets covers that of ``envelopes" ($\sim300$--$3000\,{\rm AU}$) and ``protostellar objects" ($\sim10$--$200\,{\rm AU}$) in \cite{pokhrel18}.

From the spectral index analysis (Sec.~\ref{subsec:spectral_indices}), we determined that most of our sources are dusty, though a few are hyper-compact HII regions exhibiting free-free emission. 
Some of our sources have signs of high-mass star formation such as compact HII regions (e.g., \#4 in W51-IRS2 also known as d2; Appendix.~\ref{appendix:snapshots}) or hot cores. In particular, we confirmed that 9 hot cores in the catalog of \cite{bonfand24} are found in the field of view of both protoclusters. Some of them have prominent molecular outflow, e.g., \#39 in W51-E as known as W51e2e (Appendix.~\ref{appendix:snapshots}; \citealt{goddi20}). However, the presence of protostars cannot be definitively confirmed for the rest of the sources.


Combining these facts together, the high-res sources are mostly dust emission from prestellar or protostellar objects smaller than $\sim100$--$1000\,{\rm AU}$---they are probably either prestellar dust without central protostars or dust disks/envelopes around a protostar. In the latter case, the high-res sources are likely to be Stage 0/I sources with non-negligible envelope mass $M_{env}>0.1\,M_\odot$ \citep{crapsi08, richardson24}. We will use the term ``pre/protostellar objects (PPOs)" to refer to the high-res sources.


\subsection{Temperature constraints through modified blackbody model}
\label{subsec:temp}


\begin{figure*}
    \centering
    \includegraphics[scale=0.35]{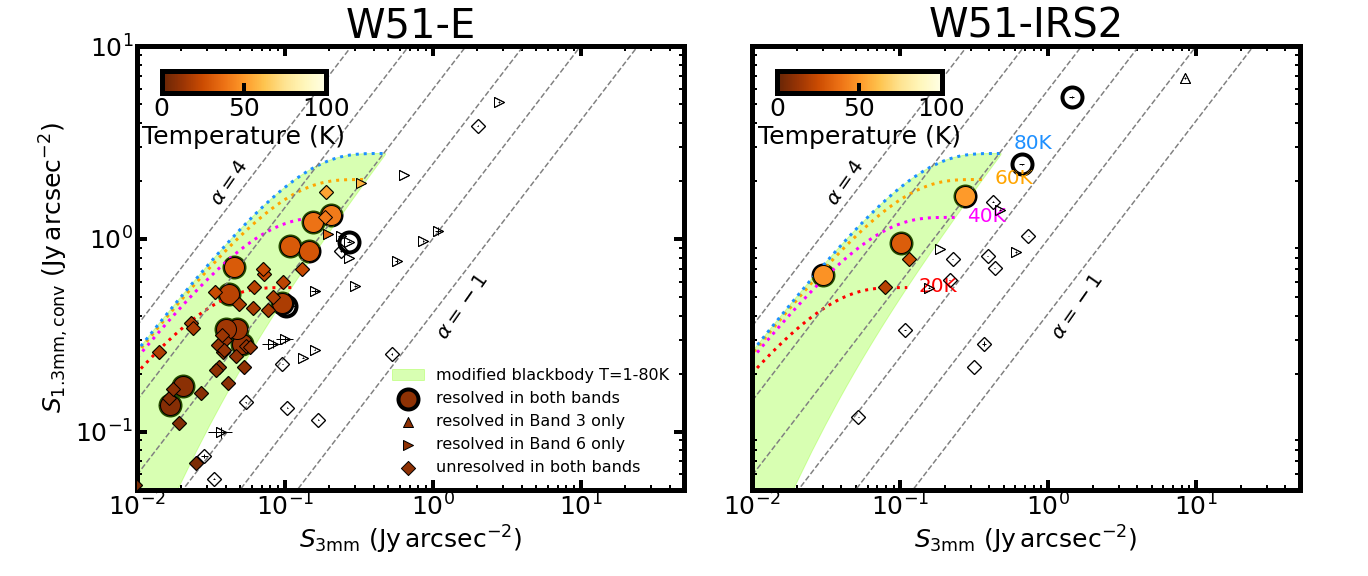}
    \caption{The surface brightness of PPOs at $1.3\,{\rm mm}$ and $3\,{\rm mm}$ observations in W51-E (left) and W51-IRS2 (middle). For $1.3\,{\rm mm}$ the surface brightness is measured from the $1.3\,{\rm mm}$ image convolved to the $3\,{\rm mm}$ image beam. Models of MBB surface brightness with surface density ranging $\Sigma=10^{21}$--$10^{27}\,{\rm cm^{-2}}$ are presented as dashed curves with colors denoting MBB temperature $T=20$, 40, 60, $80\,{\rm K}$, following Eq.~\ref{eq:mbb}. The surface brightness parameter space that MBBs with $T=5$--$400\,{\rm K}$ can occupy is shaded with green colors. We can derive the MBB temperature of the PPOs lying within the green area, and they are color-coded based on the derived MBB temperature. The PPOs outside the green area have empty symbols. Depending on which band the PPOs are resolved, they are represented as different shapes: circle (resolved in both $1.3\,{\rm mm}$ and $3\,{\rm mm}$), triangle pointing up (resolved only in $3\,{\rm mm}$) and pointing right (resolved only in $1.3\,{\rm mm}$), and diamond (unresolved in both bands). Since the upper limits of the size are imposed uniformly for unresolved sources, the derived temperature is a lower limit.  The dashed diagonal lines are spectral index $\alpha=-1$, 0, 1, 2, 3, 4 measured from the integrated fluxes. The brightness temperature corresponding to the surface brightness $1\,{\rm Jy/arcsec^2}$ is $27\,{\rm K}$ at $1.3\,{\rm mm}$ and $153\,{\rm K}$ at $3\,{\rm mm}$.
      }
    \label{fig:temp}
\end{figure*}
\begin{figure}
    \centering
    \includegraphics[scale=0.4]{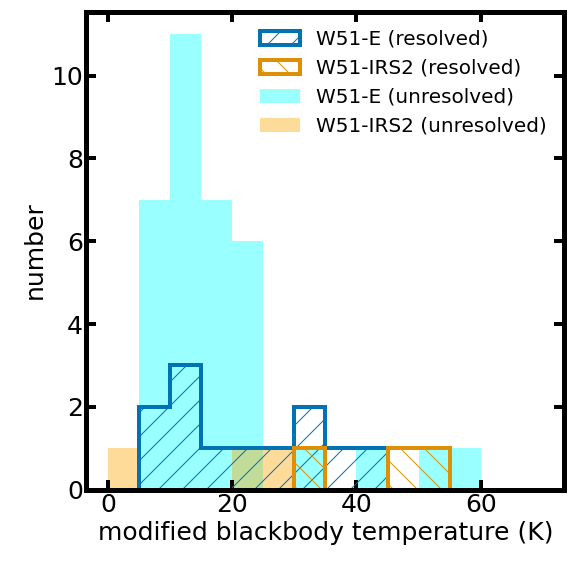}
    \caption{The dust temperatures estimated by the MBB in Fig.~\ref{fig:temp}. The MBB temperatures for unresolved sources are the lower limits.}
    \label{fig:temp_hist}
\end{figure}

\begin{figure*}
    \centering
    \includegraphics[scale=0.4]{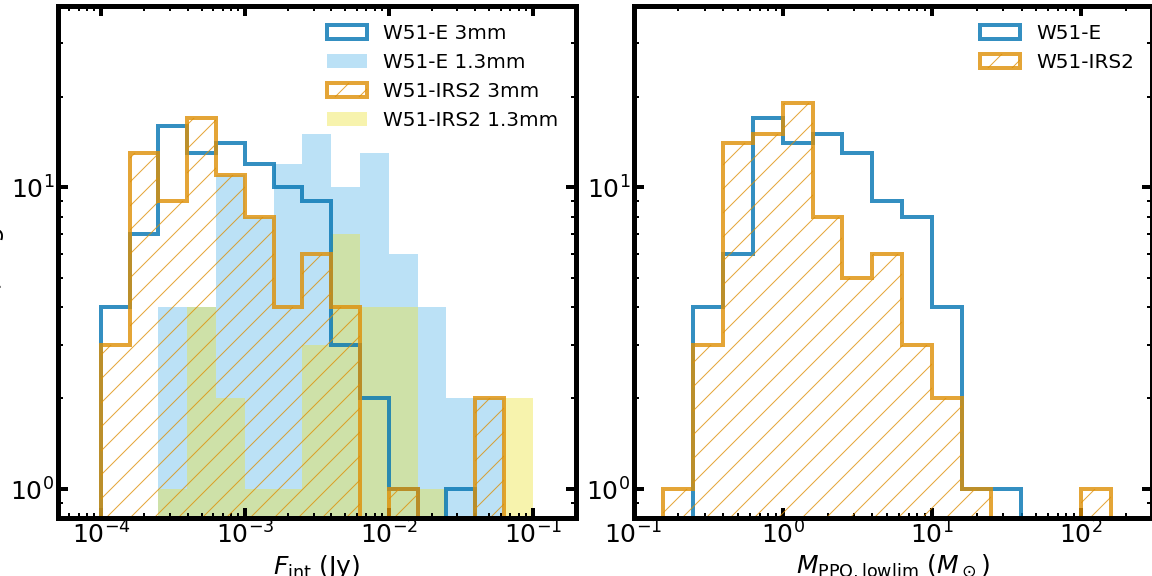}
    \caption{\textit{Left:} The histogram of integrated fluxes of PPOs except sources contaminated by free-free emission at $1.3\,{\rm mm}$ and $3\,{\rm mm}$. \textit{Right:} the lower limit of PPO masses using $3\,{\rm mm}$ integrated flux by assuming constant dust temperature 40 K ($M_{\rm PPO, lowlim}$) is displayed. }
    \label{fig:flux_mass}
\end{figure*}


We used the integrated flux rather than the peak flux to account for the flux emitted by the whole surface of PPOs. The integrated flux is measured by using Gaussian fitting, \texttt{TGIF} \citep[][]{tgif}, illustrated in Appendix~\ref{appendix:flux}. 

Dust temperature is an essential quantity for estimating dust masses from dust emission. Due to the incomplete sampling of dust SED in IR observation at the sub-arcsecond resolution, it is challenging to estimate the dust temperature of PPOs. One approach for dust temperature is to adopt a simple PPO model using modified blackbody radiation (MBB). 
Following \citet{budaiev24}, who analyzed similar-resolution data of the Sgr B2 region in the same ALMA bands, we reproduce the ALMA Band 3 and Band 6 flux of the PPO model consisting of an optically thick core and an optically thin envelope. The model radiates MBB intensity following the equation,
\begin{equation}
\label{eq:mbb}
    B_\nu = \frac{2h\nu^3}{c^2}\frac{1}{e^{h\nu/k_BT}-1}(1-e^{-\kappa_\nu \Sigma})
\end{equation}
where $\nu$ is frequency, $h$ is the Planck constant, $c$ is the speed of light, $k_B$ is the Boltzmann constant, $T$ is dust temperature of optically thick part, e.g. protostars or disk, $\kappa_\nu$ is dust opacity, and $\Sigma$ is column density of optical thin part of PPO. Here, we adopted the dust opacity $\kappa_{1.3\,{\rm mm}}=0.0083\,{\rm cm^2\,g^{-1}}$ and $\kappa_{3\,{\rm mm}}=0.0017\,{\rm cm^2\,g^{-1}}$ from the dust grain with thin ice mantles models with an age of $10^5\,{\rm yr}$ and gas density of $10^{6}\,{\rm cm^{-3}}$ \citep{ossenkopf94}. We allow for $\Sigma$ and $T$ to vary across grids ranging $\Sigma=10^{21}$--$10^{27}\,{\rm g\,cm^{-2}}$ and $T=5$--$400\,{\rm K}$ to account for various conditions of PPOs.

Fig.~\ref{fig:temp} shows the dust temperature estimated from the MBB surface brightness. We converted the observed flux density into the surface brightness in order to match those with the predicted MBB surface brightness regardless of their sizes. Here, $1.3\,{\rm mm}$ surface brightness is estimated from  $1.3\,{\rm mm}$ image convolved to the common beam, that is, $3\,{\rm mm}$ image beam to take into account different angular resolutions. For unresolved sources (Sec.~\ref{subsec:sizes}), we assign their sizes as the minimum size of the resolved PPOs and convert their flux density to surface brightness. In this case, the derived temperature should be a lower limit as the size is an upper limit. A significant number of cross-matched PPOs (26 out of 73 in W51-E and 17 out of 23 in W51-IRS2) do not have a viable temperature estimate as they lie outside of the range of surface brightness at both wavelengths spanned by the model. Even though the sources are resolved at both bands, there are still some sources of which the spectral index is too low to estimate the MBB temperature in Fig.~\ref{fig:temp}. These are the cases where the dust emission is contaminated by free-free emission or the optically thick dust emission at each band is derived from different temperature as explained in Sec.~\ref{subsec:spectral_indices}.
We display the spectral index measured from the integrated flux in Fig.~\ref{fig:temp}. This reflects the averaged optical depth in the whole area of PPOs whereas the spectral index using the peak flux in Sec.~\ref{subsec:spectral_indices} more accurately describes the optical depth of the central region of PPOs. 

The inferred dust temperature from MBB is distributed over $5$--$60\,{\rm K}$ (Fig.~\ref{fig:temp_hist}). Note that, this includes a number of dust temperature lower limits from unresolved sources. 

\subsection{ Constraints to PPO masses}
\label{subsec:masses}
In the previous section, we found that a large portion of our sample is dominated by optically thick dust emission, so column densities (or masses) are lower limits. Uncertainties in $T_{\rm dust}$ also affect mass estimates \citep[e.g.][]{richardson24}{}{}.   


From the dust continuum flux, the dust mass can be inferred by assuming optically thin dust emission \citep[][]{hildebrand83}{}{},
\begin{equation}
    M=\frac{S_\nu D^2}{\kappa_\nu B_\nu(T)}
    \label{eq:mass}
\end{equation}
where $S_\nu$ is flux density, $D$ is a distance, $\kappa_\nu$ is dust opacity, and $B_\nu(T)$ is Planck function at specific temperature.
For optically thin sources, this is a good estimate of the total mass if we have a measurement of the temperature. For optically thick sources, this mass estimate is a lower limit since we do not detect light from all of the mass.  
We expect that most of the objects in our sample are Stage 0/I, in which a substantial dust envelope surrounds a central protostar. Under this assumption, the mass we measure is that of the dust envelope, not the star.

To constrain the PPO masses, we adopt $40\,{\rm K}$ dust temperature. This is motivated by the recent YSO model study that demonstrated that the choice of dust temperature $T=40\,{\rm K}$, warmer than the commonly adopted temperature $T=20\,{\rm K}$, is more accurate in inferring the true mass with optically thin dust emission assumption \citep{richardson24}. Although the average of the MBB temperatures in W51-E and W51-IRS2 is slightly lower than $40\,{\rm K}$, this is a reasonable choice considering that a large fraction of MBB temperature estimates are lower limits for unresolved PPOs. When PPOs are associated with hot cores, we assign the same temperature as hot core temperatures ($100\,{\rm K}$ or $300\,{\rm K}$; Sec.~\ref{subsec:frag_coremass}) instead of $40\,{\rm K}$ considering their heating effects. In W51-E, 4 PPOs (\#30, \#31, \#32, \#33) are associated with a $100\,{\rm K}$ hot core and 2 (\#39, \#40) with a $300\,{\rm K}$ hot core, while in W51-IRS2, 16 PPOs (\#0, \#5, \#7, \#10, \#14, \#17, \#20, \#21, \#27, \#39, \#46, \#47, \#48, \#55, \#56, \#74) correspond spatially to $100\,{\rm K}$ hot cores and 6 (\#11, \#13, \#25, \#26, \#36, \#38) to a $300\,{\rm K}$ hot core. 

In Fig.~\ref{fig:flux_mass}, we display the integrated flux in $1.3\,{\rm mm}$ and $3\,{\rm mm}$. The fluxes of free-free contaminated sources are excluded because the mass estimation could be significantly deviated from true mass when the dust emission is contaminated by the free-free emission. Fig.~\ref{fig:flux_mass} includes a larger sample than Fig.~\ref{fig:temp} as Fig.~\ref{fig:flux_mass} displays even sources detected in only one band. In the middle and right panel of Fig.~\ref{fig:flux_mass}, we display the lower limits of mass with constant temperature. We use $3\,{\rm mm}$ flux density to derive the mass because the lower opacity in $3\,{\rm mm}$ than $1.3\,{\rm mm}$ allows us to obtain the information from the inner part of PPOs. The mass lower limits from the constant temperature model span $\sim0.1$--$200\,M_\odot$. 


We would like to caution readers against the interpretation that the distribution of the mass lower limits mimics the PPO mass function or the IMF. Several major uncertain factors--temperature, the central stellar mass, and the mass ratio of optically thin/thick dust in fragments--make the predictions of the total PPO mass almost opaque \citep[e.g.][]{padoan23}. Therefore, we will not attempt to build a PPO mass function in this paper. For a more accurate mass estimation, alternative ways such as using a Keplerian disk are required in future studies.

\section{Core Fragmentation}
\label{sec:fragmentation}
In this section, we link cores to PPOs. We will use the term \textit{fragment} to refer to PPOs that are spatially associated to cores. We inspect how the characteristics of core fragmentation, such as the number of fragments and fragmentation efficiency, vary with core masses and temperature. 

\subsection{Overview of fragmentation}
\label{subsec:overview_frag}

We associate our identified PPOs from Section~\ref{sec:YSO candidate_characterization} with cores by noting when a PPO lies within a boundary of cores.  The core boundary is defined as core FWHM extended slightly further by the size of the ALMA-IMF beam. The core FWHM is provided by the ALMA-IMF core catalog \citep[][]{louvet24} where \texttt{getsf} was used (Sec.~\ref{subsec:obs_core}).

\begin{figure*}
    \centering
    \includegraphics[scale=1]{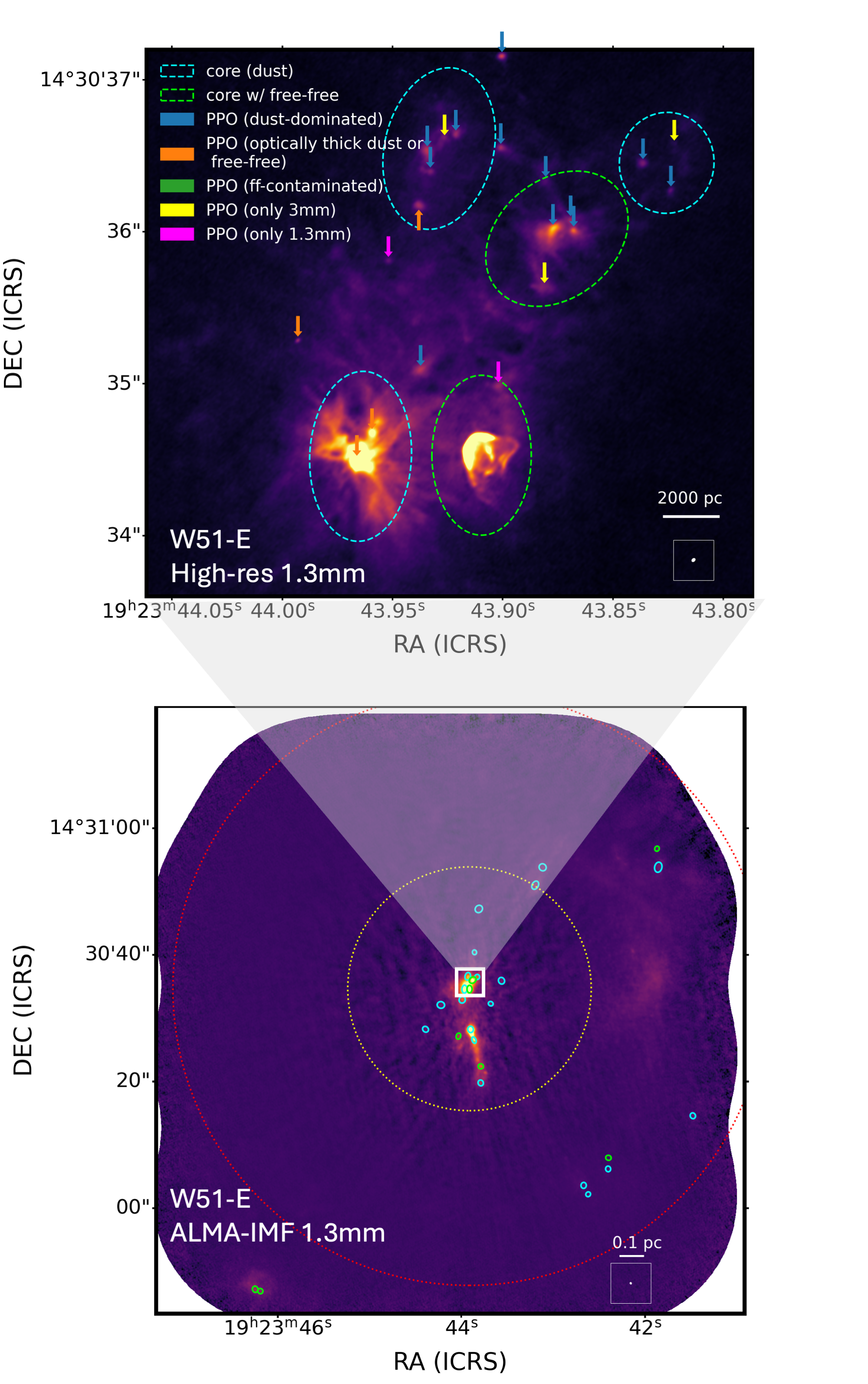}
    \caption{Examples of core fragmentation in W51-E. The cyan ellipses are dust cores and green ellipses are cores contaminated with free-free emission. The size of the ellipses represents the boundary of cores defined as the sum of the FWHM of the core and the FWHM of the synthesized beam. The background image of the bottom is an ALMA-IMF $1.3\,{\rm mm}$ continuum map. The field of view of the high-resolution image is displayed with yellow ($1.3\,{\rm mm}$) and red ($3\,{\rm mm}$) circles on top of the ALMA-IMF map. The region enclosed by white squares in the bottom image is magnified with a background image of a high-resolution Band 6 continuum on the top. The locations of PPOs are indicated by the arrows. The PPOs detected in both bands have arrows with blue (dust-dominiated), orange (possibly free-free contaminated), and green (free-free contaminated) colors. The yellow and magenta arrow indicates an object identified only in $3\,{\rm mm}$ and $1.3\,{\rm mm}$, respectively.  }
    \label{fig:frag_w51e}
\end{figure*}
\begin{figure*}
    \centering
    \includegraphics[scale=1.2]{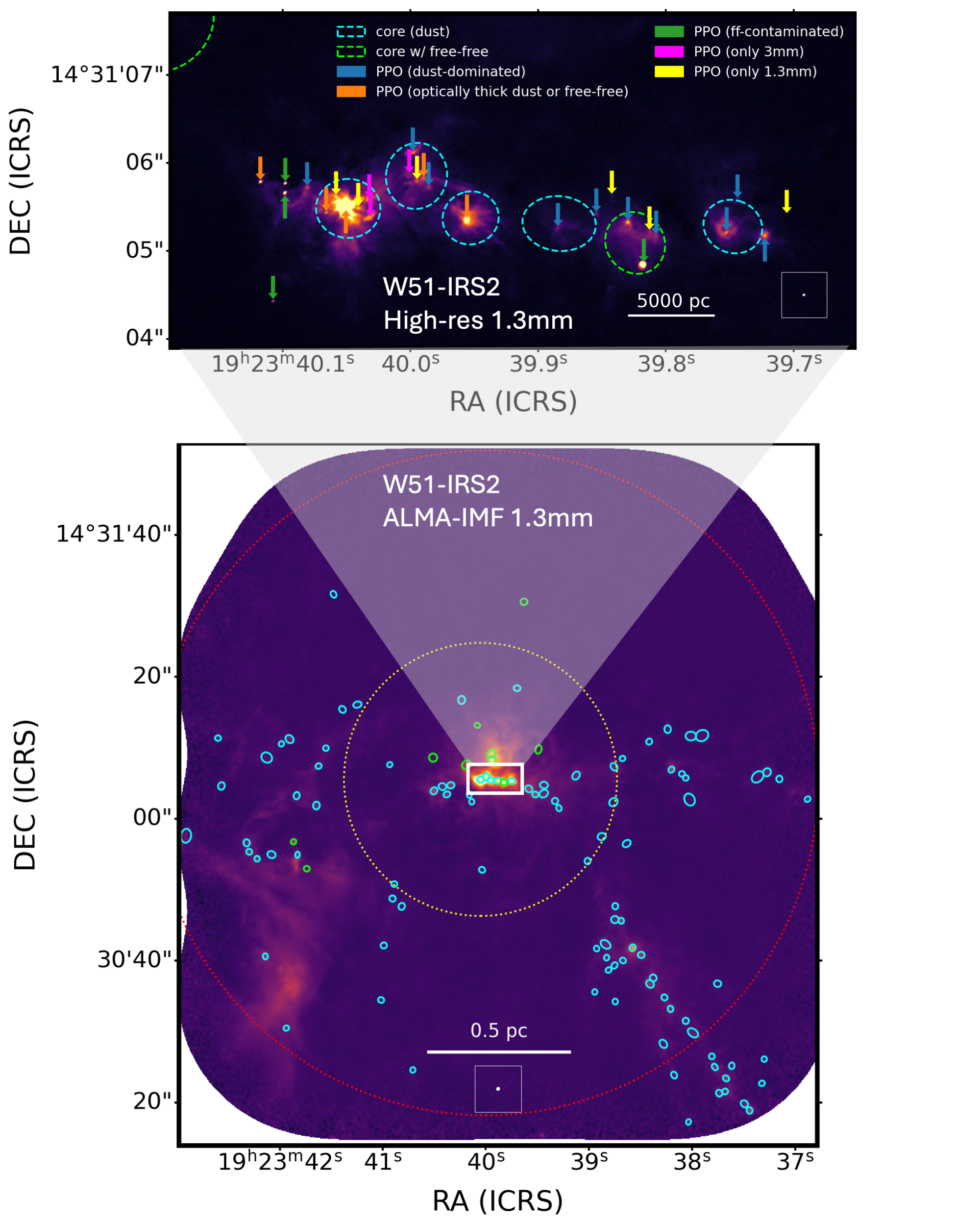}
    \caption{The same as Fig.~\ref{fig:frag_w51e} for W51-IRS2 region.} 
    \label{fig:frag_w51n}
\end{figure*}

Fig.~\ref{fig:frag_w51e} and Fig.~\ref{fig:frag_w51n} present examples of core fragmentation into PPOs in W51-E and W51-IRS2. In each image, the locations of PPOs and cores are plotted over the continuum image. In W51-E, most of the cores and PPOs are concentrated in the small region ($\sim0.4\,{\rm pc} \times \sim0.7\,{\rm pc}$) at the central dust structure. On the other hand, cores and PPOs in W51-IRS2 are more widely distributed along the neighboring dust filamentary structure. Matching the positions of cores and fragments reveals that not all the cores harbor fragments inside their boundaries; some cores have single or multiple fragments, but others do not. In this paper, we categorize cores without fragments as \textit{unfragmented} and cores containing fragments as \textit{fragmented}. 

\begin{figure}
    \centering
    \includegraphics[scale=0.36]{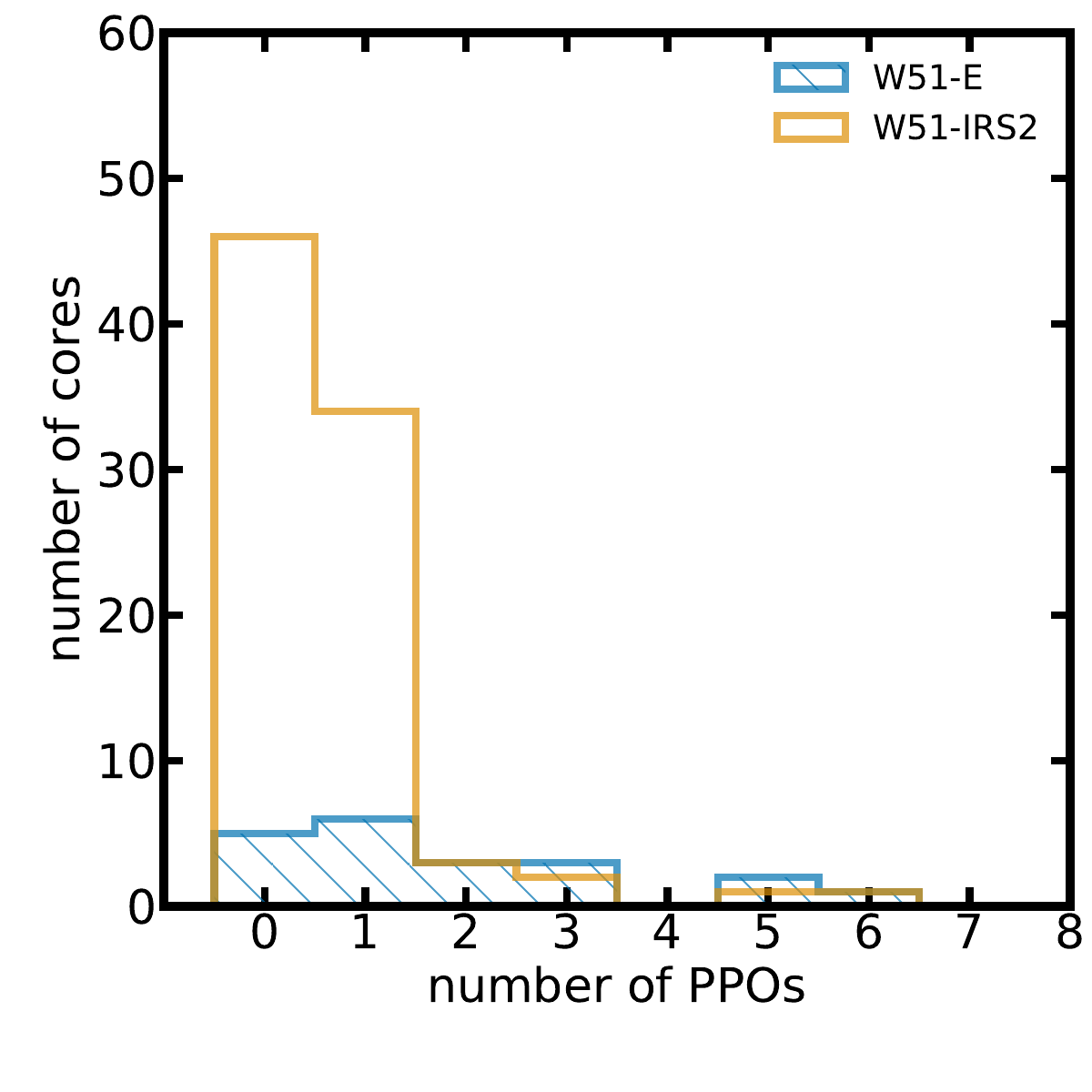}
    \caption{Histogram of the number of core fragments from W51-E (blue) and W51-IRS2 (orange). The cores without free-free emission in the same field of view of high-resolution $3\,{\rm mm}$ images are used in the analysis.}
    \label{fig:frag_count}
\end{figure}


The number of core fragments is displayed in Fig.~\ref{fig:frag_count}. We selected only those cores found in the same field of view as the high-resolution $3\,{\rm mm}$ image as it has a larger field of view than $1.3\,{\rm mm}$ image. This results in selecting 20 and 87 cores in W51-E and W51-IRS2 (group A in Fig.~\ref{fig:core_subsample}), respectively. In W51-E and W51-IRS2, 15/20 and 41/87 cores have fragments, respectively. In particular, W51-E has only five unfragmented cores. This is probably due to the higher completeness limit for the ALMA-IMF maps in W51-E ($3.86\,M_\odot$) than W51-IRS2 ($1.64\,M_\odot$) \citep{louvet24}. Another possible reason will be discussed in Sec.~\ref{subsec:yso_outside_cores}. 

Among the fragmented cores, the number of fragments is not uniform. Within 15 and 41 fragmented cores in W51-E and W51-IRS2, 6 and 34 cores exhibit a single fragment, and 9 and 7 cores have multiple fragments, respectively. 
The most highly fragmented cores have 6 fragments in both regions.

It is worth noting that fragments of massive cores are particularly surrounded by dust structures. For example, \#10 and \#11 in W51-IRS2, \#32 and \#39 in W51-E show similar morphology connected or surrounded by dust lanes (see Appendix.~\ref{appendix:snapshots}). They are all the most massive fragments in their parent cores. In fact, three of them (\#11 in W51-IRS2 and \#32, \#39 in W51-E) were suggested to have accretion flow in earlier study \citep{goddi20}. They are probably the progenitors of high-mass stars that are actively accreting material from cores. The temperatures of parent cores for \#11 in W51-IRS2 and \#39 in W51-E are assigned as $300\,{\rm K}$ (Sec.~\ref{subsec:frag_coremass}).
\begin{figure*}
    \centering
    \includegraphics[scale=0.28]{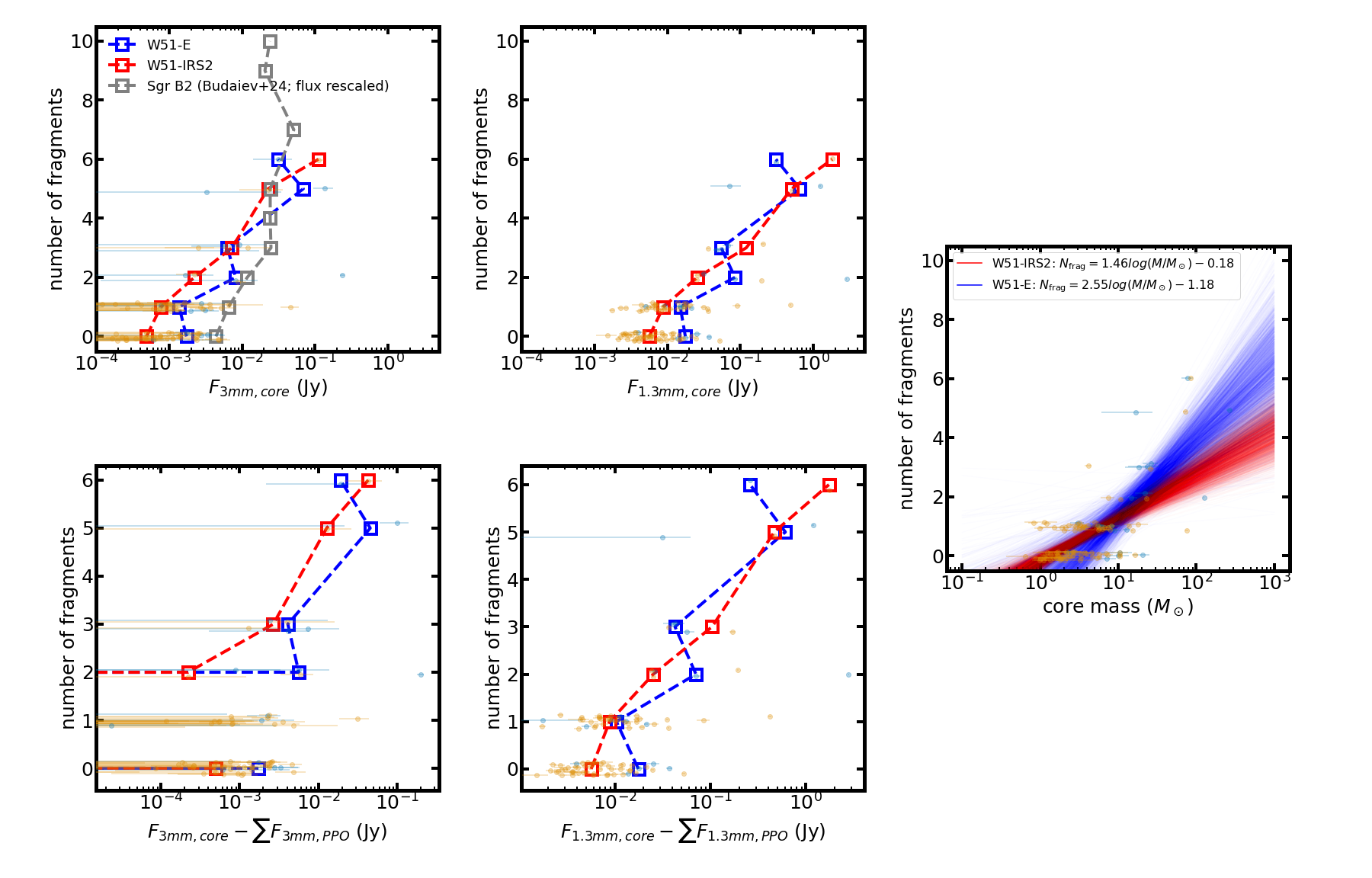}
    \caption{Number of core fragments as a function of core fluxes (\textit{upper row}), core fluxes subtracted by their fragment fluxes (\textit{lower row}) and masses (\textit{right}). Each cyan (W51-E) and orange (W51-IRS2) data point with an error bar represents the flux or the mass of the core versus the number of fragments. For core fluxes, we use the integrated flux measured from ALMA-IMF 1.3mm and 3mm continuum data. The flux measured from another high-mass star-forming region, Sgr B2 \citep{budaiev24}, is rescaled to the distance of W51, $5.4\,{\rm kpc}$, and the rescaled flux and the number of fragments are displayed for comparison. For the number of PPOs, a small random deviation on the y-axis is given to differentiate each other visually. The medians of fluxes and masses for each PPO number bin on the y-axis are plotted with square marks and connected with dashed lines. The horizontal dashed lines and some absence of squares in the lower left panel are due to the negative medians of core residual flux at $3\,{\rm mm}$ within the range of uncertainties.   Each solid line represents 100 MCMC runs of the linear fits on $\log M_{\rm core}$-$N_{\rm frag}$. The legend displays the median of the slope and the y-intercept of these MCMC runs.}
 
    \label{fig:frag_flux}
\end{figure*}

\begin{figure}
    \centering
    \includegraphics[scale=0.32]{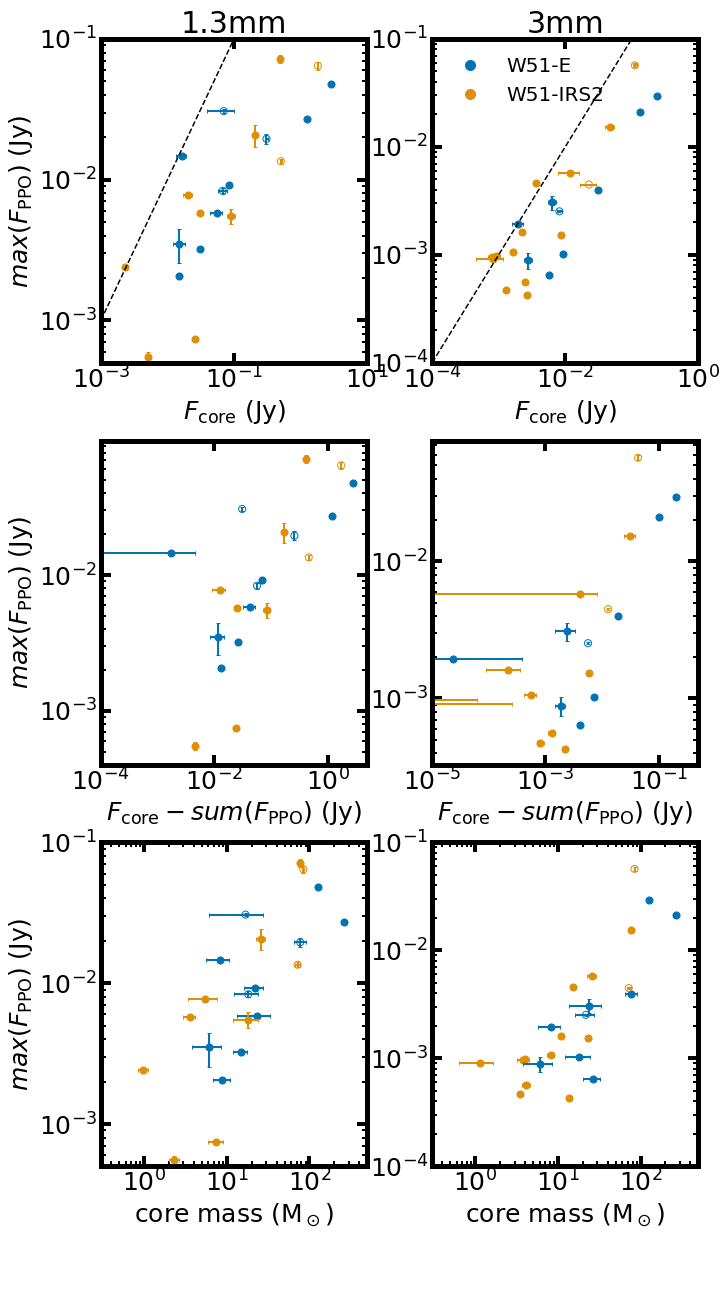}
    \caption{The fluxes of the brightest PPOs associated with each core as a function of $1.3\,{\rm mm}$ and $3\,{\rm mm}$ core fluxes (\textit{top}), core fluxes subtracted by the sum of their fragment fluxes (\textit{middle}), and core masses (\textit{bottom}). The error bars denote the error of the core fluxes and that of the integrated PPO fluxes. If the integrated flux of more than one PPO is missing in the plotted band, the PPO fluxes are represented as empty circles; otherwise, they are shown as filled circles.  The dashed lines in the top row indicate the 1-to-1 line.}
    \label{fig:ysofluxmax}
\end{figure}

\begin{figure}
    \centering
    \includegraphics[scale=0.34]{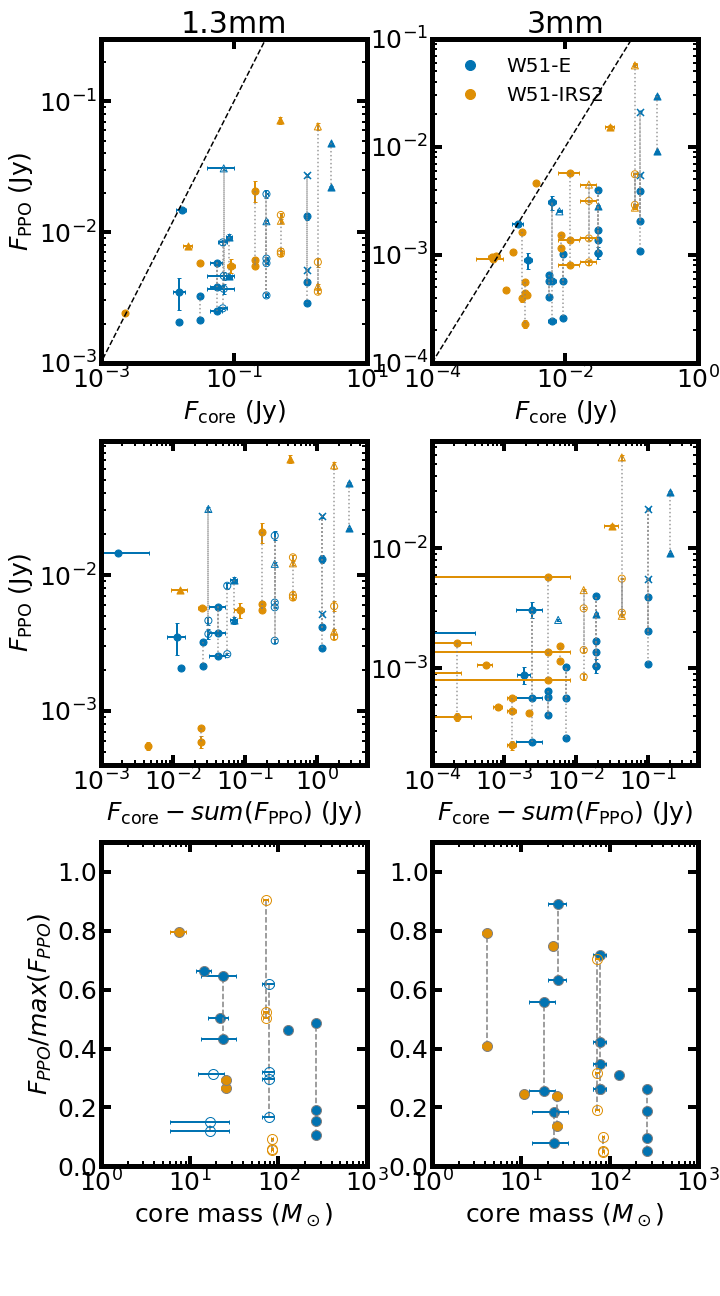}
    \caption{ \textit{Top} and \textit{middle}: The $1.3\,{\rm mm}$ and $3\,{\rm mm}$ fluxes of all PPOs as a function of their parent core fluxes (\textit{top}) and core envelope fluxes (\textit{middle}). \textit{Bottom}: The PPO flux ratios normalized to the brightest PPO flux.  For clarity, the flux of the brightest PPO in each core is omitted. Only cores in which there are at least 2 fragments are displayed in this plot. In all panels, the fluxes of PPO associated with the same parent core are connected with the vertical dashed lines.} 
    \label{fig:ysofluxall}
\end{figure}

\subsection{Mass dependence of fragmentation}
\label{subsec:frag_coremass}

 How each core fragments into individual stars is one of the keys to understanding the evolution of the CMF into the IMF. In this section, we analyze the number of PPOs in each core as a function of core mass.
 
 The core masses were computed by \cite{louvet24} who considered the optical thickness of $1.3\,{\rm mm}$ dust emission \citep[Eq.~6 of][]{pouteau22},
 \begin{equation}
 \label{eq:opt_thick_mass}
     M=-\frac{\Omega_{\rm 1.3mm, beam}d^2}{\kappa_{\rm 1.3mm}}\frac{S^{\rm int}_{\rm 1.3mm}}{S^{\rm peak}_{\rm 1.3mm}}\ln(1-\frac{S^{\rm peak}_{\rm 1.3mm}}{\Omega_{\rm beam}B_{\rm 1.3mm}(T_{\rm dust})})
 \end{equation}
 where $\Omega_{\rm 1.3mm, beam}$ is the solid angle of the $1.3\,{\rm mm}$ ALMA-IMF image beam, $d=5.4\,{\rm kpc}$ is the distance to W51, $\kappa_{\rm 1.3mm}$ is the dust opacity at $1.3\,{\rm mm}$, $S^{\rm int}_{\rm 1.3mm}$ is $1.3\,{\rm mm}$ integrated flux density, $S^{\rm peak}_{\rm 1.3mm}$ is $1.3\,{\rm mm}$ peak flux, and $B_{\rm 1.3mm}(T_{\rm dust}$) is the blackbody radiation intensity at $1.3\,{\rm mm}$. For the core temperature in W51-E and W51-IRS2, \cite{louvet24} used the Bayesian PPMAP \citep{marsh15, dellova24} method to infer core dust temperature from SED fitting except for hot core candidates. For 1 W51-E and 5 W51-IRS2 cores associated with hot cores candidates found in \cite{bonfand24} where ${\rm CH_3OCHO}$ map is used, $100\,{\rm K}$ dust temperature is adopted. In addition, $300\,{\rm K}$ dust temperature is specifically assigned for cores, \#2 in W51-E and \#1 in W51-IRS2, motivated by the molecular gas temperature models in \cite{ginsburg17} and \cite{goddi20}. Note that, we used the same dust opacity with the one used in the PPO mass lower limits for consistency. This results in by a factor of $\sim1.3$ larger core masses compared to those estimation in \cite{louvet24} where the dust opacity is computed as $\kappa=0.1(\nu/1000\,{\rm GHz})^{1.5}\,{\rm cm^2\,g^{-1}}$.

\subsubsection{Cores with high fragment numbers are preferentially massive}
\label{subsubsec:nfrag_coremass}
We find that \textit{cores with high fragment numbers are preferentially massive.}
Fig.~\ref{fig:frag_flux} shows the number of PPOs as a function of core fluxes and masses measured at $1.3\,{\rm mm}$ and $3\,{\rm mm}$. Since the flux of a core implicitly includes the contributions from its fragments, we also present the core flux with the summed fluxes of its fragments subtracted. This residual flux indicates the emission from the core envelope, independent of the fluxes from the embedded fragments. We use core group A for $1.3\,{\rm mm}$ and core mass plot, and group C for $3\,{\rm mm}$ plots.  

The number of fragments shows positive correlation with core fluxes at $1.3\,{\rm mm}$ and $3\,{\rm mm}$ in general. The Spearman correlation coefficients between the logarithmic flux and the number of fragments of W51-E and W51-IRS2 are 0.733 (p-value=$2.37\times10^{-4}$) and 0.466 (p-value=$5.42\times10^{-6}$) at $1.3\,{\rm mm}$ and 0.669 (p-value=$3.32\times10^{-3}$) and 0.312 (p-value=$9.94\times10^{-3}$) at $3\,{\rm mm}$, respectively. This relation persists even when the sum of fragment flux is removed from the core flux. The Spearman coefficients of W51-E and W51-IRS2 are 0.663 (p-value=$1.42\times10^{-3}$) and 0.483 (p-value=$2.37\times10^{-6}$) at $1.3\,{\rm mm}$, and 0.618 (p-value=$1.41\times10^{-2}$) and 0.277 (p-value=$3.86\times10^{-2}$) at $3\,{\rm mm}$. Even though the residual fluxes have weaker correlation with the number of fragments, p-values are still low ($<0.05$), indicating that the trend is not an output of the interdependence of cores and fragment fluxes. At $3\,{\rm mm}$, the core residual flux sometimes becomes negative as the sum of the PPO flux can exceed the core flux within uncertainties, which is not physical (Sec.~\ref{subsec:frag_eff}). This mainly reduces the degree of the correlation between the core residual flux and the number of fragments. 

Overall, the Spearman test gives lower coefficients and p-values at the same time for W51-IRS2 than W51-E. The lower coefficient can be explained by the substantial number of unfragmented cores (58\%) in W51-IRS2 compared to that in W51-E (25\%), which contributes to the large scatter at the bottom along the horizontal direction. On the other hand, the lower p-values are simply attributed to the larger size of the total core samples (N=99 for $1.3\,{\rm mm}$) than that of W51-E (N=20 for $1.3\,{\rm mm}$). 

All the cores with multiple fragments ($N_{\rm frag}\geq2$) are more massive than $\sim4\,M_\odot$. However, massive cores do not all have a high number of fragments. Instead, massive cores with $M_{\rm core}\gtrsim10\,M_\odot$ have a wide range of the number of fragments from $N_{\rm frag}=0$ to 6. The general trend in W51 is also observed in another high-mass star-forming region, Sgr B2 \citep{budaiev24}.
This trend is therefore likely to be a common characteristic of core fragmentation in high-mass star-forming regions.


We provide a fit model of the relationship between core mass and the number of fragments. We estimated the linear fit on $\log M_{\rm core}$-$N_{\rm frag}$ by using MCMC runs. We utilized the Python package \texttt{linmix} \citep{kelly07} to run 100 MCMC for W51-E and W51-IRS2 data, respectively.  We obtain the medians of slopes and intercepts, 1.46 and -0.18 in W51-IRS2, and 2.55 and -1.18 in W51-E. The shaded regions in the right panel of Fig.~\ref{fig:flux_mass} display 1$\sigma$ range of slopes (1.85 to 3.16 in W51-E and 1.23 to 1.68 in W51-IRS2) and intercepts (-0.33 to -0.02 in W51-E and -1.98 to -0.28 in W51-IRS2) of the linear fits produced by MCMC.



\subsubsection{Bright PPOs reside in bright cores at $1.3\,{\rm mm}$ and $3\,{\rm mm}$}
\label{subsubsec:maxfragflux_coreflux}

We also found that \textit{the bright PPOs, or possible massive stars, are preferentially formed in the bright (massive) cores.} When we relate the fluxes of the brightest PPO in each core to the core fluxes (Fig.~\ref{fig:ysofluxmax}), there is a clear correlation between the two quantities in both regions and both wavelengths. This holds when the fluxes of brightest PPOs are compared to the residual flux of cores ($F_{core}-sum(F_{PPO})$).  In Eq.~\ref{eq:mbb}, a large amount of flux is produced when temperature and/or column density is high. Both conditions are usually found in massive PPOs where massive dust disks or envelopes and significant heating from massive protostars are expected. Therefore, it is reasonable to assume that the brightest PPOs are the most massive ones. This trend was also found at a larger scale, e.g. clump-to-core fragmentation ($\sim0.3$--$1\,{\rm pc}$ to $\sim0.06$--$0.1\,{\rm pc}$; e.g. \citealt[][]{csengeri17,lin19}; c.f. \citealt[][]{morii23}).

\subsubsection{Fragments within a given core have a range of fluxes}
\label{subsubsec:epsilon_coremass}

 Another finding from PPO fluxes is that \textit{the flux distribution between fragments is not uniform} (Fig.~\ref{fig:ysofluxall}).  The top and middle panel of Fig.~\ref{fig:ysofluxall} show a wide range of flux distribution of all the fragments as a function of core flux or core residual flux. The lower panels of Fig.~\ref{fig:ysofluxall} provide more details about the flux distribution by showing the flux ratio of PPOs normalized to the flux of the brightest PPO. We only select cores with $N_{frag}\ge2$. Cores in both regions show a broad range of flux distribution between 0 to 1. 
This result suggests that masses of core fragments cannot be simply predicted by a single characteristic mass, e.g., Jeans mass. A possible explanation is the different mass growth after the initial fragmentation in each fragment, which will be discussed in Sec.~\ref{subsubsec:lowMJ}.




\subsection{Fragmentation efficiency and thermal suppression of fragmentation}
\label{subsec:frag_eff}

During core fragmentation, not all the core mass is transferred to fragments. The fraction of core mass contained in fragments has been usually referred to as the core formation efficiency \citep[e.g.][]{bontemps10,louvet14, palau15, diazgonzalez23}{}{}. By analogy, but using the flux instead of mass, we define the fragmentation efficiency ($\epsilon_{\rm frag}$) by comparing the flux of fragments and the parent core. 

\begin{figure}
    \centering
    \includegraphics[scale=0.34]{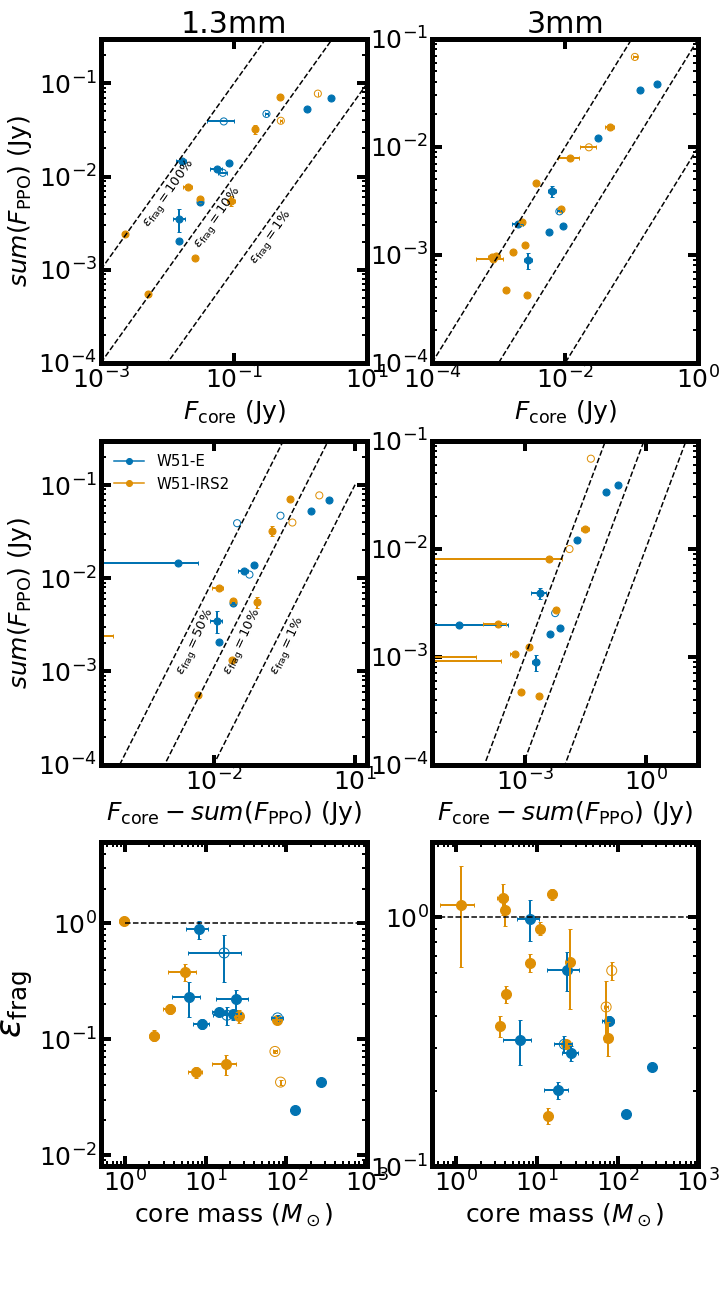}
    \caption{ \textit{Top and Middle}: The sum of fragment fluxes as a function of core fluxes (\textit{top}) and core envelope fluxes (\textit{middle}) in $1.3\,{\rm mm}$ and $3\,{\rm mm}$. If the integrated flux of more than one PPO is missing in the plotted band, the PPO fluxes are represented as empty circles; otherwise, they are shown as filled circles. The diagonal dashed lines mark the fragmentation efficiency $\epsilon_{frag}=1, 0.1, 0.01$ in the \textit{top} row and $\epsilon_{frag}=0.5, 0.1, 0.01$ in the \textit{middle} row. \textit{Bottom}: Fragmentation efficiencies ($\epsilon_{\rm frag}$) as a function of core masses. The fragmentation efficiency is defined as the ratio between the sum of all the PPO fluxes and core fluxes in each band. The horizontal dashed line marks $\epsilon_{\rm frag}=1$. Some cores with $\epsilon > 1$ are not physically plausible and are instead attributed to measurement uncertainties. }
    \label{fig:epsilon_frag}
\end{figure}

We compare the sum of PPO fluxes ($sum(F_{\rm PPO})$) within a core with the flux of the parent core ($F_{\rm core}$) in the upper panels of Fig.~\ref{fig:epsilon_frag}. The ratio between two quantities indicates the fragmentation efficiency as denoted by the dashed lines. A few sources have $\epsilon_{\rm frag}>1$, which is unphysical, and it is mainly a result of the measurement errors. 
In the lower panels of Fig.~\ref{fig:epsilon_frag}, we compare $\epsilon_{\rm frag}$ and core masses. The data points are absent in the right upper (high $\epsilon_{frag}$, high core mass) and lower left (low $\epsilon_{frag}$, low core mass) corners. The absence in the lower left corner can be attributed to either the detection limits in our images or the intrinsic anti-correlation between core mass and $\epsilon_{frag}$. On the other hand, the upper right corner cannot be explained by the observational effects, suggesting that massive cores may have low fragmentation efficiency.

\begin{figure*}
    \centering
    \includegraphics[scale=0.28]{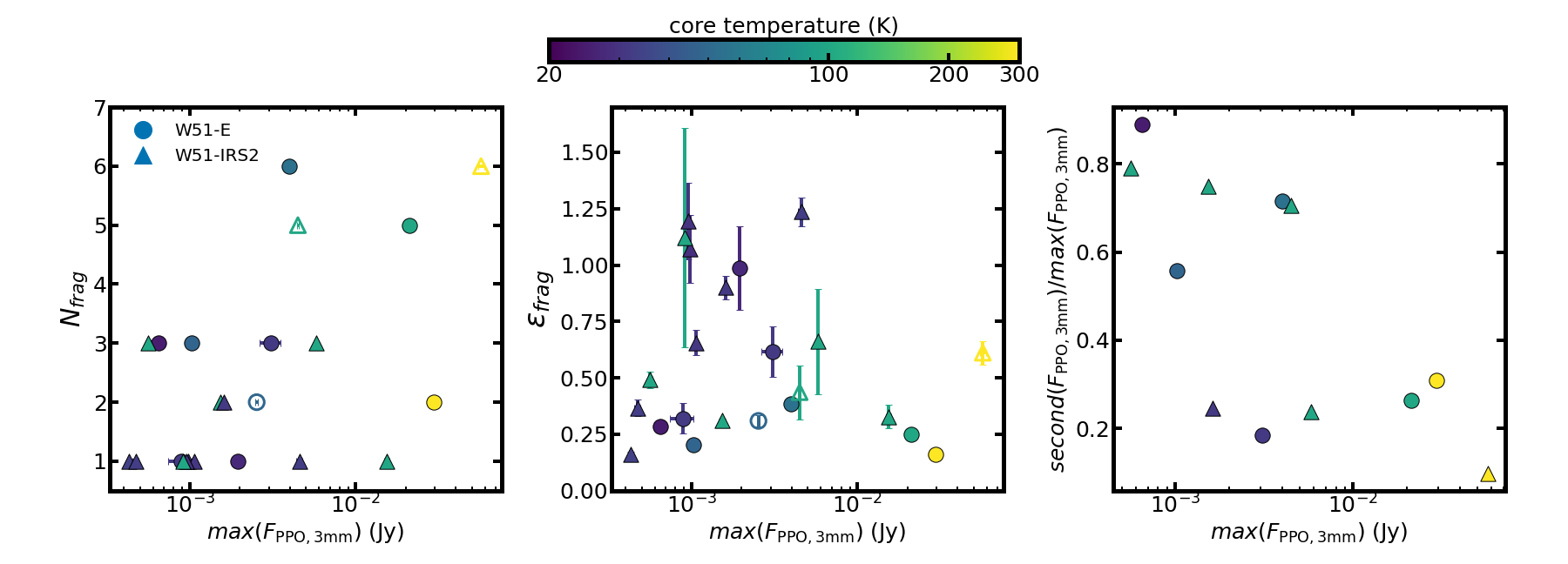}
    \caption{The fragmentation efficiency ($\epsilon_{\rm frag}$; left), the number of fragments ($N_{\rm frag}$; middle) and the ratio between the brightest and the second brightest PPOs ($second(F_{\rm PPO})/max(F_{\rm PPO})$; right) as a function of the brightest PPO flux ($max(F_{\rm PPO})$) at $3\,{\rm mm}$ in W51-E and W51-IRS2. The color of each data point denotes core temperature. The filled and empty circles denote the complete and incomplete samples as same as Fig.~\ref{fig:ysofluxall}. }
    \label{fig:epsilon_nfrag}
\end{figure*}

We find a suggestive trend that the cores hosting faint PPOs are more likely to exhibit high fragmentation efficiency, although the sample is not statistically large enough. In Fig.~\ref{fig:epsilon_nfrag}, $\epsilon_{\rm frag}$ (left panel) and $N_{\rm frag}$ (middle panel) as a function of the brightest PPO flux at $3\,{\rm mm}$ is shown. The color of each data point represents the core temperature. Overall, cores forming brighter PPOs have higher temperatures, suggesting that a massive protostar might be the main heating source. In particular, there are four cores hosting bright PPO $max(F_{PPO,3mm})\gtrsim10^{-2}\,{\rm Jy}$ have fragmentation efficiency limited to $\epsilon_{\rm frag}\lesssim0.7$, while cores with faint PPOs span a broader range. Furthermore, these four cores showing possible signatures of strong thermal feedback are all massive ($\sim100$--$300\,M_\odot$) and associated with hot core candidates identified in \cite{bonfand24}. The broad range of $\epsilon_{\rm frag}$ for other cores is probably because some of these cores still have ongoing mass accretion from the core gas reservoir without feedback of massive protostar. However, we cannot rule out that this trend is due to the low number of samples ($N=4$) hosting brightest PPOs with $max(F_{PPO,3mm})\gtrsim10^{-2}\,{\rm Jy}$. To test if the trend is statistically meaningful, we conduct Fisher's exact test with a 2x2 contingency table of four core groups based on threshold of fragmentation efficiency, $\epsilon_{\rm frag}>0.7$ and the flux of the brightest PPO, $max(F_{PPO,3mm})\gtrsim10^{-2}\,{\rm Jy}$. The test for a possible trend that cores hosting bright PPO exhibit lower fragmentation efficiency yields a p-value of 1.0, indicating no statistical significance. Conversely, the inverse test—whether cores with less bright PPO have higher fragmentation efficiency—gives a p-value of 0.27, suggesting that the observed trend can be made from random sampling by 27\% chance. Therefore, more samples are required to examine our hypothesis.




The number of fragments ($N_{\rm frag}$) does not have a clear correlation with $max(F_{\rm PPO})$. Rather, $N_{\rm frag}$ is more closely correlated with the core masses as in Fig.~\ref{fig:frag_flux}. This may simply reflect the fact that protostellar heating is not strong enough to regulate the fragmentation in the whole core.  



Lastly, we inspect the impact of the suppression of fragmentation on the growth of massive stars. In the right panel of Fig.~\ref{fig:epsilon_nfrag}, we compare $3\,{\rm mm}$ flux of the brightest PPOs and the second brightest PPOs. We found that there is a more drastic difference in fluxes between the brightest PPOs and the second brightest PPOs when a core hosts a massive protostar. In particular, the cores with $max(F_{PPO,3mm})\gtrsim10^{-2}\,{\rm Jy}$ show a high mass ratio between the most massive and the second massive fragment ($second(F_{PPO, 3mm})/max(F_{PP0,3mm})\lesssim0.4$). 
This can be interpreted as the process by which massive protostars grow---radiative feedback from massive protostars effectively regulates the accretion flow from fragmentation \citep{krumholz07a} and hinders the growth of other neighboring protostars. This idea was previously suggested as ``enforced isolation" in \cite{ginsburg17} where massive stars grow in isolation without fragmentation in the vicinity. However, again, the number of samples is too small (N=3) to rule out the null hypothesis.

\section{Discussion}\label{sec:discussion}

In this section, we compare the number of fragments predicted from Jeans mass with the observed number of fragments to assess the impact of thermal pressure in supporting cores.  We also classify our core catalog into unfragmented and fragmented sources and compare the physical properties of these populations (Sec.~\ref{subsec:prestellar_vs_protostellar}). Finally, we attempt to characterize the PPOs found outside of core boundaries. 





\subsection{Jeans analysis}
\label{subsec:jeans}
\begin{figure*}
    \centering
    \includegraphics[scale=0.45]{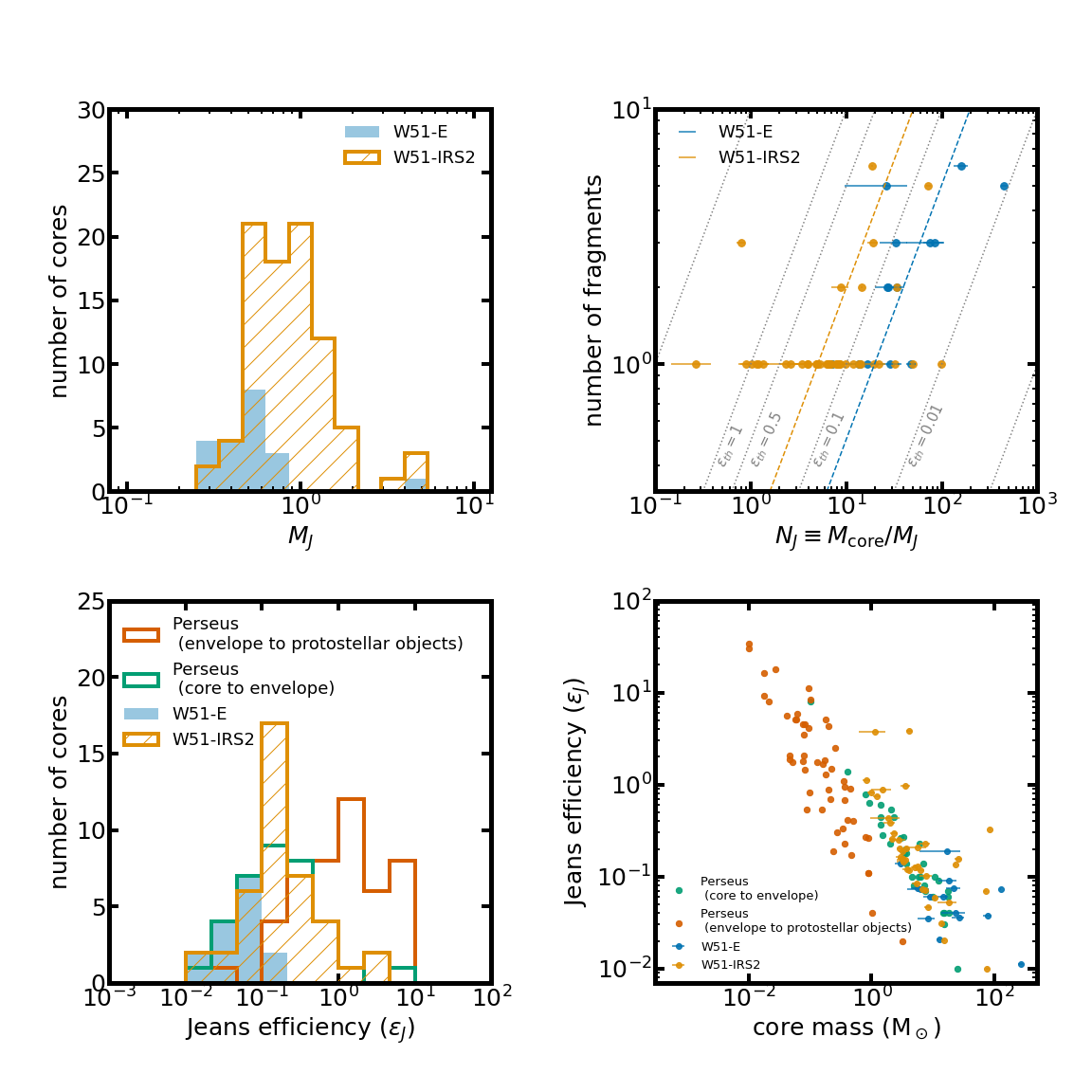}
    \caption{\textit{Upper left}: Thermal Jeans masses of cores in W51-E and W51-IRS2. \textit{Upper right}: Jeans number and number of fragments of cores. The ratio of the actual number of fragments and Jeans number is Jeans efficiency indicated as gray dashed lines. The average Jeans efficiency in each region is plotted as colored thin dashed lines. \textit{Lower left}: Histogram of Jeans efficiency. For comparison, Jeans efficiency of envelopes in Perseus \citep{pokhrel18} is also displayed in green. The averages of Jeans efficiency in W51-E and W51-IRS2 are 5\% and 20\%, respectively, which are lower than the average in Perseus ($\sim40\%$ for core to envelope and $\sim50\%$ for envelope to protostellar object). \textit{Lower right}: Core masses and Jeans efficiency of W51-E, W51-IRS2, and Perseus \citep{pokhrel18}. }
    \label{fig:jeans}
\end{figure*}

\begin{figure}
    \centering
    \includegraphics[scale=0.4]{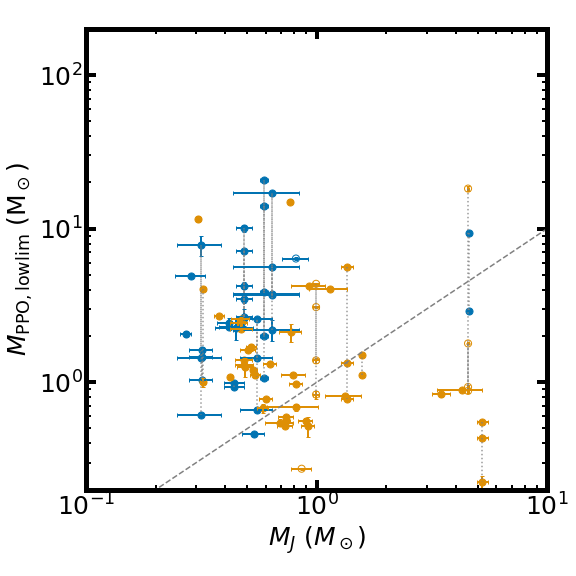}
    \caption{Comparison between the Jeans masses of cores and the PPO masses associated with each core estimated from constant temperature model ($T=40\,{\rm K}$). The PPOs associated with the same core are connected with the vertical dashed lines. A majority of PPOs exhibit a lower mass limit exceeding the Jeans mass of the core. }
    \label{fig:jeansYSO}
\end{figure}

The number of potential fragments can be simply predicted by assuming thermal/non-thermal Jeans fragmentation of cores. For instance, \cite{pokhrel18}
utilized the Jeans number defined as the object mass divided by Jeans mass ($N_J\equiv M_{\rm core}/M_J$) to infer the expected number of fragments.  We also measure the Jeans efficiency, representing the ratio of actual fragments to the Jeans number ($\epsilon_J\equiv N_{\rm frag}/N_J$). This ratio therefore tells us how effectively thermal fragmentation, if present, operates on each core.

Following the description in \cite{pokhrel18}, we measured Jeans masses of cores using the equation \citep{binney87},
\begin{equation}
    M_J = \frac{\pi^{5/2}}{6G^{3/2}}c_s^3\rho^{-1/2}
    \label{eq:Jeans}
\end{equation}
where $G$ is gravitational constant, $\rho$ is density, and $c_s$ is the sound speed. The density of the core is derived from the FWHM size of cores estimated by \texttt{getsf} and the mass estimated by Eq.~\ref{eq:opt_thick_mass}. The sound speed is computed based on the PPMAP temperature of the cores. We note that only thermal pressure is considered in the analysis.
Thermal Jeans masses of cores measured from Eq.~\ref{eq:Jeans} have a range of $\sim0.2$--$6\,M_\odot$  (Fig.~\ref{fig:jeans}). Interestingly, Jeans masses of cores in W51-IRS2 (on average $1.03\,M_\odot$) are systematically higher than those in W51-E (on average $0.69\,M_\odot$) because of their relatively low density ($\rho_{avg}=2.9\times10^{-17}\,{\rm g\,cm^{-3}}$) compared to those in W51-E ($\rho_{avg}=1.0\times10^{-17}\,{\rm g\,cm^{-3}}$). 

Based on the Jeans masses, the Jeans number and the number of fragments in W51-E and W51-IRS2 are estimated and shown in the upper right panel of Fig.~\ref{fig:jeans}. Note that, in the log-scale plot, unfragmented cores (with $N_{frag}=0$) are not displayed. Most of the Jeans number are below unity, meaning that fewer fragments are observed than what is expected from thermal Jeans masses.

 The lower left panel of Fig.~\ref{fig:jeans} shows the Jeans efficiency of both regions. The average Jeans efficiency is estimated to be 0.05 and 0.20 in W51-E and W51-IRS2, respectively, which are far below unity. The Jeans efficiency tends to be lower in high-mass cores (lower right panel of Fig.~\ref{fig:jeans}). This explains the relatively lower $\epsilon_J$ in W51-E compared to W51-IRS2 since the average mass of cores used in this analysis is higher in W51-E ($34\,M_\odot$) than W51-IRS2 ($7\,M_\odot$).

In fact, a substantial number of fragments are more massive than the thermal Jeans mass (Fig.~\ref{fig:jeansYSO}). We derived the (lower limits of) PPO mass based on constant temperature $T=40\,{\rm K}$ in Sec.~\ref{subsec:masses}. A majority of fragment masses are up to $\sim20$ times higher than Jeans masses of their parent cores, indicating again that thermal Jeans masses are not sufficient to explain the masses of core fragments.

\subsubsection{Comparison to other studies}
\label{subsubsec:comp_to_others}
Since we followed the Jeans analysis in \cite{pokhrel18}, we can directly compare our results with the Perseus region (Fig.~\ref{fig:jeans}).  We compare core-to-PPO fragmentation in W51 with the two different spatial scales of fragmentation, core-to-envelope and envelope-to-protostellar objects fragmentation in \cite{pokhrel18}. The spatial scales of cores and PPOs in this study are $\sim2000$--$6000\,{\rm AU}$ and $\lesssim100$--$500\,{\rm AU}$ (Sec.~\ref{subsec:sizes}), respectively. On the other hand, core samples in \cite{pokhrel18} have sizes of $0.05$--$0.1\,{\rm pc}$ which is a factor of 2 larger than our core samples. The envelopes and protostellar objects have $\sim300$--$3000\,{\rm AU}$, $10$--$200\,{\rm AU}$, respectively, and thus our PPOs are in between these two scales. 

The average Jeans efficiencies in Perseus cores and envelopes were reported as 40\% and 50\%, respectively, which are larger than the average Jeans efficiencies in W51-E (5\%) and W51-IRS2 (20\%). \cite{pokhrel18} discussed that there might be a mass dependency of Jeans efficiency; this trend is confirmed in the lower right panel of Fig.~\ref{fig:jeans} where Perseus and the two W51 protoclusters both display massive cores with low Jeans efficiency. 

Prior to this study, Jeans analysis of W51-IRS2 at core-to-PPOs scale was carried out by \cite{tang22} where the same long base-line ALMA visibility data of W51-IRS2 was used but short base-line data was not combined. In \cite{tang22}, the "small-scale ensembles" were defined as a group of fragments at a scale of $\sim4000\,{\rm AU}$ which is almost the same with the size of cores in this study. 
 For each small-scale ensemble, Jeans numbers were not high enough to explain the observed number of fragments, but they reached a different conclusion that thermal fragmentation is dominant in W51-IRS2 by assuming core formation efficiency (CFE) of 24\% from the equation, $N_J=M_{core}{\rm CFE}/M_J$. In other words, thermal pressure would explain the observed number of fragments if 76\% of core (``small-scale ensemble" in their study) mass is not fragmenting. However, uniform CFE is not applicable to our data given the broad range of fragmentation efficiency (Fig.~\ref{fig:epsilon_frag}). 
 
 Moreover, different fragment identification scheme results in different number of fragments. \cite{tang22} used $12\sigma$ contour where $\sigma=0.11\,{\rm mJy/beam}$ and the peaks were verified by \texttt{dendrogram} with paramters of \textit{min\_value}=$1.32\,{\rm mJy/beam}$ and \textit{min\_delta}=$0.011\,{\rm mJy/beam}$. Notably, \textit{min\_delta} is much lower than the value adopted in this study, $1.5\sigma\sim0.1{\rm mJy/beam}$, resulting in higher number of fragments identified in their analysis. In \#11 PPO in W51-IRS2, also known as W51north (SE1 in \citealt{tang22}), 20 fragments were identified in \cite{tang22}, whereas 6 PPOs are found in this study. In particular, there is a huge difference in the fragment counts at the central continuous structure around W51north, where 10 fragments were reported in \cite{tang22}, in contrast to a single PPO recognized in this study. We take a more conservative approach as the peaks in the continuous structure may represent transient substructure rather than embedded fragments. The higher abundance of the fragments within the core results in an overestimation of the required CFE. The overestimation can be amplified by the fact that the short baseline data were not included; the missing flux at large scale would lead to decreasing $M_{core}/M_J\propto\rho^{3/2}$, and thereby raising the required CFE to reconcile with the observed fragment count. 

 The different fragment counts around W51north may suggest that the result is dependent on the choice of \texttt{dendrogram} parameters. However, the sum of all the fragment masses only accounts for $\sim20$\% of the core mass in \cite{tang22},  indicating that our conclusion regarding the low fragmentation efficiency of massive cores—presented in Fig.~\ref{fig:epsilon_frag}—remains valid despite differences in the number of identified fragments.

\subsubsection{Discussion on insufficient Jeans mass}
\label{subsubsec:lowMJ}
Several possibilities may account for the low efficiency of fragmentation in W51A: unresolved smaller fragments, underestimated Jeans mass, insufficient thermal pressure, and growth of PPO mass.


\cite{offner23} listed candidates for additional physics halting the cascade of fragmentation, such as angular momentum and magnetic support, and concluded that these limit the minimum scale of core fragmentation to $\sim100\,{\rm AU}$. This scale is comparable to the sizes of most of our PPOs at $1.3\,{\rm mm}$ (Sec.~\ref{subsec:sizes}), implying that we are most likely observing the final products of core fragmentation. In other words, each individual object in our sample is not likely to be composed of multiple spatially-unresolved fragments that resulted from core fragmentation. We therefore exclude the possibility that we have misidentified a collection of multiple fragments as a single object, which would have resulted in the measured mass exceeding the Jeans mass ($M>M_J$).  

 Another explanation is that the Jeans masses of cores evaluated from current physical properties might be underestimated. Two factors could lead to underestimated Jeans mass: the core temperature is estimated at low resolution, and the current density may have increased from the initial density \citep[e.g.][]{xu24}. The core temperature used in our Jeans masses is derived from PPMAP analysis \citep[][]{marsh15, dellova24} with $2.5"$ resolution, which is larger than the typical size of cores ($\sim1"$). Thus, it may underestimate core temperature in the cores heated by central protostars \citep{motte25}. However, cores at the time of the initial fragmentation, before the formation of the central protostar, are less likely influenced by protostellar heating, and thus the core temperature from low resolution may play a minor role. On the other hand, there is a chance that the current density is overestimated if the cores evolved significantly from the initial fragmentation stage. Indeed, several fragmentation models suggested that the parent structures continue to collapse while they form dense fragments within themselves \citep[e.g.][]{guszejnov15, guszejnov16, vazquez-semadeni19}. Since Jeans mass has a negative power of density, the higher density of evolved cores could result in the underestimation of Jeans mass.   

 In particular, the growth in masses by gas accretion from the parent core could be another reason for the observed $M_{\rm frag}>M_{\rm J}$. Indeed, \cite{goddi20} suggested that multi-directional accretion flows exist around hot cores in W51 such as W51e2e (\#39 in W51-E), W51e8 (\#32 in W51-E), and W51north (\#11 in W51-IRS2) which are all fragments of cores identified in this study. If we assume that the initial fragment masses are solely determined by the Jeans masses, the mass lower limits of fragments $M_{\rm frag}\simeq0.1$--$200\,M_\odot$ in the constant temperature model (Fig.~\ref{fig:jeansYSO}) means that many fragments should accrete most of their masses. The mass infall rates estimated by \cite{goddi20} for the three hot cores based on the outflow rates are $\sim1$--$5\times10^{-3}\,M_\odot/{\rm yr}$. Assuming constant infall rates and 50\,{\rm K} dust temperature, the mass lower limits of the three fragments, 57, 42, and $92\,M_\odot$, require the ages for gas infall $t\sim1$--$5\times10^{4}\,{\rm yr}$. 
 Hence, the scenario that almost all the masses of fragments are acquired through accretion cannot be ruled out. 

 Given that the Jeans masses used in this study only consider thermal pressure, fragments more massive than their Jeans masses in Fig.~\ref{fig:jeansYSO} may imply that thermal pressure is insufficient to support the fragment against gravity. If this is true, we need other types of support, for example, turbulence, magnetic support, and rotation to account for lower Jeans efficiency but none of them has been confirmed as a robust agent for non-thermal support at a scale of $\sim100\,{\rm AU}$.

 While none of these explanations provide conclusive evidence, our finding remains open to multiple, particularly the last two interpretations -- the non-thermal pressure and the mass growth after the initial fragmentation. The suggested explanations might not be exclusive to each other. For example, the magnetic stabilization in accreting flow around massive PPOs can aid further growth of massive PPOs \citep[][]{koch22}.

\subsection{Unfragmented/fragmented cores}
\label{subsec:prestellar_vs_protostellar}
As illustrated in Sec.~\ref{subsec:overview_frag}, we classified our core datasets into unfragmented and fragmented cores by the presence of fragments. 

\begin{figure*}
    \centering
    \includegraphics[scale=0.28]{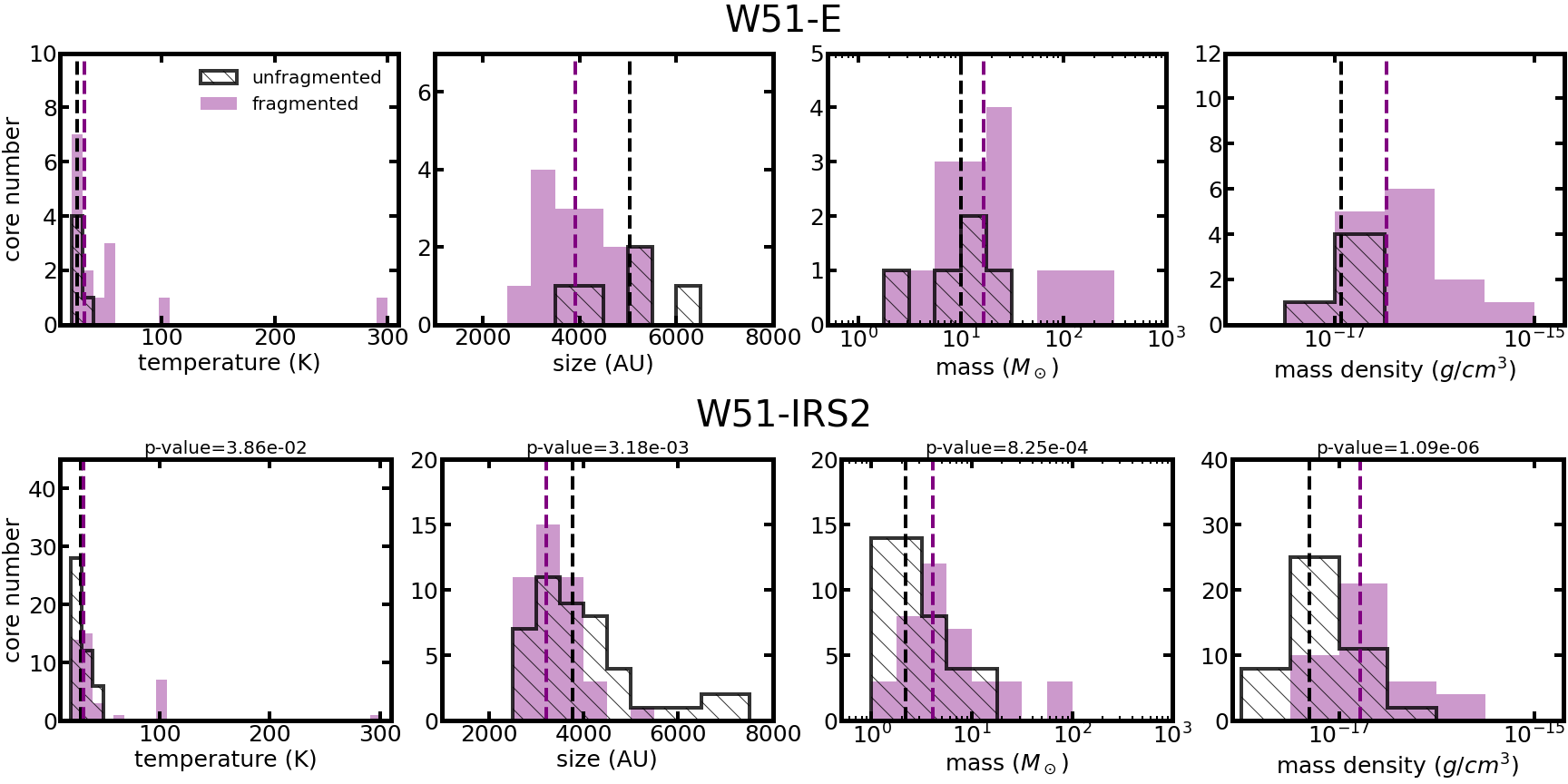}
    \caption{Histogram of temperature, size, mass, and mass density of unfragmented and fragmented cores from W51-E (top), and W51-IRS2 (bottom). The medians of each quantity are denoted by the vertical dashed lines. The p-values from the two-sample KS test for W51-IRS2 cores are displayed on the bottom row.}
    \label{fig:proto_vs_pre}
\end{figure*}

 We present the temperature, size, and mass distribution of unfragmented and fragmented cores without free-free emission  (Fig.~\ref{fig:proto_vs_pre}). For the core temperature and mass, we illustrated the derivation in Sec.~\ref{subsec:frag_coremass}. The size of cores is an average of FWHM of \texttt{getsf} source in major and minor axis, $\theta_{\rm avg}\equiv({a_{\rm FWHM}\times b_{\rm FWHM}^2})^{\frac{1}{3}}$. In W51-E, there are only 5 unfragmented cores and thus, W51-E is not suitable for unfragmented/fragmented core comparison. For this reason, we limit our comparison to the cores in W51-IRS2. 
 
 In W51-IRS2, the difference between core temperatures is small. Although the Kolmogorov-Smirnov (KS) test between two populations yields a p-value of $3.86\times10^{-2}$, the medians of core temperature are $29\,{\rm K}$ and $31\,{\rm K}$ in unfragmented and fragmented cores, respectively. 
 This is probably because either the PPOs associated with the fragmented cores have not yet formed a protostellar system or the protostellar luminosity is not strong enough to heat the dusty region over $2.5''$ PPMAP resolution (equivalent to the physical scale $\sim10^4\,{\rm AU}$).

 We find that the median of the size distribution is larger for unfragmented cores than for fragmented cores in W51-IRS2. On the other hand, the median mass of unfragmented cores is lower than that of fragmented cores. 
 Moreover, systematic differences exist in mass density between unfragmented and fragmented cores--fragmented cores are denser than unfragmented cores. We evaluate the p-value from the two-population KS test, $3.18\times10^{-3}$, $8.25\times10^{-4}$, and $1.09\times10^{-6}$ for size, mass, and mass density, respectively. Therefore, those three physical quantities of the two populations are statistically distinct from each other.

 Multiple interpretations of smaller, more massive, and denser fragmented cores are possible. Assuming that cores are independent self-gravitating objects generated from the density structure by turbulence \citep[e.g.][]{mckeetan02,mckee03}{}{} and that all the cores are formed at the same time, this trend can be simply a result of the faster collapse of initial denser cores due to their shorter free-fall time ($t_{ff}\propto\rho^{-1/2}$). Another view is that cores grow in mass--more massive fragmented cores are the result of net increase in mass as they evolve \citep[e.g.][]{nony23, armante24}. In fact, gas infall onto dense cores has been found using dense gas kinematics from molecular line emission such as DCN, ${\rm N_2H^+}$, and ${\rm H^{13}CO^{+}}$ \citep[e.g.][]{sanhueza21, Yang23, alvarez-gutierrez24, sandoval-garrido24, Pan24}. If this is true, more massive fragmented cores are evidence of the clump-fed model where mass accreting to protostars originates from spatial scales larger than cores. This clump-fed model was earlier supported by more massive protostellar cores than prestellar cores in other star-forming regions \citep{nony23, armante24, morii24}. While two different scenarios are possible, detailed studies of the lifetime and gas infall at each phase of cores are needed to differentiate the core evolution models. 



\subsection{PPOs outside core boundaries}
\label{subsec:yso_outside_cores}
\begin{figure*}
    \centering
    \includegraphics[scale=0.4]{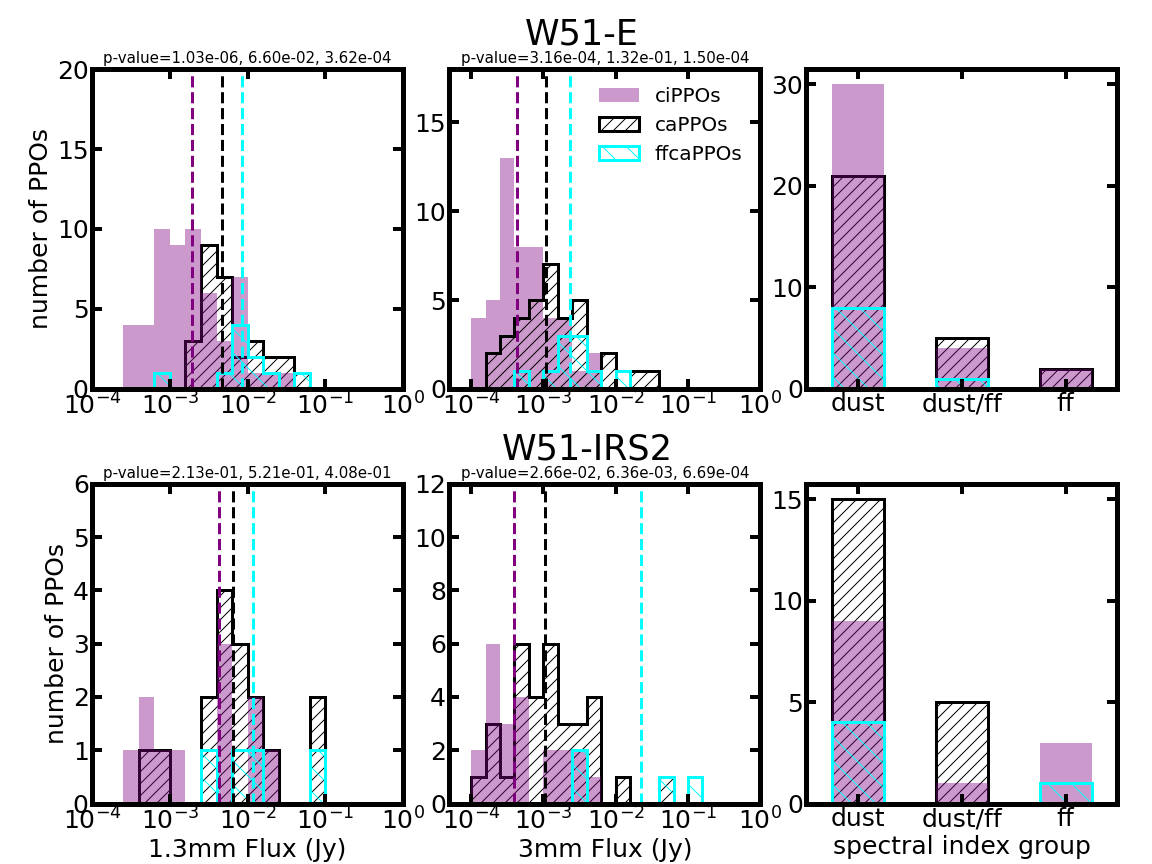}
    \caption{Comparison of $1.3\,{\rm mm}$ and $3\,{\rm mm}$ flux, and spectral index group, dust, dust/ff (optically thick dust emission or possibly free-free contaminated), and ff (free-free contaminated), defined in Sec.~\ref{subsec:spectral_indices}, of ciPPOs, caPPOs, and ffcaPPOs in W51-E (top row) and W51-IRS2 (bottom). The core samples used in this plot have masses higher than the completeness limit of W51-E, $3.86\,M_\odot$. The median of each quantity is denoted by vertical dashed lines. The p-values of the two-sample KS test for ciPPOs/caPPOs, caPPOs/ffcaPPOs, and ciPPOs/ffcaPPOs are displayed at the top of each panel. }
    \label{fig:YSO-comp}
\end{figure*}

\subsubsection{Different core-independent PPO fractions between the two regions}

The spatial correspondence between PPOs and cores in Fig.~\ref{fig:frag_w51e} and Fig.~\ref{fig:frag_w51n} shows that some PPOs are not associated with cores. A fraction of these PPOs were spatially related to free-free contaminated cores that were removed from the core catalog, and thus they need to be treated separately from other PPOs that are associated with/independent of cores without free-free contamination. We classify our PPOs into three groups: core-associated PPOs (caPPOs), core-independent PPOs (ciPPOs), and free-free-contaminated core-associated PPOs (ffcaPPOs). In W51-E and W51-IRS2, we found 37 and 57 caPPOs, 69 and 27 ciPPOs, and 12 and 9 ffcaPPOs, respectively. A much higher fraction of PPOs (58\%) in W51-E is unrelated to cores than those in W51-IRS2 (29\%).

One explanation for the different number fraction of ciPPOs is the poorer completeness limit of the ALMA-IMF image in W51-E as explained in Sec.~\ref{subsec:overview_frag}. Even if a PPO exists without any common envelope at the core scale, we would detect a point source at the same position in the ALMA-IMF image only if the noise level is sufficiently low. When we consider the cores of which masses are above the completeness limit of the W51-E ALMA-IMF image, $3.86\,M_\odot$, the number of caPPOs decreases from 37 to 36 in W51-E, while that of W51-IRS2 declines from 57 to 37. Also, the number of ffcaPPOs changes from 12 to 11 in W51-E and from 9 to 5 in W51-IRS2. This results in the change of the fractions of caPPOs in W51-IRS2 from 61\% to 54\% while that of W51-E remains almost the same at 31\%. Simply assuming Poisson error, the difference is almost comparable to the uncertainty range (9\% from W51-E and 10\% from W51-IRS2). Thus, the different caPPO fraction between the two regions is mainly attributed to their different completeness limits. We will discuss whether other possible causes, e.g., an intrinsically different population, can explain the non-detection of cores in the next section.


\subsubsection{Attempts to characterize core-independent PPOs}

We further investigate the three different PPO groups, ciPPOs, caPPOs, and ffcaPPOs by comparing several physical properties between them (Fig.~\ref{fig:YSO-comp}). In this analysis, we used mass-limited core samples with masses higher than the W51-E completeness limit ($M>3.86\,M_\odot$) to eliminate possible observational effects. In W51-E, ciPPOs are distinctly fainter than other populations at both the $1.3\,{\rm mm}$ and $3,{\rm mm}$. In W51-E, the p-values for the two-sample KS tests are lower than 0.05 for the comparison between caPPOs/ffcaPPOs and ciPPOs at both bands, but the comparison between caPPOs and ffcaPPOs is not. This could be due to the insufficient number of samples in ffcaPPOs. For the same reason, the p-values of the two-sample KS test in W51-IRS2 are generally higher than 0.05 for all comparisons at $1.3\,{\rm mm}$.


The higher fluxes of caPPOs than ciPPOs in W51-E can be explained by the previous argument that the faint PPOs are less likely to be identified in the low-resolution image. In other words, caPPOs can be faint sources that are not just detected in the ALMA-IMF image. 

Another possible factor for the non-detection in the ALMA-IMF image is a lack of gas in the reservoir at the core scale. Besides observational effects, there are several possibilities to make some PPOs appear independent of cores: PPOs rapidly escaped from their birthplace through, e.g., dynamical interaction \citep[e.g.][]{reipurth01}, or PPOs formed earlier in an accreting flow that has subsequently dispersed \citep[e.g.][]{girichidis12}, or PPOs at the end phase of mass accretion from cores--most of the mass in these cores has already been transferred to PPOs \citep[e.g.][]{fernandez-lopez11}. In the last case, we expect that some of those PPOs finishing accretion could already contain massive stars creating hyper-compact HII regions. Higher fluxes of ciPPOs than caPPOs at both $1.3\,{\rm mm}$ and $3\,{\rm mm}$ fluxes might indicate that ciPPOs have less amount of cold dust envelope mass because they are at a later stage where the envelope has been disrupted.

We attempt to characterize ciPPOs by investigating the spectral index groups (Sec.~\ref{subsec:spectral_indices}) of the two PPO populations  (Fig.~\ref{fig:YSO-comp}), but we do not find any common trend between the two regions. In the W51-E and W51-IRS2 regions, the fractions of ciPPOs exhibiting free-free contamination are 5.6\% and 23.1\%, respectively. In contrast, the corresponding fractions for caPPOs are 7.2\% in W51-E and 0.0\% in W51-IRS2. Therefore, it is still uncertain whether the ciPPOs and caPPOs are intrinsically different populations.

\section{Summary and Conclusions}
\label{sec:conclusions}
\subsection{Summary}
In the first half of this paper, we characterize the compact sources from the long-baseline ALMA image--pre/protostellar objects (PPOs) in the W51 high-mass star-forming region. The following is a summary of the characterization of PPOs.
\begin{itemize}
    \item After cross-matching candidate sources in $1.3\,{\rm mm}$ and $3\,{\rm mm}$, we identified 118 compact sources in W51-E and 93 in W51-IRS2.

    \item We measure the spectral indices of PPOs between $1.3\,{\rm mm}$ and $3\,{\rm mm}$. Based on the spectral index, we classified our PPOs into three groups: dusty sources, sources with optically thick dust emission or free-free contamination, and sources dominated by free-free emission. In W51-E, 59 out of 70 sources, and in W51-IRS2, 13 out of 23 sources, are dominated by dust emission, exhibiting spectral indices $\alpha > 2$. The second group consists of 10 sources in W51-E and 6 in W51-IRS2, with spectral indices in the range $1.5 < \alpha < 2$; these are likely to be either optically thick dust emitters or sources with possible free-free contamination. The remaining 4 sources in W51-E and 4 in W51-IRS2 are certainly contaminated by free-free emission. In particular, the spectral index distribution peaks around 2, indicating that many PPOs are optically thick at mm wavelength (Sec.~\ref{subsec:spectral_indices}).

    \item We measure the integrated flux of PPOs using the 2D Gaussian model at $1.3\,{\rm mm}$ and $3\,{\rm mm}$. The measured fluxes are distributed over $10^{-4}$ to $10^{-1}\,{\rm Jy}$. Assuming a modified blackbody radiation (MBB) model, we infer the dust temperature of PPOs for which fluxes at both $1.3\,{\rm mm}$ and $3\,{\rm mm}$ are available (47 in W51-E and 6 in W51-IRS2), ranging $T=5$--$60\,{\rm K}$. We also estimate the mass lower limits of PPOs from the flux density of dust emission and constant temperature $T=40\,{\rm K}$. The estimated mass lower limits range from 0.1 to $200\,M_\odot$ for the constant temperature model.

    \item We estimate the source sizes by deconvolving the 2D Gaussian fit from the image beam. The sources that are too small to be deconvolved are regarded as unresolved sources, and 
approximately 70\% at $3\,{\rm mm}$ and 20--30\% at $1.3\,{\rm mm}$ are unresolved. The sizes of the resolved sources are distributed over $200$--$1000\,{\rm AU}$ at $3\,{\rm mm}$ and $100-500\,{\rm AU}$ at $1.3\,{\rm mm}$.

\end{itemize}

The PPO catalogs in the long-baseline images are linked to the cores extracted from the ALMA-IMF short-baseline images by spatial correspondence. When we eliminated free-free contaminated cores, we found that 15 out of 20 cores in W51-E and 41 out of 87 cores in W51-IRS2 have PPOs inside their FWHM boundaries. Here, we briefly review our findings about core fragmentation in Sec.~\ref{subsec:frag_coremass}, Sec.~\ref{subsec:frag_eff}, and Sec.~\ref{subsec:jeans}.
\begin{itemize}
\item We found that some cores fragment while others do not. The number of cores with 0, 1, and $>1$ fragments are 5, 6, 9 (25\%, 30\%, 45\%) in W51-E and 46, 34, 7 (53\%, 39\%, 8\%) in W51-IRS2, respectively.
    \item A high degree of fragmentation (i.e., $N_{\rm frag}\geq3$) is only found in massive cores with $M_{\rm core} \gtrsim4\,M_\odot$ (Fig.~\ref{fig:frag_flux}).
    \item Bright PPOs, which likely represent progenitors of massive stars, are preferentially found in massive cores (Fig.~\ref{fig:ysofluxmax}).
    \item The distribution of fragment fluxes within a core is not uniform. The difference between the flux of the brightest source and the others in a given core is larger in more massive cores (Fig.~\ref{fig:ysofluxall}).
    \item We do not find massive cores with high fragmentation efficiency ($\epsilon_{frag}\gtrsim0.7$).   
    (Fig.~\ref{fig:epsilon_frag}).
    \item  We find a suggestive trend that massive cores harboring bright (or massive) PPOs are warmer and have lower fragmentation efficiency, but the number of samples (N=4) is not large enough to draw a statistically significant conclusion.  
   
    \item Jeans efficiency, which is the ratio between Jeans number and the observed number of core fragments, tends to decrease with core masses (Fig.~\ref{fig:jeans}).
    \item Most PPOs have mass lower limits that exceed the Jeans mass of their parent cores (Fig.~\ref{fig:jeans} and Fig.~\ref{fig:jeansYSO}). 
\end{itemize}
Also, we further discuss the link between the cores and PPOs in Sec.~\ref{subsec:prestellar_vs_protostellar} and Sec.~\ref{subsec:yso_outside_cores} as follows.
\begin{itemize}
    \item We classify our cores into unfragmented and fragmented cores with the presence of PPOs inside their boundaries (Fig.~\ref{fig:proto_vs_pre}). We find that the fragmented cores are systematically smaller in size, more massive, and denser than the unfragmented cores. 
    \item We found that a larger fraction of PPOs in W51-E (58\%) is not spatially associated with cores than in W51-IRS2 (29\%). This is mainly due to the observational effect that a higher completeness limit of the W51-E ALMA-IMF image reduces the sensitivity to detecting low-resolution counterparts of PPOs at the same locations.
    
\end{itemize}


\subsection{Conclusions}

Some of our findings provide important constraints on high-mass star formation models. 
We found that cores hosting bright PPOs, the progenitors of high-mass stars, are more massive, warmer, and contain more fragments compared to the average population.
Our results do not necessarily indicate that high-mass stars form in initially massive cores, since the current core mass could be a result of accretion from larger scale \citep[e.g.][]{padoan20}. Nevertheless, it is obvious that the core should be massive at the time when the core has high-mass PPOs that are actively accreting material. 

We find evidence for suppressed fragmentation by protostellar heating in more massive cores.  This evidence includes warmer core temperature and the presence of bright PPOs at higher masses. The heated gas is prone to accrete to a central high-mass protostar rather than making new fragments. In this way, the high-mass protostar is able to gain more mass. However, the number of core fragments is more likely to be controlled by the core masses (Fig.~\ref{fig:frag_flux} and ~\ref{fig:epsilon_nfrag}) since the protostellar heating is limited to the local area around the protostar. Furthermore, this result supports the idea that the higher degree of the multiplicity of massive stars in part comes from fragmentation on core scales \citep[see the recent review of][]{offner23}.  

Core fragmentation influences the mapping between the CMF and the IMF (see \citealt{offner14})
The trend that massive cores possess more PPOs is generally expected to steepen the slope of the CMF unless the initial CMF slope is too flat (e.g., $\alpha<-1$; Thomasson et al. in prep). Other trends that can possibly change the CMF slope, e.g, different fragmentation efficiencies or mass ratios between fragments as a function of core mass, are not clearly confirmed in this study.
The CMF slope change we expect from the $N_{frag}$--$M_{core}$ relation is limited to the evolution of the CMF to the PPO mass function, and thus the emerging IMF slope still can be affected by other processes, e.g, protostellar jet \citep[e.g.][]{hennebelle24} and dynamical ejection in cluster environment \citep[e.g.][]{oh15,oh16}.
\begin{acknowledgments}

This paper makes use of the following ALMA data: ADS/JAO.ALMA\#2017.1.01355.L, ADS/JAO.ALMA\#2015.1.01596.S, ADS/JAO.ALMA\# 2017.1.00293.S. ALMA is a partnership of ESO (representing its member states), NSF (USA) and NINS (Japan), together with NRC (Canada), MOST and ASIAA (Taiwan), and KASI (Republic of Korea), in cooperation with the Republic of Chile. The Joint ALMA Observatory is operated by ESO, AUI/NRAO and NAOJ. The National Radio Astronomy Observatory is a facility of the National Science Foundation operated under cooperative agreement by Associated Universities, Inc.
This research made use of astrodendro, a Python package to compute \texttt{dendrogram}s of Astronomical data (http://www.dendrograms.org/). The figures in this paper make use of the Python visualization tools, \texttt{Matplotlib} \citep{hunter07} and \texttt{Seaborn} \citep{waskom21}. The authors acknowledge University of Florida Research Computing for providing computational resources and support that have contributed to the research results reported in this publication. URL: http://www.rc.ufl.edu.

AG acknowledges support from the NSF under grants AAG 2008101 and 2206511 and CAREER 2142300. R.G.M. acknowledges support from UNAM-PAPIIT project IN105225. 
AG and FL acknowledges the support of the programme national "Physique et Chimie du Milieu Interstellaire" (PCMI) and programme national  "Physique Stellaire" (PNPS) of the CNRS/INSU with INC/INP co-funded by CEA and CNES. AS gratefully acknowledges support by the Fondecyt Regular (project
code 1220610), and ANID BASAL project FB210003. LB gratefully acknowledges support by the ANID BASAL project FB210003. PS was partially supported by a Grant-in-Aid for Scientific Research (KAKENHI Number JP22H01271 and JP23H01221) of JSPS. PS was supported by Yoshinori Ohsumi Fund (Yoshinori Ohsumi Award for Fundamental Research). FM, FL, BT, MVM, and AG  have received funding from the European Research Council (ERC) via the ERC Synergy Grant \textsl{ECOGAL} (grant 855130) and from the French Agence Nationale de la Recherche (ANR) through the project \textsl{COSMHIC} (ANR-20-CE31-0009). G.B. acknowledges support from the PID2023-146675NB-I00 (MCI-AEI-FEDER, UE) program.
\end{acknowledgments}

%

\vspace{5mm}
\facilities{ALMA}


\software{CASA (version 5.7.0) \citep{mcmullin07}, CARTA 
\citep{harris20, carta}, \texttt{numpy}
\citep{carta21}, \texttt{scipy} \citep{scipy20}, \texttt{astropy} \citep{astropy13,astropy18},  \texttt{astrodendro} 
\citep{rosolowsky08}, \texttt{dendrocat}, \texttt{radio\_beam}, \texttt{lifeline} \citep{davidson-pilon19}, \texttt{TGIF} \citep{tgif}          }

\bibliography{sample631}{}
\bibliographystyle{aasjournal}


\appendix

\section{PPO Flux measurement with a 2D Gaussian model}
\label{appendix:flux}

In this section, we illustrate the procedure of the flux measurement of fragments from the continuum images. The modeling process consists of three steps: sub-pixel offset adjustment, local background noise estimation, and the flux measurement from the 2D Gaussian model. The whole process of measuring the flux and size of sources is available in Python package \texttt{TGIF} (Two d Gaussian In Fitting). The source code of TGIF can be found at Github page (\url{https://github.com/tyoo37/TGIF}) and Zenodo (\url{https://zenodo.org/records/13973837}).

To model each source with a 2D Gaussian, we first need to find the peak position. By default, the peak of the \texttt{dendrogram} source is the center of the brightest image pixel. However, the discrete representation of a continuous light profile with pixels has a limitation in marking the precise peak position. For this reason, we adjusted the sub-pixel offset of the real peak position in the image. The adjustment process is achieved by fitting the one-dimensional Gaussian model to the image profile near the peak. We extracted the 1D image profile along the major and minor axis of the image beam.
For modeling the 1D Gaussian curve, we utilized \texttt{curve\_fit} of \texttt{scipy} \citep{scipy20} Python package. When we fit the 1D Gaussian model, we measure the offset between the previous peak and the peak of the model and adjust the peak position by the offset. We repeat this process until the offset is measured to be smaller than 1\% of the pixel size.

After adjusting the sub-pixel offset of the peak position, we conduct an initial 2D Gaussian fitting before the local background subtraction. We use the linear chi-square method to minimize the residual function defined as the difference between the image and the 2D Gaussian model. We use a small cutout of the image centered on the peak of the source to compute the residual function to minimize the effect of the surrounding structure. In addition, we give more weight to the pixels producing the negative residuals (${\rm image} - {\rm model} < 0$) in the residual function in order to prevent the case where the fitting function overestimates the size of the source due to the neighboring structure. For the pixels with the negative residual, the weight in each pixel is computed by the median absolute deviation of the local background $\sigma$ multiplied by the exponential of the offset in each pixel such that $\sigma_{\rm weighted}=\sigma\exp(\lambda |x_{\rm image}-x_{\rm model}|)$ where $x_{\rm model}$ and $x_{\rm image}$ are the pixel values of the image and the model, and $\lambda=100$ is a penalty factor. Then $\sigma$ is used to evaluate the chi-square residual function to be minimized $\chi^2=\sum[(x_{\rm image}-x_{\rm model})^2/\sigma^2]$. This is one of the main philosophies of this method to accurately measure the flux and size of the source which are often surrounded by the diffuse emission as seen in Fig.~\ref{fig:frag_w51e} and Fig.~\ref{fig:frag_w51n}. We utilize the minimizer tool \texttt{LMFIT} \citep{lmfit} to optimize the 2D Gaussian model. The 2D Gaussian model has four free parameters: standard deviation along the major axis ($\sigma_{\rm major}$), and minor axis ($\sigma_{\rm minor}$), position angle ($\theta$), and the base flux density ($F_{\rm base}$) which is the flux density that the 2D Gaussian converges as the offset from the center goes to infinity. Other parameters, the positions and the height of the peak, are fixed as those of the image. 

\begin{figure}[H]
    \centering
    \includegraphics[scale=0.22]{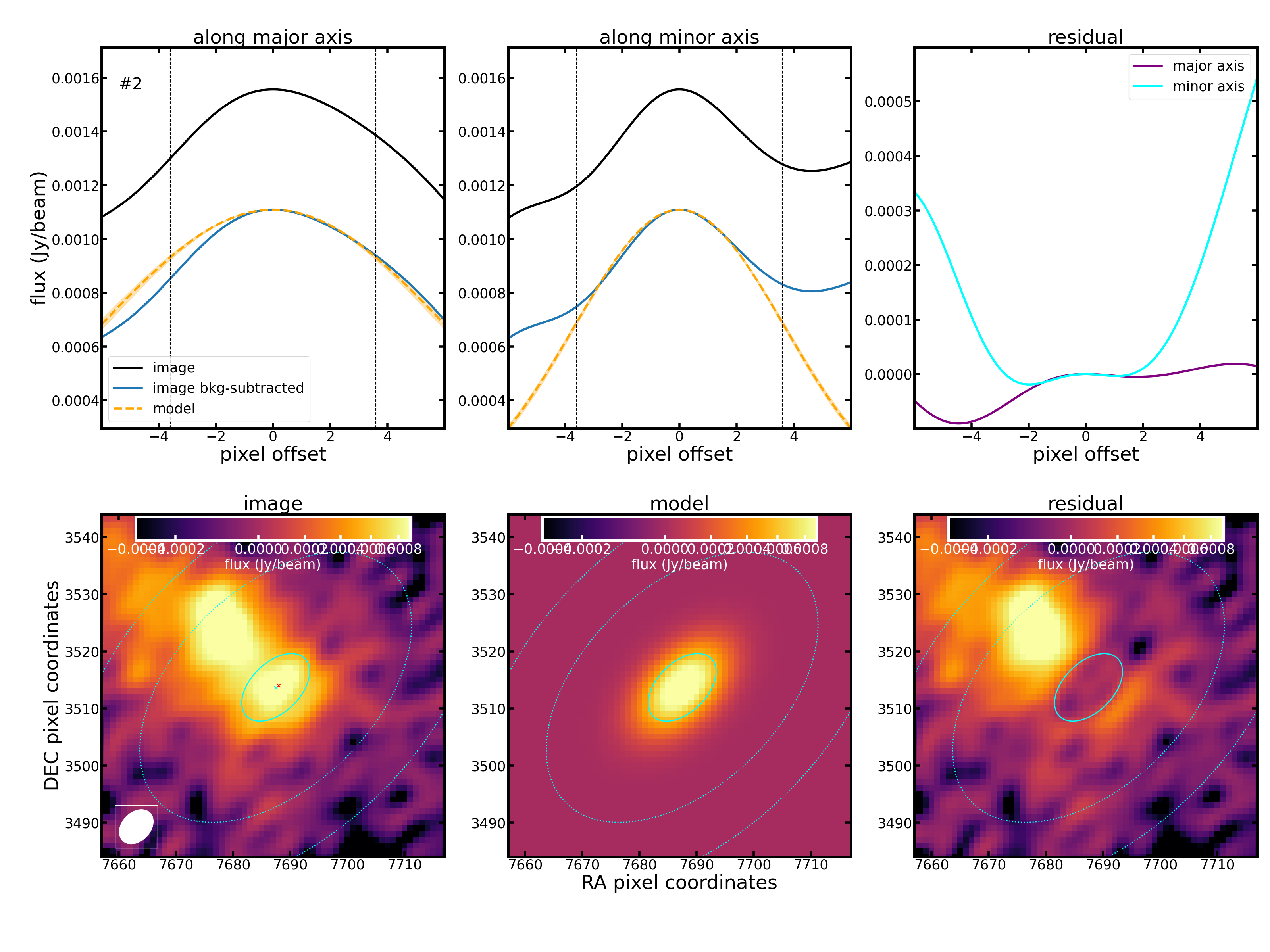}
    \caption{An example of 2D Gaussian modeling for the ALMA $1.3\,{\rm mm}$ flux measurement of PPOs (\#2) in W51-E. \textit{Upper row}: the 1D profile of the continuum image and the model along the major axis (left) and minor axis (center). The solid lines indicate the image profile before (black) and after background subtraction (blue). The orange dashed lines represent the model profile fitted to the background-subtracted image profile. The residual, the image profile subtracted by the model, is presented on the upper right panel. \textit{Bottom row}: the cutout of the original image (lower left), the model (lower center), and the residual image (lower right). The cyan solid ellipse at the center indicates the FWHM of the 2D Gaussian model whereas the cyan dashed ellipses are the annulus where the local background value is evaluated. The size of the synthesized beam of the image is displayed in the left lower corner.  }
    \label{fig:flux_appendix}
\end{figure}

The local background value is then evaluated from the elliptical annulus around the peak of each source (lower panels in Fig.~\ref{fig:flux_appendix}). The annulus is enclosed by two ellipses, each representing the FWHM of the two Gaussian model obtained from the initial fitting extended by factors of 2 and 3, respectively. The local background value is then determined by the median of the pixel values in the annulus. To minimize the contamination from known sources, we masked out the regions with $2\times$ the beam size centered on each cataloged peak. 

Next, we carry out the 2D Gaussian fitting again from the local background-subtracted image. Once the 2D Gaussian model was obtained, we computed the integrated flux from the volume of the model $V=2\pi A\sigma_{\rm major}\sigma_{\rm minor}$, where $A$ is a peak flux density. We also measured the physical size of the source by deconvolving the FWHM along both axes with the beam with the aid of \texttt{radio\_beam} Python package.

\section{The completeness test}
\label{appendix:completeness}

The completeness limit of the PPO detection is achieved by injecting the synthetic sources on the regular grids of the image as it was done in the ALMA-IMF project \citep[e.g.][]{pouteau22}{}{}. Given that the $3\,{\rm mm}$ flux is used for mass estimation, we only test the $3\,{\rm mm}$ continuum image. The synthetic sources are created by 2D Gaussian with the median sizes of our samples ($\sim100\,{\rm AU}$). We injected synthetic sources with 11 different fluxes corresponding to mass values ranging from $0.05\,M_\odot$ to $1\,M_\odot$ at $50\,{\rm K}$ temperature using optically thin dust emission (Eq.~\ref{eq:mass}). To take into account the different probability of detection at different background levels, we placed each synthetic source on the grids with the size of 100 pixels ($\sim0''.7$). In total, we created 2210 and 2086 sources on the images of W51-E and W51-IRS2, respectively. 

\begin{figure}
    \centering
    \includegraphics[scale=0.4]{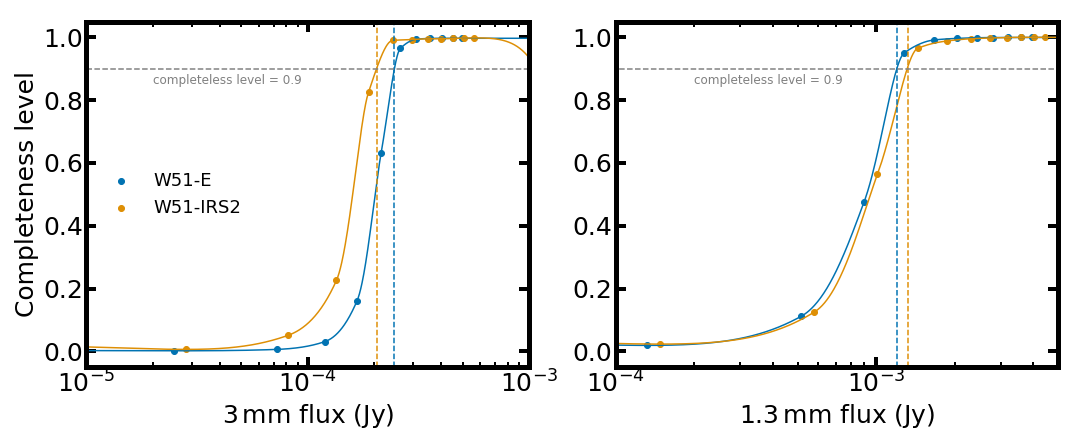}
    \caption{The completeness test result. We define the completeness limit mass as the mass reaching the completeness level = 0.9. Each point is generated by testing if the synthetic sources are detected in the \texttt{dendrogram} or not. The interpolation between points is done with PCHIP interpolator.}
    \label{fig:completeness}
\end{figure}

The completeness level of synthetic sources is displayed in Fig.~\ref{fig:completeness}. The completeness level is defined as the fraction of synthetic sources identified in the \texttt{dendrogram} with the same parameters used in finding our PPO samples. We define the mass reaching a 0.9 completeness level as a completeness limit. By interpolating the result produced at discrete masses using the PCHIP interpolator, we obtain the completeness limits of W51-E and W51-IRS2, $0.25\,{\rm mJy}$ and $0.20\,{\rm mJy}$, respectively, in Band 3 and $1.2\,{\rm mJy}$ and $1.3\,{\rm mJy}$, in Band 6. Assuming $40\,{\rm K}$ dust temperature, inserting the Band 3 fluxes into Eq.~\ref{eq:mass} produces to 0.631 $M_\odot$ and 0.505 $M_\odot$ in W51-E and W51-IRS2, respectively.

\section{PPO catalog}
\label{appendix:ppo_tab}
We present PPO catalogs containing the sky coordinates and their photometric fluxes. The full machine-readable form of the catalogs is available at \url{https://zenodo.org/records/16235156}. 

\begin{deluxetable}{ccccccc}
\tablecaption{Basic properties of PPO catalog in W51-E for first 30 entries}
\tablehead{
ID & RA & Dec
 & 
$F_{\rm peak, 1.3mm}$ & $F_{\rm peak, 3mm}$ & Spectral Index & Spectral Index Group \\
 & (deg) & (deg) & (${\rm Jy/beam}$) & (${\rm Jy/beam}$)  &  & }
 \startdata
0 & 290.93232 & 14.50541 & $1.31\times10^{-3}$ & $2.62\times10^{-4}$ & $2.9\pm0.1$ & dust-dominated \\
1 & 290.93237 & 14.50546 & $1.41\times10^{-3}$ & $3.01\times10^{-4}$ & $2.4\pm0.1$ & dust-dominated \\
2 & 290.93246 & 14.50546 & $1.11\times10^{-3}$ & $6.54\times10^{-4}$ & $2.3\pm0.0$ & dust-dominated \\
3 & 290.93246 & 14.50551 & $1.44\times10^{-3}$ & $1.55\times10^{-4}$ & $3.2\pm0.1$ & dust-dominated \\
4 & 290.93225 & 14.50594 & $3.8\times10^{-3}$ & $9.09\times10^{-4}$ & $2.8\pm0.0$ & dust-dominated \\
5 & 290.93258 & 14.50593 & $7.24\times10^{-4}$ & $2.28\times10^{-4}$ & $2.4\pm0.1$ & dust-dominated \\
6 & 290.93261 & 14.50596 & $7.97\times10^{-4}$ & $2.12\times10^{-4}$ & $2.4\pm0.1$ & dust-dominated \\
7 & 290.93128 & 14.50607 & $1.31\times10^{-3}$ & $1.17\times10^{-3}$ & $0.6\pm0.1$ & free-free dominated \\
8 & 290.93236 & 14.50621 & $3.81\times10^{-3}$ & $1.75\times10^{-3}$ & $2.2\pm0.0$ & dust-dominated \\
9 & 290.93248 & 14.50622 & $2.02\times10^{-3}$ & $1.07\times10^{-3}$ & $2.3\pm0.0$ & dust-dominated \\
10 & 290.93243 & 14.50624 & $2.32\times10^{-3}$ & $1.03\times10^{-3}$ & $2.2\pm0.0$ & dust-dominated \\
11 & 290.93258 & 14.5065 & $1.64\times10^{-3}$ & $4.14\times10^{-4}$ & $2.6\pm0.0$ & dust-dominated \\
12 & 290.93261 & 14.5065 & $5.75\times10^{-4}$ & $5.29\times10^{-4}$ & $1.8\pm0.1$ & optically thick or possibly ff \\
13 & 290.93263 & 14.50699 & $7.69\times10^{-4}$ & $4.89\times10^{-4}$ & $2.2\pm0.0$ & dust-dominated \\
14 & 290.93079 & 14.507 & $7.32\times10^{-4}$ & $2.02\times10^{-4}$ & $2.3\pm0.1$ & dust-dominated \\
15 & 290.93341 & 14.50702 & $6.24\times10^{-4}$ & $1.65\times10^{-4}$ & $2.7\pm0.1$ & dust-dominated \\
16 & 290.93372 & 14.50703 & $1.31\times10^{-3}$ & $4.32\times10^{-4}$ & $2.3\pm0.1$ & dust-dominated \\
17 & 290.93343 & 14.50704 & $6.4\times10^{-4}$ & $1.48\times10^{-4}$ & $2.5\pm0.1$ & dust-dominated \\
18 & 290.9333 & 14.50714 & $6.9\times10^{-4}$ & $2.34\times10^{-4}$ & $2.5\pm0.1$ & dust-dominated \\
19 & 290.93325 & 14.50726 & $1.51\times10^{-3}$ & $1.82\times10^{-4}$ & $3.2\pm0.1$ & dust-dominated \\
20 & 290.93272 & 14.50736 & $9.12\times10^{-4}$ & $3.07\times10^{-4}$ & $2.7\pm0.0$ & dust-dominated \\
21 & 290.93275 & 14.50738 & $3.47\times10^{-3}$ & $2.54\times10^{-3}$ & $1.9\pm0.0$ & optically thick or possibly ff \\
22 & 290.93281 & 14.50742 & $9.0\times10^{-4}$ & $5.03\times10^{-4}$ & $2.4\pm0.0$ & dust-dominated \\
23 & 290.93347 & 14.50755 & $1.53\times10^{-3}$ & $1.22\times10^{-3}$ & $1.8\pm0.0$ & optically thick or possibly ff \\
24 & 290.93315 & 14.50755 & $6.81\times10^{-4}$ & $1.36\times10^{-4}$ & $2.6\pm0.2$ & dust-dominated \\
25 & 290.93274 & 14.50757 & $2.14\times10^{-3}$ & $4.27\times10^{-4}$ & $3.0\pm0.0$ & dust-dominated \\
26 & 290.93289 & 14.50758 & $3.03\times10^{-3}$ & $1.29\times10^{-3}$ & $2.5\pm0.0$ & dust-dominated \\
27 & 290.93497 & 14.50759 & $4.57\times10^{-4}$ & $8.6\times10^{-5}$ & $2.5\pm0.3$ & dust-dominated \\
28 & 290.93291 & 14.50764 & $1.63\times10^{-3}$ & $2.01\times10^{-3}$ & $1.8\pm0.0$ & optically thick or possibly ff \\
29 & 290.93294 & 14.50771 & $3.05\times10^{-3}$ & $3.62\times10^{-3}$ & $1.1\pm0.0$ & free-free dominated \\
\enddata
\end{deluxetable}

\begin{deluxetable}{ccccccc}
\tablecaption{Basic properties of PPO catalog in W51-IRS2 for first 30 entries}
\tablehead{
ID & RA & Dec
 & 
$F_{\rm peak, 1.3mm}$ & $F_{\rm peak, 3mm}$ & Spectral Index & Spectral Index Group \\
 & (deg) & (deg) & (${\rm Jy/beam}$) & (${\rm Jy/beam}$)  &  & }
 \startdata
0 & 290.91465 & 14.5176 & $1.9\times10^{-3}$ & $3.68\times10^{-4}$ & $2.9\pm0.1$ & dust-dominated \\
1 & 290.91496 & 14.51782 & $2.29\times10^{-3}$ & $9.51\times10^{-4}$ & $1.9\pm0.0$ & optically thick or possibly ff \\
2 & 290.91711 & 14.5179 & $2.71\times10^{-3}$ & $1.68\times10^{-3}$ & $0.7\pm0.0$ & free-free dominated \\
3 & 290.9182 & 14.51792 & $4.13\times10^{-4}$ & $8.46\times10^{-5}$ & $2.3\pm0.3$ & dust-dominated \\
4 & 290.91591 & 14.51801 & $2.1\times10^{-2}$ & $4.21\times10^{-2}$ & $0.2\pm0.0$ & free-free dominated \\
5 & 290.91551 & 14.5181 & $2.8\times10^{-3}$ & $1.71\times10^{-3}$ & $2.2\pm0.0$ & dust-dominated \\
6 & 290.91586 & 14.5181 & $1.15\times10^{-3}$ & $6.06\times10^{-4}$ & $2.3\pm0.0$ & dust-dominated \\
7 & 290.91564 & 14.51811 & $2.01\times10^{-3}$ & $7.79\times10^{-4}$ & $2.7\pm0.0$ & dust-dominated \\
8 & 290.91619 & 14.51813 & $9.91\times10^{-4}$ & $5.11\times10^{-4}$ & $2.3\pm0.0$ & dust-dominated \\
9 & 290.91596 & 14.51815 & $3.2\times10^{-3}$ & $1.38\times10^{-3}$ & $2.3\pm0.0$ & dust-dominated \\
10 & 290.91648 & 14.51815 & $4.17\times10^{-3}$ & $3.89\times10^{-3}$ & $1.8\pm0.0$ & optically thick or possibly ff \\
11 & 290.91688 & 14.51819 & $6.24\times10^{-3}$ & $8.41\times10^{-3}$ & $1.8\pm0.0$ & optically thick or possibly ff \\
12 & 290.91606 & 14.51817 & $1.36\times10^{-3}$ & $2.41\times10^{-4}$ & $2.7\pm0.1$ & dust-dominated \\
13 & 290.91694 & 14.51818 & $2.37\times10^{-3}$ & $1.78\times10^{-3}$ & $1.8\pm0.0$ & optically thick or possibly ff \\
14 & 290.9156 & 14.51822 & $7.86\times10^{-4}$ & $3.92\times10^{-4}$ & $2.2\pm0.1$ & dust-dominated \\
15 & 290.917 & 14.51826 & $1.86\times10^{-3}$ & $4.72\times10^{-4}$ & $2.8\pm0.0$ & dust-dominated \\
16 & 290.91708 & 14.51824 & $8.9\times10^{-3}$ & $3.62\times10^{-3}$ & $1.5\pm0.0$ & free-free dominated \\
17 & 290.91661 & 14.51826 & $1.16\times10^{-3}$ & $2.48\times10^{-4}$ & $2.9\pm0.0$ & dust-dominated \\
18 & 290.91707 & 14.51827 & $6.68\times10^{-3}$ & $2.32\times10^{-3}$ & $1.4\pm0.0$ & free-free dominated \\
19 & 290.91716 & 14.51827 & $6.31\times10^{-3}$ & $1.91\times10^{-3}$ & $1.8\pm0.0$ & optically thick or possibly ff \\
20 & 290.91662 & 14.51828 & $1.8\times10^{-3}$ & $1.69\times10^{-3}$ & $1.9\pm0.0$ & optically thick or possibly ff \\
21 & 290.91666 & 14.51837 & $2.96\times10^{-3}$ & $1.42\times10^{-3}$ & $2.1\pm0.0$ & dust-dominated \\
22 & 290.92055 & 14.51877 & $1.39\times10^{-3}$ & $2.87\times10^{-4}$ & $2.1\pm0.1$ & dust-dominated \\
23 & 290.92035 & 14.51407 & - & $1.29\times10^{-4}$ & - & - \\
24 & 290.91722 & 14.5173 & - & $1.12\times10^{-4}$ & - & - \\
25 & 290.9168 & 14.51816 &-  & $1.87\times10^{-3}$ & - & - \\
26 & 290.9168 & 14.51822 & - & $7.9\times10^{-4}$ & - & - \\
27 & 290.91667 & 14.5183 & -  & $3.15\times10^{-4}$ & - & - \\
28 & 290.9115 & 14.51868 & - & $1.92\times10^{-4}$ & - & - \\
29 & 290.91916 & 14.51875 & - & $1.93\times10^{-4}$ & - & - \\
\enddata
\end{deluxetable}

\begin{deluxetable}{ccccccc}
\rotate
\tablecaption{Photometric properties of PPO catalog in W51-E for first 30 entries}
\tablehead{
ID & $F_{\rm int, 1.3mm}$ & $F_{\rm int, 3mm}$ & $T_{\rm MBB}$ & $M_{\rm lowlim}$ & $a_{\rm 1.3mm}\times b_{\rm 1.3mm}$ & $a_{\rm 3mm}\times b_{\rm 3mm}$  \\
 & (${\rm Jy}$) & (${\rm Jy}$) & (K) & ($M_\odot$)  & (${\rm AU}\times{\rm AU}$) & (${\rm AU}\times{\rm AU}$) }
\startdata
0 & $3.8\times10^{-3}\pm 1.2\times10^{-4}$ & $6.4\times10^{-4}\pm 2.3\times10^{-5}$ & 16 & $1.6\times10^{0}\pm 5.8\times10^{-2}$ & $160\times 75$ & - \\
1 & $2.5\times10^{-3}\pm 4.7\times10^{-5}$ & $5.7\times10^{-4}\pm 5.7\times10^{-6}$ & - & $1.4\times10^{0}\pm 1.4\times10^{-2}$ & - & - \\
2 & $5.8\times10^{-3}\pm 1.9\times10^{-4}$ & $3.1\times10^{-3}\pm 4.6\times10^{-4}$ & - & $7.8\times10^{0}\pm 1.2\times10^{0}$ & $290\times 115$ & $546\times 292$ \\
3 & $3.8\times10^{-3}\pm 1.3\times10^{-4}$ & $2.4\times10^{-4}\pm 3.3\times10^{-6}$ & 18 & $6.1\times10^{-1}\pm 8.3\times10^{-3}$ & - & - \\
4 & $1.5\times10^{-2}\pm 1.8\times10^{-4}$ & $2.0\times10^{-3}\pm 2.7\times10^{-5}$ & 53 & $5.0\times10^{0}\pm 6.8\times10^{-2}$ & $215\times 98$ & - \\
5 & $1.3\times10^{-3}\pm 7.9\times10^{-5}$ & $4.0\times10^{-4}\pm 2.7\times10^{-5}$ & 11 & $1.0\times10^{0}\pm 6.7\times10^{-2}$ & - & - \\
6 & $1.2\times10^{-3}\pm 1.2\times10^{-5}$ & $3.0\times10^{-4}\pm 1.6\times10^{-5}$ & - & $7.6\times10^{-1}\pm 4.1\times10^{-2}$ & - & - \\
7 & $2.3\times10^{-3}\pm 8.2\times10^{-5}$ & $1.8\times10^{-3}\pm 6.8\times10^{-6}$ & - & $4.5\times10^{0}\pm 1.7\times10^{-2}$ & - & - \\
8 & $2.3\times10^{-2}\pm 5.0\times10^{-4}$ & $3.5\times10^{-3}\pm 4.9\times10^{-5}$ & 58 & $8.7\times10^{0}\pm 1.2\times10^{-1}$ & $242\times 200$ & - \\
9 & $1.2\times10^{-2}\pm 4.5\times10^{-4}$ & $2.9\times10^{-3}\pm 1.9\times10^{-5}$ & - & $7.2\times10^{0}\pm 4.9\times10^{-2}$ & $276\times 158$ & - \\
10 & $1.4\times10^{-2}\pm 5.3\times10^{-4}$ & $2.5\times10^{-3}\pm 1.1\times10^{-4}$ & - & $6.4\times10^{0}\pm 2.7\times10^{-1}$ & $294\times 144$ & - \\
11 & $6.5\times10^{-3}\pm 2.2\times10^{-4}$ & $7.6\times10^{-4}\pm 5.2\times10^{-6}$ & 23 & $1.9\times10^{0}\pm 1.3\times10^{-2}$ & $184\times 142$ & - \\
12 & $2.1\times10^{-3}\pm 1.3\times10^{-5}$ & $1.4\times10^{-3}\pm 1.7\times10^{-5}$ & - & $3.5\times10^{0}\pm 4.2\times10^{-2}$ & $202\times 94$ & - \\
13 & $3.4\times10^{-3}\pm 2.4\times10^{-4}$ & $1.7\times10^{-3}\pm 1.2\times10^{-5}$ & - & $4.2\times10^{0}\pm 3.1\times10^{-2}$ & $241\times 119$ & - \\
14 & $2.5\times10^{-3}\pm 4.9\times10^{-5}$ & $3.7\times10^{-4}\pm 8.5\times10^{-6}$ & 10 & $9.5\times10^{-1}\pm 2.1\times10^{-2}$ & $182\times 104$ & - \\
15 & $1.4\times10^{-3}\pm 4.4\times10^{-5}$ & $4.1\times10^{-4}\pm 3.7\times10^{-6}$ & 12 & $1.0\times10^{0}\pm 9.4\times10^{-3}$ & - & - \\
16 & $6.1\times10^{-3}\pm 8.3\times10^{-5}$ & $8.7\times10^{-4}\pm 9.9\times10^{-6}$ & 18 & $2.2\times10^{0}\pm 2.5\times10^{-2}$ & $298\times 67$ & - \\
17 & $1.4\times10^{-3}\pm 6.7\times10^{-5}$ & $2.8\times10^{-4}\pm 3.6\times10^{-6}$ & 8 & $7.2\times10^{-1}\pm 9.1\times10^{-3}$ & - & - \\
18 & $1.9\times10^{-3}\pm 1.3\times10^{-4}$ & $1.7\times10^{-3}\pm 1.3\times10^{-4}$ & 9 & $4.2\times10^{0}\pm 3.3\times10^{-1}$ & $153\times 78$ & $899\times 255$ \\
19 & $5.2\times10^{-3}\pm 1.5\times10^{-4}$ & $3.5\times10^{-4}\pm 3.7\times10^{-6}$ & 24 & $8.9\times10^{-1}\pm 9.3\times10^{-3}$ & $198\times 85$ & - \\
20 & $3.7\times10^{-3}\pm 3.0\times10^{-4}$ & $8.7\times10^{-4}\pm 1.3\times10^{-4}$ & - & $2.2\times10^{0}\pm 3.3\times10^{-1}$ & $205\times 127$ & - \\
21 & $3.1\times10^{-2}\pm 9.3\times10^{-4}$ & $6.7\times10^{-3}\pm 3.7\times10^{-5}$ & - & $1.7\times10^{1}\pm 9.3\times10^{-2}$ & $298\times 270$ & - \\
22 & $4.6\times10^{-3}\pm 3.2\times10^{-4}$ & $1.5\times10^{-3}\pm 4.9\times10^{-5}$ & 17 & $3.7\times10^{0}\pm 1.2\times10^{-1}$ & $251\times 148$ & $391\times 126$ \\
23 & $8.9\times10^{-3}\pm 2.4\times10^{-4}$ & $4.3\times10^{-3}\pm 1.1\times10^{-4}$ & - & $1.1\times10^{1}\pm 2.8\times10^{-1}$ & $249\times 185$ & $394\times 244$ \\
24 & $1.4\times10^{-3}\pm 3.6\times10^{-5}$ & $2.6\times10^{-4}\pm 2.7\times10^{-6}$ & 5 & $6.6\times10^{-1}\pm 6.8\times10^{-3}$ & - & - \\
25 & $6.7\times10^{-3}\pm 1.7\times10^{-4}$ & $7.5\times10^{-4}\pm 4.4\times10^{-6}$ & 24 & $1.9\times10^{0}\pm 1.1\times10^{-2}$ & $185\times 67$ & - \\
26 & $2.7\times10^{-2}\pm 1.4\times10^{-3}$ & $4.1\times10^{-3}\pm 7.3\times10^{-5}$ & 40 & $1.0\times10^{1}\pm 1.8\times10^{-1}$ & $376\times 212$ & $448\times 97$ \\
27 & $7.7\times10^{-4}\pm 3.0\times10^{-5}$ & $1.0\times10^{-4}\pm 1.9\times10^{-6}$ & 5 & $2.6\times10^{-1}\pm 4.8\times10^{-3}$ & - & - \\
28 & $9.1\times10^{-3}\pm 2.1\times10^{-4}$ & $6.0\times10^{-3}\pm 1.5\times10^{-4}$ & - & $1.5\times10^{1}\pm 3.7\times10^{-1}$ & $264\times 156$ & - \\
29 & $5.2\times10^{-3}\pm 5.3\times10^{-5}$ & $5.5\times10^{-3}\pm 3.2\times10^{-5}$ & - & $1.4\times10^{1}\pm 8.1\times10^{-2}$ & - & - \\
\enddata
\end{deluxetable}
\begin{deluxetable}{ccccccc}
\rotate
\tablecaption{Photometric properties of PPO catalog in W51-IRS2 for first 30 entries}
\tablehead{
ID & $F_{\rm int, 1.3mm}$ & $F_{\rm int, 3mm}$ & $T_{\rm MBB}$ & $M_{\rm lowlim}$ & $a_{\rm 1.3mm}\times b_{\rm 1.3mm}$ & $a_{\rm 3mm}\times b_{\rm 3mm}$   }
\startdata
0 & $5.8\times10^{-3}\pm 1.0\times10^{-4}$ & $8.5\times10^{-4}\pm 2.6\times10^{-5}$ & 26 & $8.3\times10^{-1}\pm 2.6\times10^{-2}$ & $174\times 71$ & - \\
1 & $7.8\times10^{-3}\pm 1.7\times10^{-4}$ & $1.6\times10^{-3}\pm 9.2\times10^{-6}$ & - & $4.1\times10^{0}\pm 2.3\times10^{-2}$ & $176\times 106$ & - \\
2 & $4.3\times10^{-3}\pm 4.0\times10^{-5}$ & $2.3\times10^{-3}\pm 6.2\times10^{-6}$ & - & $5.9\times10^{0}\pm 1.6\times10^{-2}$ & - & - \\
3 & $7.4\times10^{-4}\pm 2.7\times10^{-5}$ & $1.1\times10^{-4}\pm 1.4\times10^{-6}$ & 4 & $2.7\times10^{-1}\pm 3.5\times10^{-3}$ & - & - \\
4 & $7.7\times10^{-2}\pm 2.2\times10^{-3}$ & $1.2\times10^{-1}\pm 1.4\times10^{-3}$ & - & $3.1\times10^{2}\pm 3.5\times10^{0}$ & $165\times 144$ & $378\times 88$ \\
5 & $2.1\times10^{-2}\pm 3.6\times10^{-3}$ & $5.8\times10^{-3}\pm 5.5\times10^{-5}$ & 50 & $5.6\times10^{0}\pm 5.4\times10^{-2}$ & $276\times 235$ & $456\times 95$ \\
6 & $8.6\times10^{-3}\pm 1.7\times10^{-4}$ & $3.4\times10^{-3}\pm 9.6\times10^{-5}$ & 31 & $8.5\times10^{0}\pm 2.4\times10^{-1}$ & $293\times 221$ & $573\times 387$ \\
7 & $5.5\times10^{-3}\pm 1.5\times10^{-4}$ & $1.4\times10^{-3}\pm 4.8\times10^{-6}$ & - & $1.3\times10^{0}\pm 4.7\times10^{-3}$ & - & - \\
8 & $5.5\times10^{-3}\pm 7.0\times10^{-4}$ & $1.7\times10^{-3}\pm 1.2\times10^{-5}$ & - & $4.2\times10^{0}\pm 2.9\times10^{-2}$ & $280\times 142$ & - \\
9 & $1.5\times10^{-2}\pm 2.6\times10^{-4}$ & $3.5\times10^{-3}\pm 5.1\times10^{-5}$ & - & $8.8\times10^{0}\pm 1.3\times10^{-1}$ & $203\times 167$ & - \\
10 & $7.2\times10^{-2}\pm 4.5\times10^{-3}$ & $1.5\times10^{-2}\pm 5.7\times10^{-4}$ & - & $1.5\times10^{1}\pm 5.6\times10^{-1}$ & $542\times 315$ & $478\times 221$ \\
11 & $6.4\times10^{-2}\pm 3.9\times10^{-3}$ & $5.7\times10^{-2}\pm 2.3\times10^{-3}$ & - & $1.8\times10^{1}\pm 7.4\times10^{-1}$ & $346\times 278$ & $625\times 449$ \\
12 & $6.1\times10^{-3}\pm 9.0\times10^{-5}$ & $5.8\times10^{-4}\pm 1.5\times10^{-5}$ & 20 & $1.5\times10^{0}\pm 3.7\times10^{-2}$ & $239\times 118$ & - \\
13 & $3.8\times10^{-3}\pm 1.4\times10^{-4}$ & $2.7\times10^{-3}\pm 7.1\times10^{-5}$ & - & $8.8\times10^{-1}\pm 2.3\times10^{-2}$ & - & - \\
14 & $6.1\times10^{-3}\pm 3.4\times10^{-4}$ & $8.0\times10^{-4}\pm 8.9\times10^{-6}$ & - & $7.8\times10^{-1}\pm 8.7\times10^{-3}$ & $369\times 167$ & - \\
15 & $5.4\times10^{-3}\pm 5.0\times10^{-5}$ & $1.2\times10^{-3}\pm 1.8\times10^{-5}$ & - & $2.9\times10^{0}\pm 4.5\times10^{-2}$ & $159\times 78$ & - \\
16 & $1.8\times10^{-2}\pm 4.7\times10^{-4}$ & $5.4\times10^{-3}\pm 2.7\times10^{-5}$ & - & $1.4\times10^{1}\pm 6.9\times10^{-2}$ & - & - \\
17 & $7.1\times10^{-3}\pm 4.9\times10^{-4}$ & $8.5\times10^{-4}\pm 5.0\times10^{-5}$ & 47 & $8.3\times10^{-1}\pm 4.8\times10^{-2}$ & $237\times 214$ & $531\times 7$ \\
18 & $1.2\times10^{-2}\pm 2.9\times10^{-4}$ & $3.2\times10^{-3}\pm 5.8\times10^{-6}$ & - & $8.1\times10^{0}\pm 1.5\times10^{-2}$ & - & - \\
19 & $1.3\times10^{-2}\pm 2.6\times10^{-4}$ & $2.9\times10^{-3}\pm 9.6\times10^{-6}$ & - & $7.3\times10^{0}\pm 2.4\times10^{-2}$ & - & - \\
20 & $1.2\times10^{-2}\pm 3.9\times10^{-4}$ & $4.5\times10^{-3}\pm 6.9\times10^{-5}$ & - & $4.4\times10^{0}\pm 6.8\times10^{-2}$ & $297\times 185$ & - \\
21 & $1.4\times10^{-2}\pm 5.1\times10^{-4}$ & $3.2\times10^{-3}\pm 6.1\times10^{-5}$ & - & $3.1\times10^{0}\pm 6.0\times10^{-2}$ & $257\times 105$ & - \\
22 & $2.4\times10^{-3}\pm 5.7\times10^{-5}$ & $3.9\times10^{-4}\pm 1.9\times10^{-6}$ & - & $9.8\times10^{-1}\pm 4.7\times10^{-3}$ & - & - \\
23 & - & $2.2\times10^{-4}\pm 1.3\times10^{-5}$ & - & $5.6\times10^{-1}\pm 3.2\times10^{-2}$ & - & - \\
24 & - & $3.2\times10^{-4}\pm 1.3\times10^{-5}$ & - & $8.1\times10^{-1}\pm 3.3\times10^{-2}$ & - & - \\
25 & - & $5.6\times10^{-3}\pm 2.1\times10^{-5}$ & - & $1.8\times10^{0}\pm 6.7\times10^{-3}$ & - & - \\
26 & - & $2.9\times10^{-3}\pm 7.7\times10^{-5}$ & - & $9.3\times10^{-1}\pm 2.5\times10^{-2}$ & - & $540\times 19$ \\
27 & - & $1.4\times10^{-3}\pm 4.7\times10^{-5}$ & - & $1.4\times10^{0}\pm 4.6\times10^{-2}$ & - & $466\times 349$ \\
28 & - & $6.7\times10^{-4}\pm 3.3\times10^{-5}$ & - & $1.7\times10^{0}\pm 8.3\times10^{-2}$ & - & - \\
29 & - & $2.1\times10^{-4}\pm 2.7\times10^{-6}$ & - & $5.3\times10^{-1}\pm 6.8\times10^{-3}$ & - & - \\
\enddata
\end{deluxetable}
\clearpage

\section{PPO maps}
\label{appendix:ppo_map}
\begin{figure*}[h]
    \centering
    \includegraphics[width=1.0\linewidth]{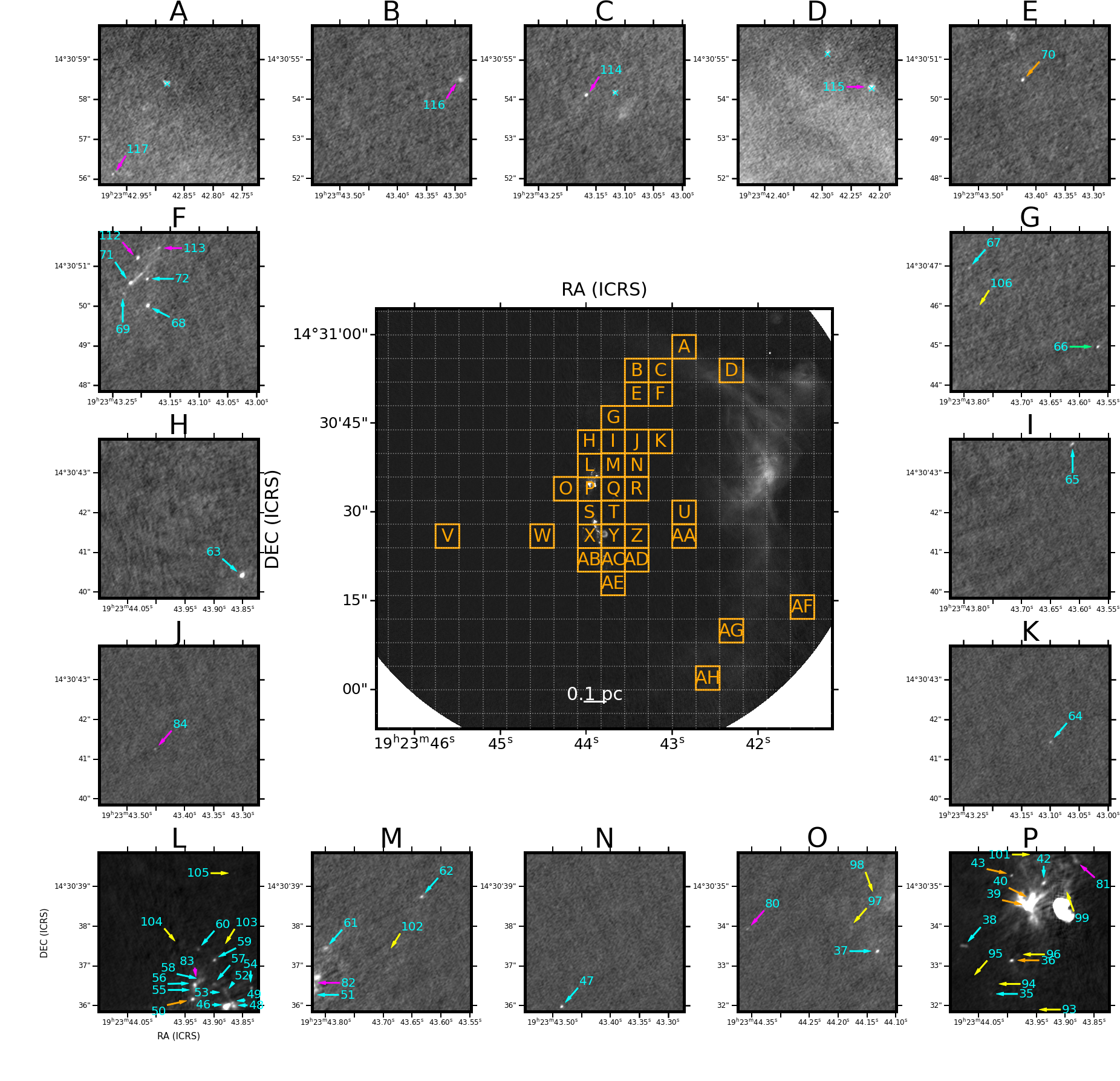}

    \caption{A map of PPOs in W51-E. In the central panel, subfields containing PPOs are labeled. The cutout images of each labeled grid are displayed on the side panels. The positions of PPOs are annotated by the arrow with their ID number and colors representing their classification (cyan: dust-dominated, orange: dust+free-free, green: free-free contaminated, magenta: only found at $1.3\,{\rm mm}$, yellow: only found at $3\,{\rm mm}$). The cyan crosses mark PPOs included in the W51-IRS2 catalog. The background image is $3\,{\rm mm}$ long baseline continuum image. The color scaling in each inset is individually adjusted for better visual clarity.}
    \label{fig:ppo_map_w51e}
\end{figure*}

\begin{figure*}
\centering

    \includegraphics[width=1.0\linewidth]{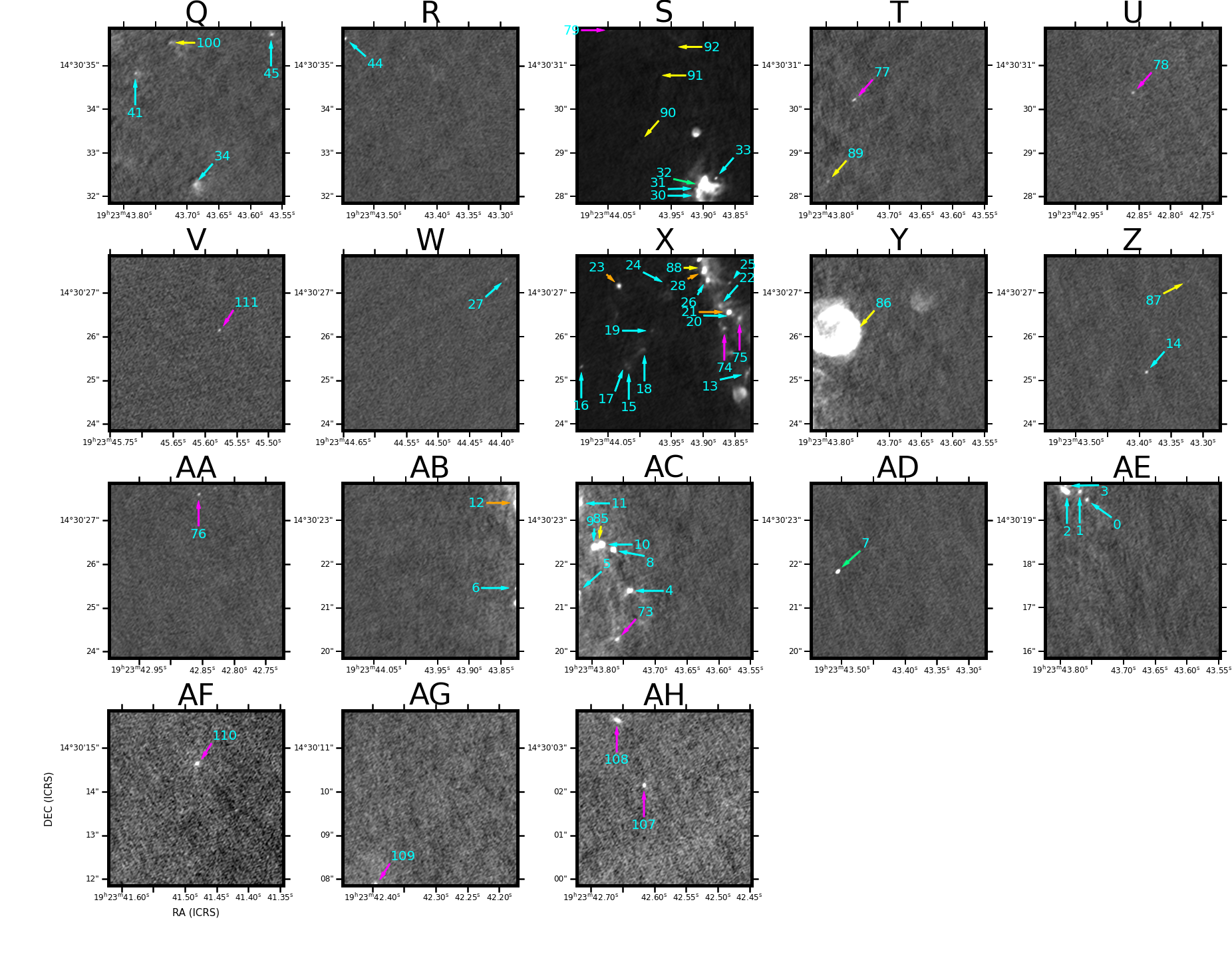}

    \caption{Fig.~\ref{fig:ppo_map_w51e} continued.}
    \label{fig:ppo_map_w51e2}
\end{figure*}

\begin{figure*}

    \centering
    \includegraphics[width=1.0\linewidth]{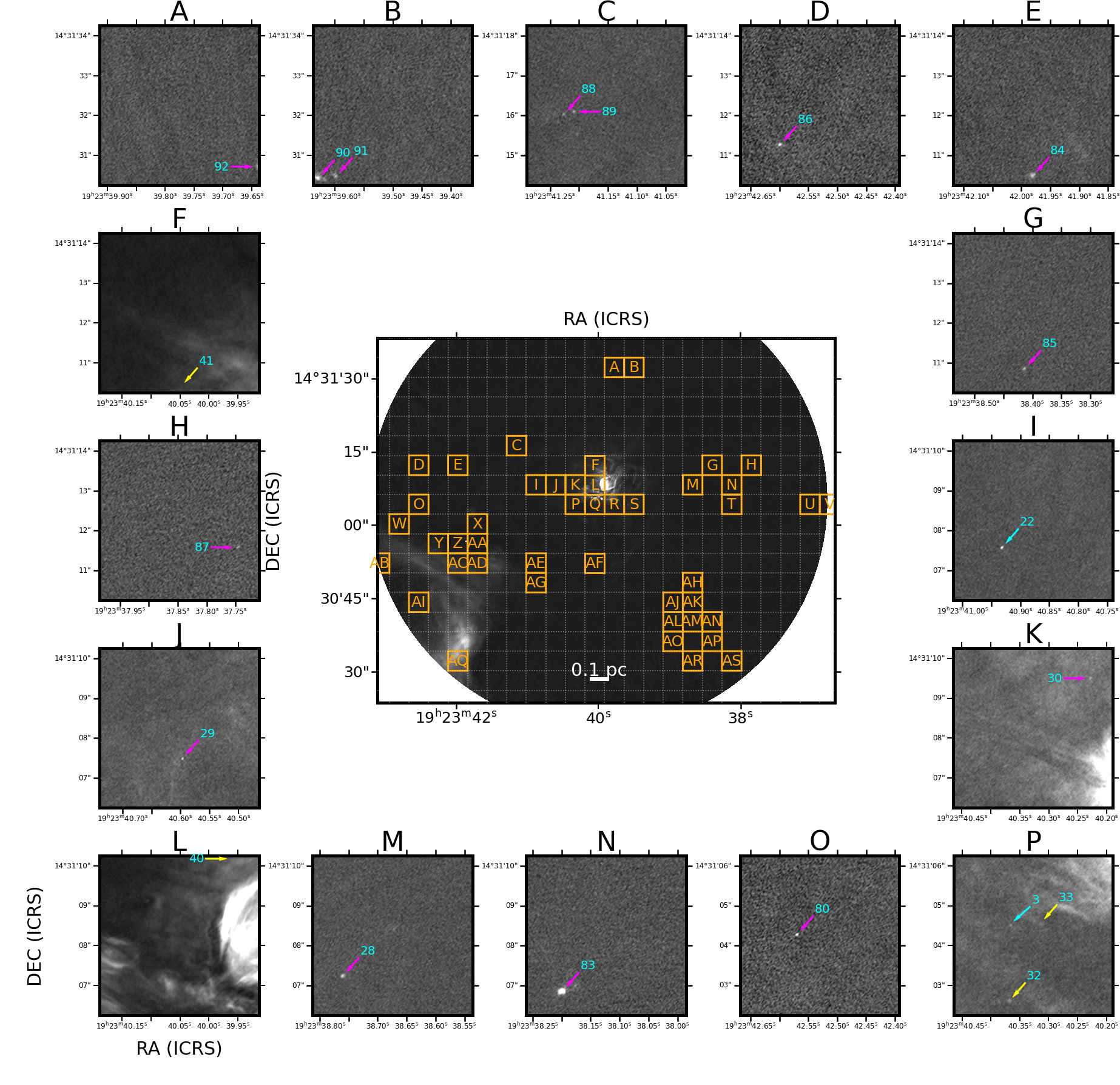}
    \caption{Same as Fig.~\ref{fig:ppo_map_w51e} but for W51-IRS2. The cyan crosses mark the PPOs in the W51-E catalog.}
    \label{fig:ppo_map_w51n}
\end{figure*}

\begin{figure*}

    \centering
    \includegraphics[width=1.0\linewidth]{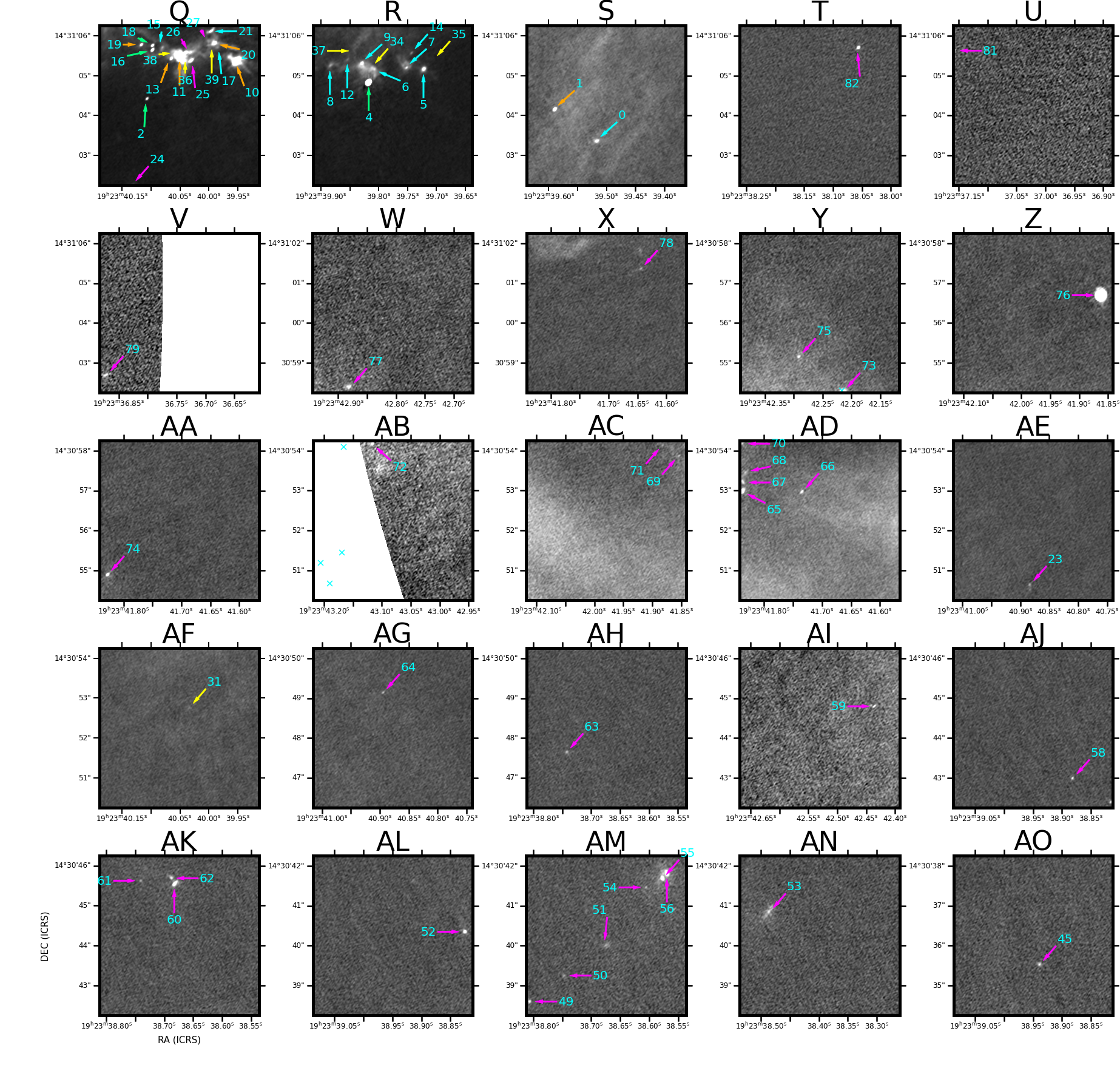}
    \caption{Fig.~\ref{fig:ppo_map_w51n} continued.}
    \label{fig:ppo_map_w51n2}
\end{figure*}

\begin{figure*}

    \centering
    \includegraphics[width=1.0\linewidth]{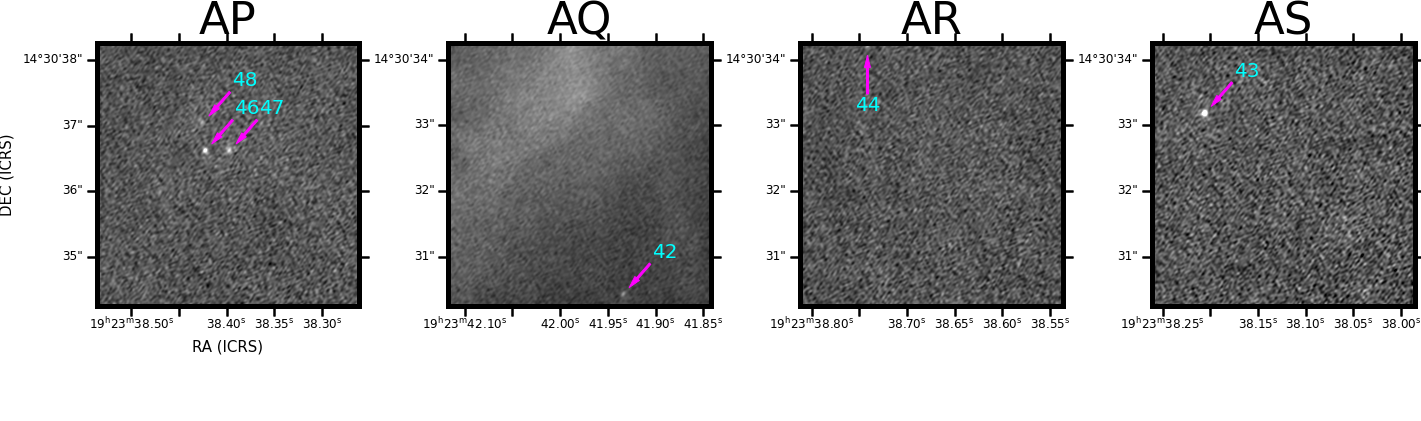}
    \caption{Fig.~\ref{fig:ppo_map_w51n} continued.}
    \label{fig:ppo_map_w51n3}
\end{figure*}

\clearpage
\section{PPO snapshots}
\label{appendix:snapshots}
\begin{figure*}[h]
    \centering
    \includegraphics[scale=0.17]{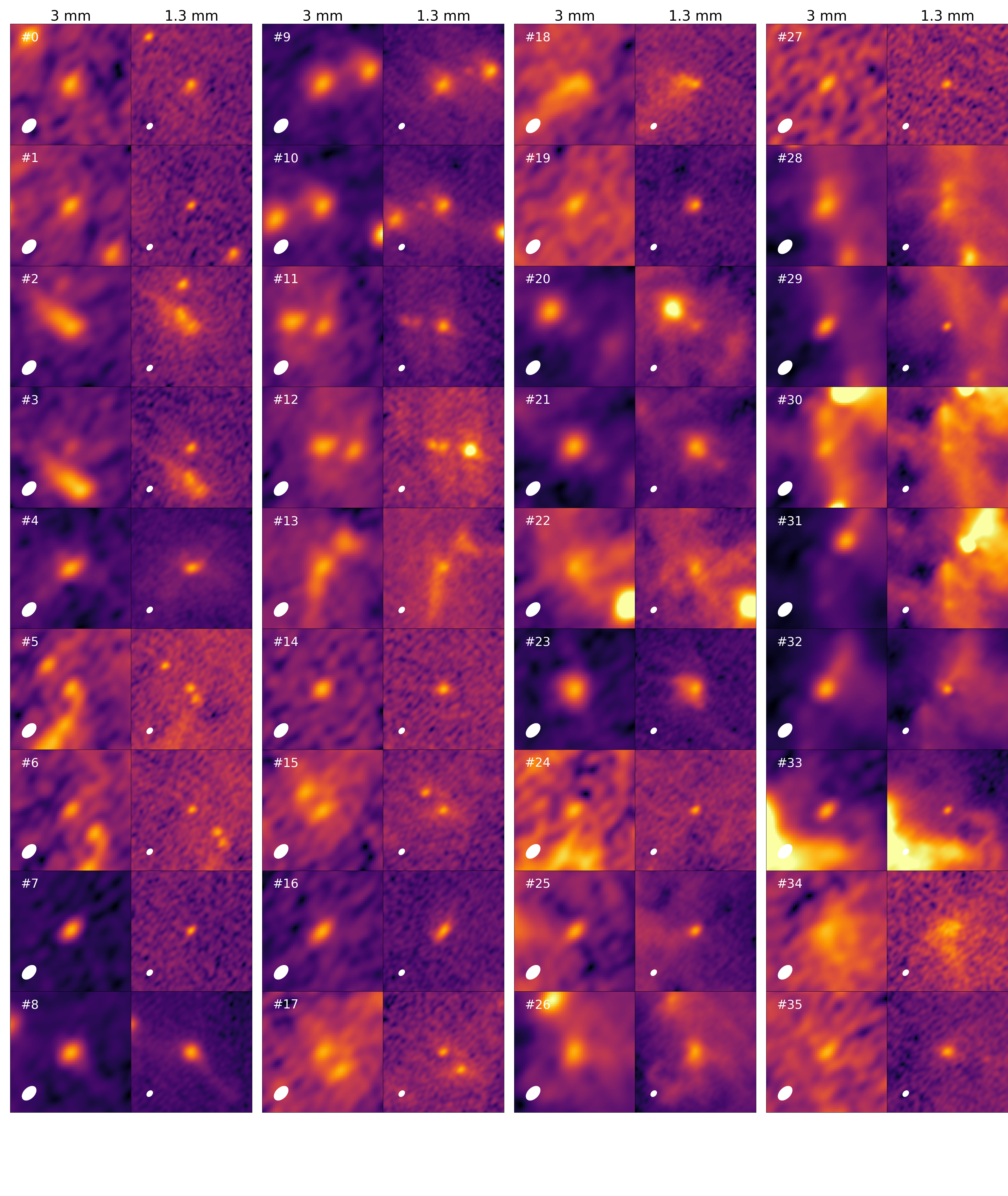}
    \caption{Snapshots of PPOs identified in W51-E $1.3\,{\rm mm}$ and $3\,{\rm mm}$ continuum image. The color scale is normalized to the flux of the central object for clear visualization. The size of the image beam is represented as a filled ellipse on the left lower corner. }
    \label{fig:snapshot1_w51e}
\end{figure*}

\begin{figure*}[h]
    \centering
    \includegraphics[scale=0.17]{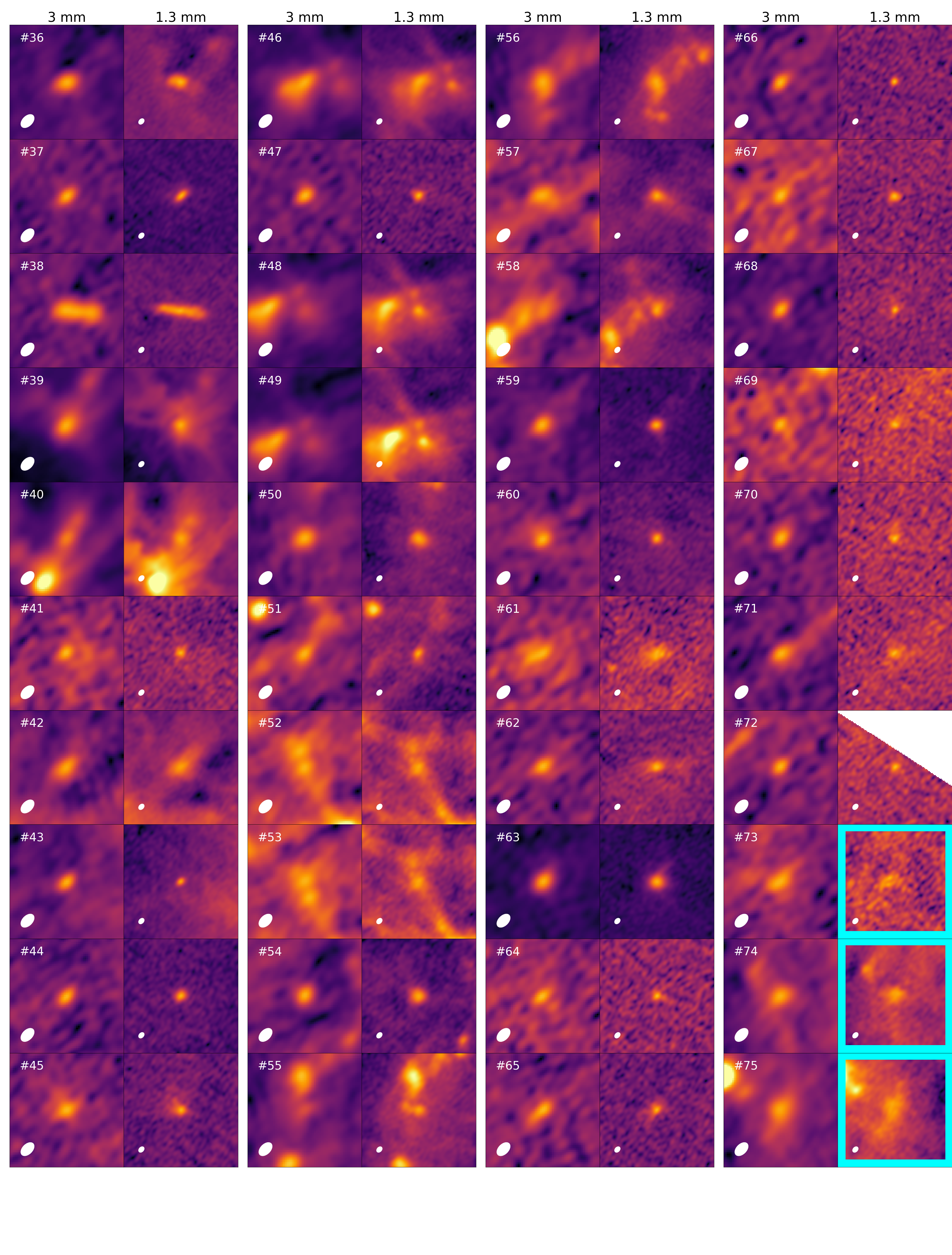}
    \caption{Fig.~\ref{fig:snapshot1_w51e} continued. The images of undetected sources in the \texttt{dendrogram} are surrounded by cyan squares.}
    \label{fig:snapshot2_w51e}
\end{figure*}

\begin{figure*}[h]
    \centering
    \includegraphics[scale=0.17]{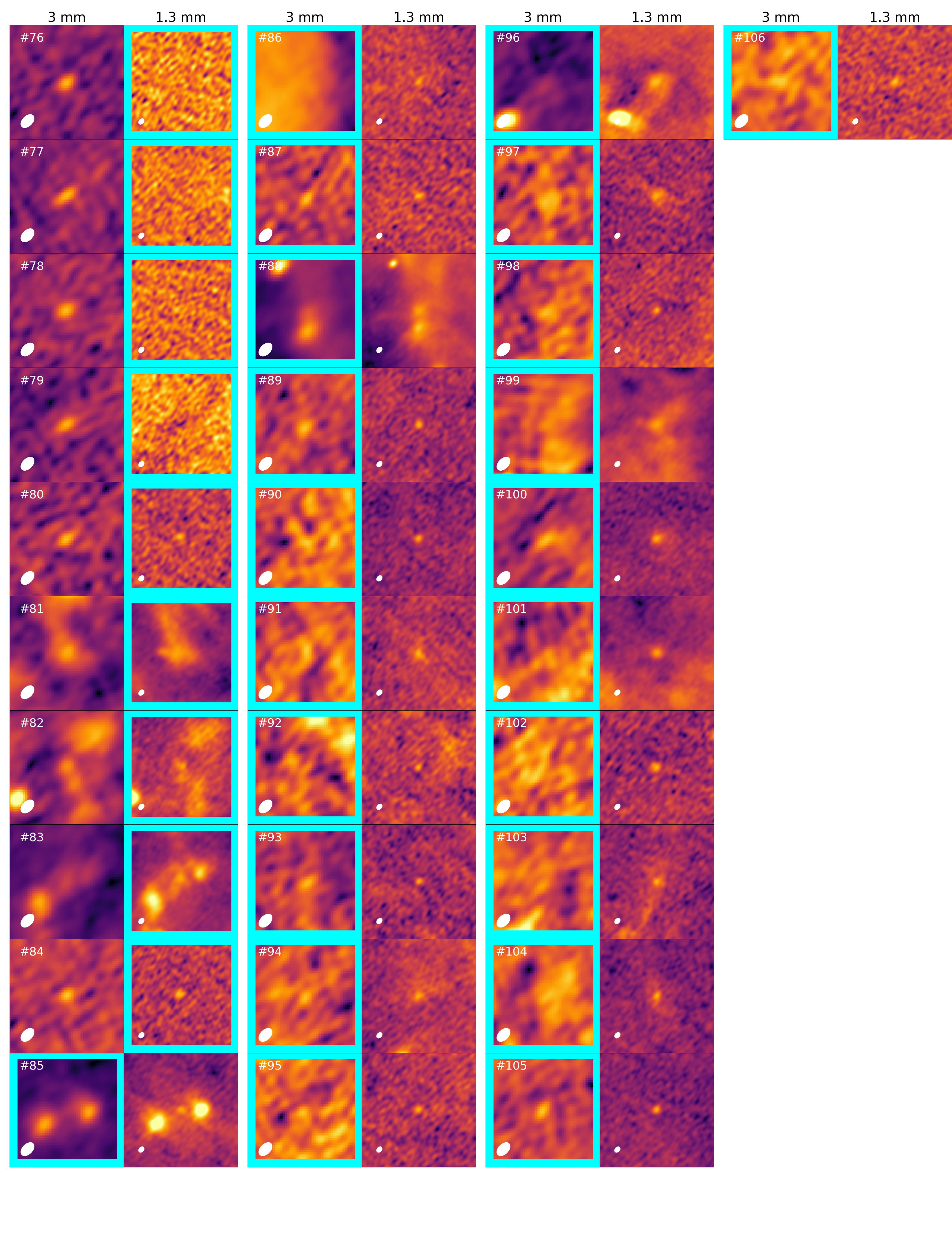}
    \caption{Fig.~\ref{fig:snapshot1_w51e} continued. }
    \label{fig:snapshot3_w51e}
\end{figure*}

\begin{figure*}[h]
    \centering
    \includegraphics[scale=0.17]{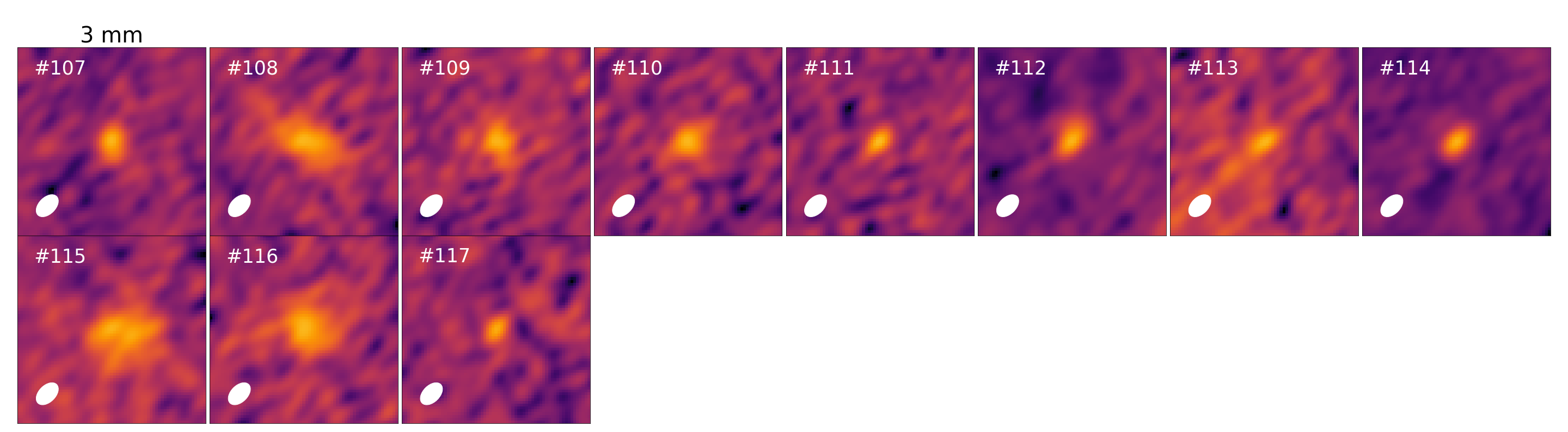}
    \caption{Fig.~\ref{fig:snapshot1_w51e} continued.  For \#107--117, the sky position of $1.3\,{\rm} mm$ counterpart is out of the field of view.}
    \label{fig:snapshot4_w51e}
\end{figure*}

\begin{figure*}[h]
    \centering
    \includegraphics[scale=0.17]{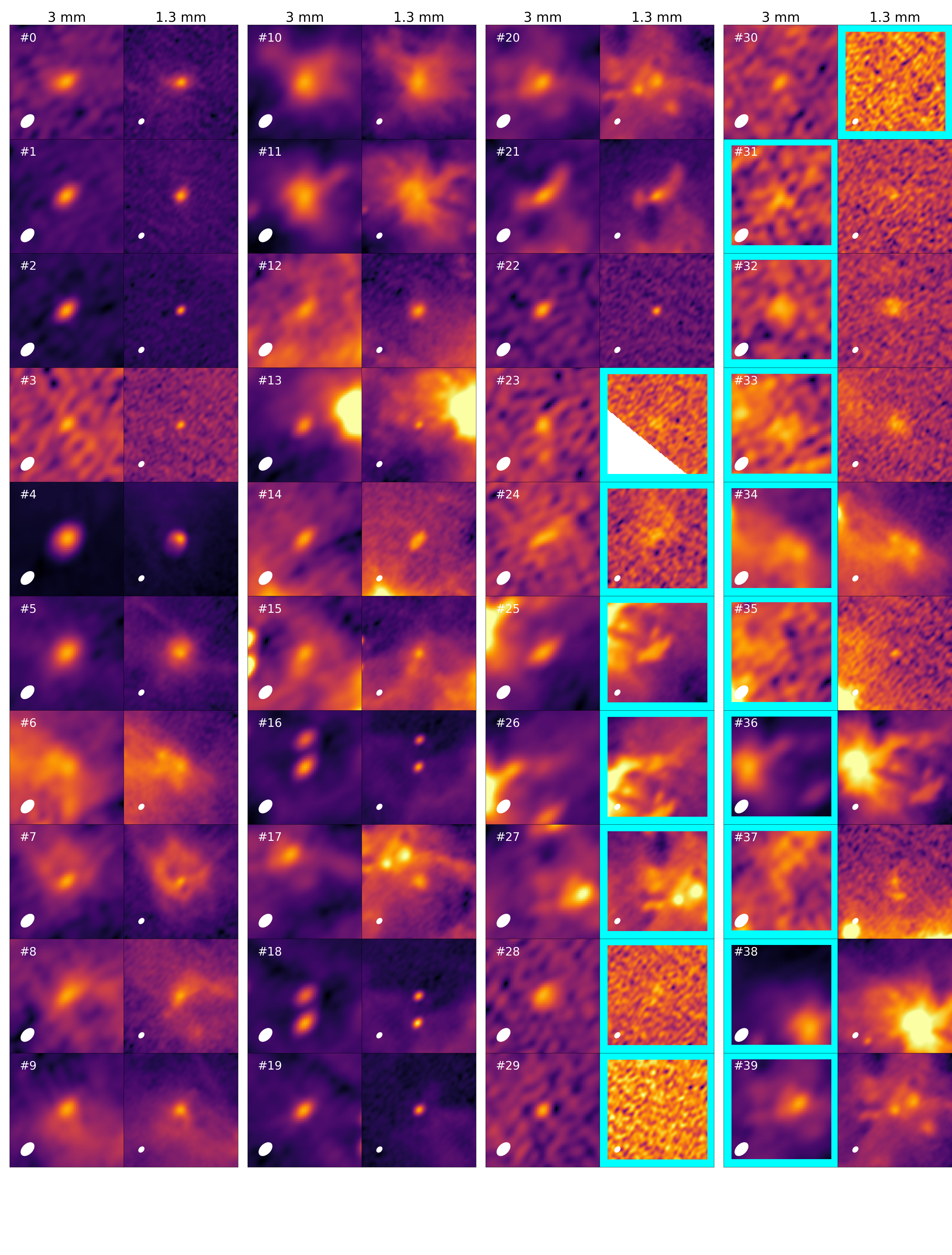}
    \caption{Snapshots of PPOs identified in W51-IRS2 $1.3\,{\rm mm}$ and $3\,{\rm mm}$ continuum image. Sources \#32--39 does not have $1.3\,{\rm mm}$ counterpart in the $1.3\,{\rm mm}$ field of view image. }
    \label{fig:snapshot1_w51n}
\end{figure*}

\begin{figure*}[h]
    \centering
    \includegraphics[scale=0.17]{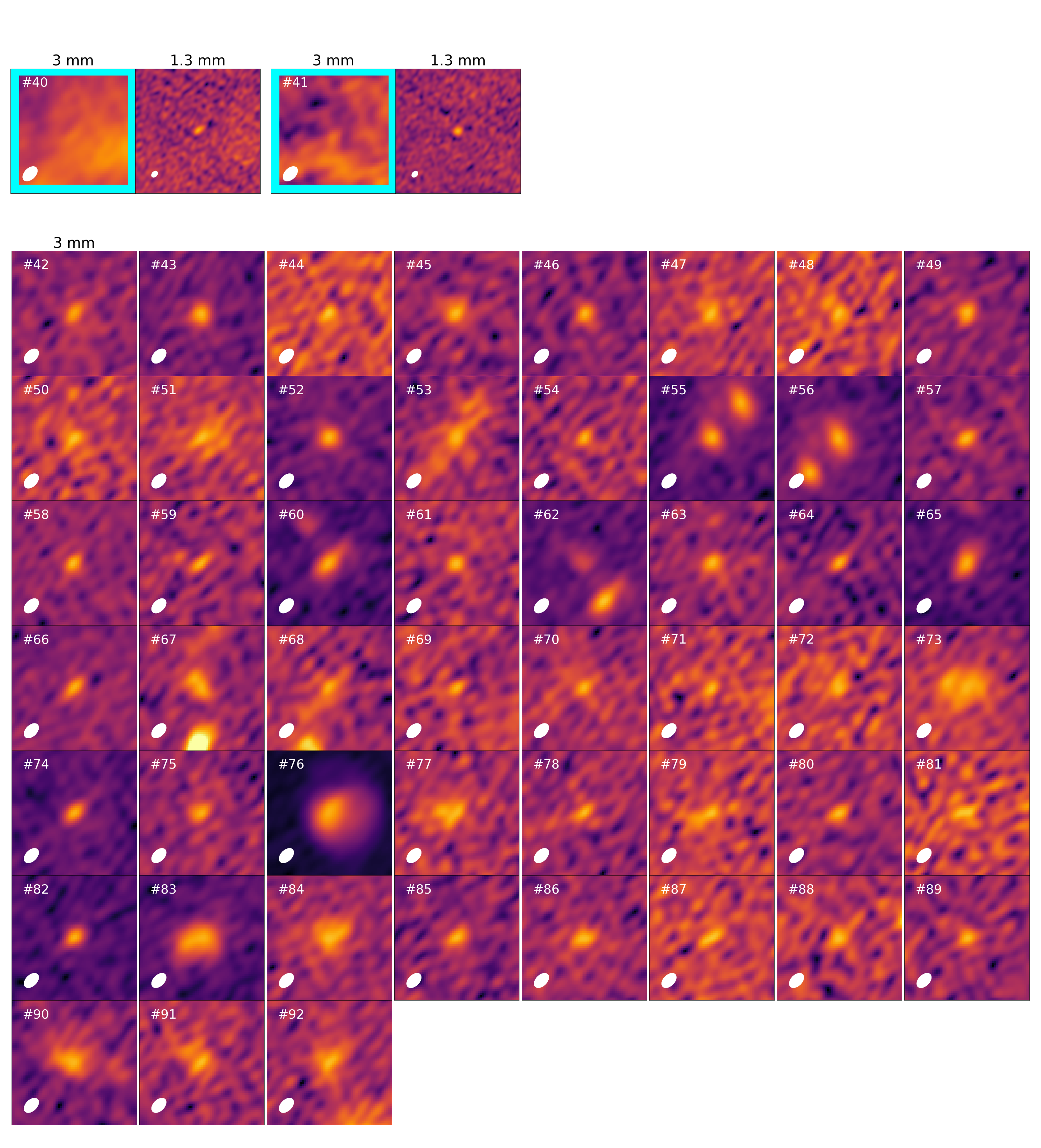}
    \caption{Fig.~\ref{fig:snapshot1_w51n} Continued. For \#42--92, the sky position of $1.3\,{\rm} mm$ counterpart is out of the field of view. }
    \label{fig:snapshot2_w51n}
\end{figure*}

\clearpage
\section{Ambiguous and low S/N source snapshots}
\label{appendix:ambiguous}
\begin{figure*}[h]
    \centering
    \includegraphics[scale=0.17]{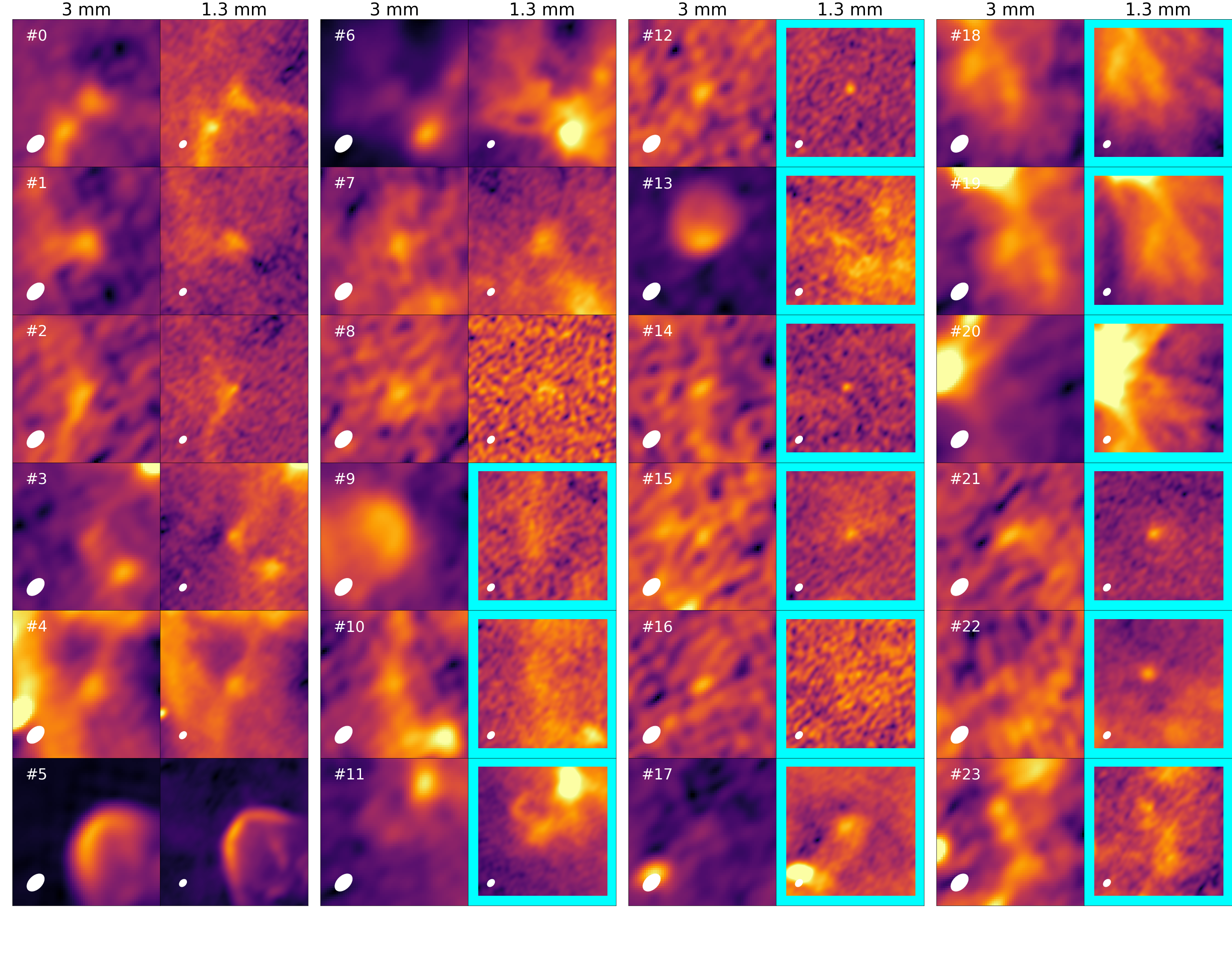}
    \caption{Snapshots of \textit{ambiguous} sources that were not selected in the final catalog of PPOs in W51-E. The cyan square indicates the absence of the sources in \textit{ambiguous} catalog. }
    \label{fig:snapshot1_ambiguous_w51e}
\end{figure*}

\begin{figure*}[h]
    \centering
    \includegraphics[scale=0.17]{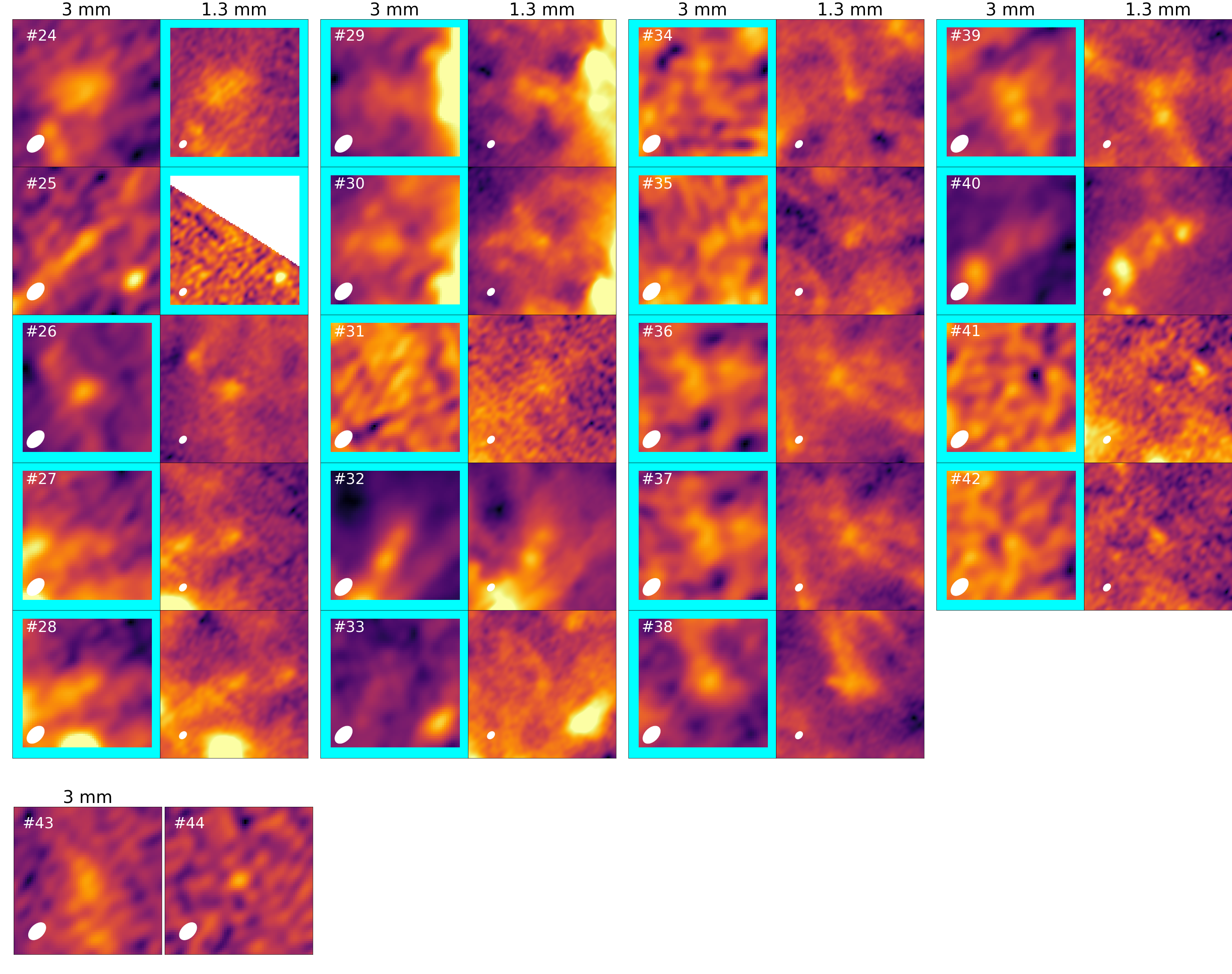}
    \caption{Fig.~\ref{fig:snapshot1_ambiguous_w51e} continued. The sources \#43 and \#44 are outside of the field of view of the $1.3\,{\rm mm}$ continuum image.}
    \label{fig:snapshot2_ambiguous_w51e}
\end{figure*}

\begin{figure*}[h]
    \centering
    \includegraphics[scale=0.17]{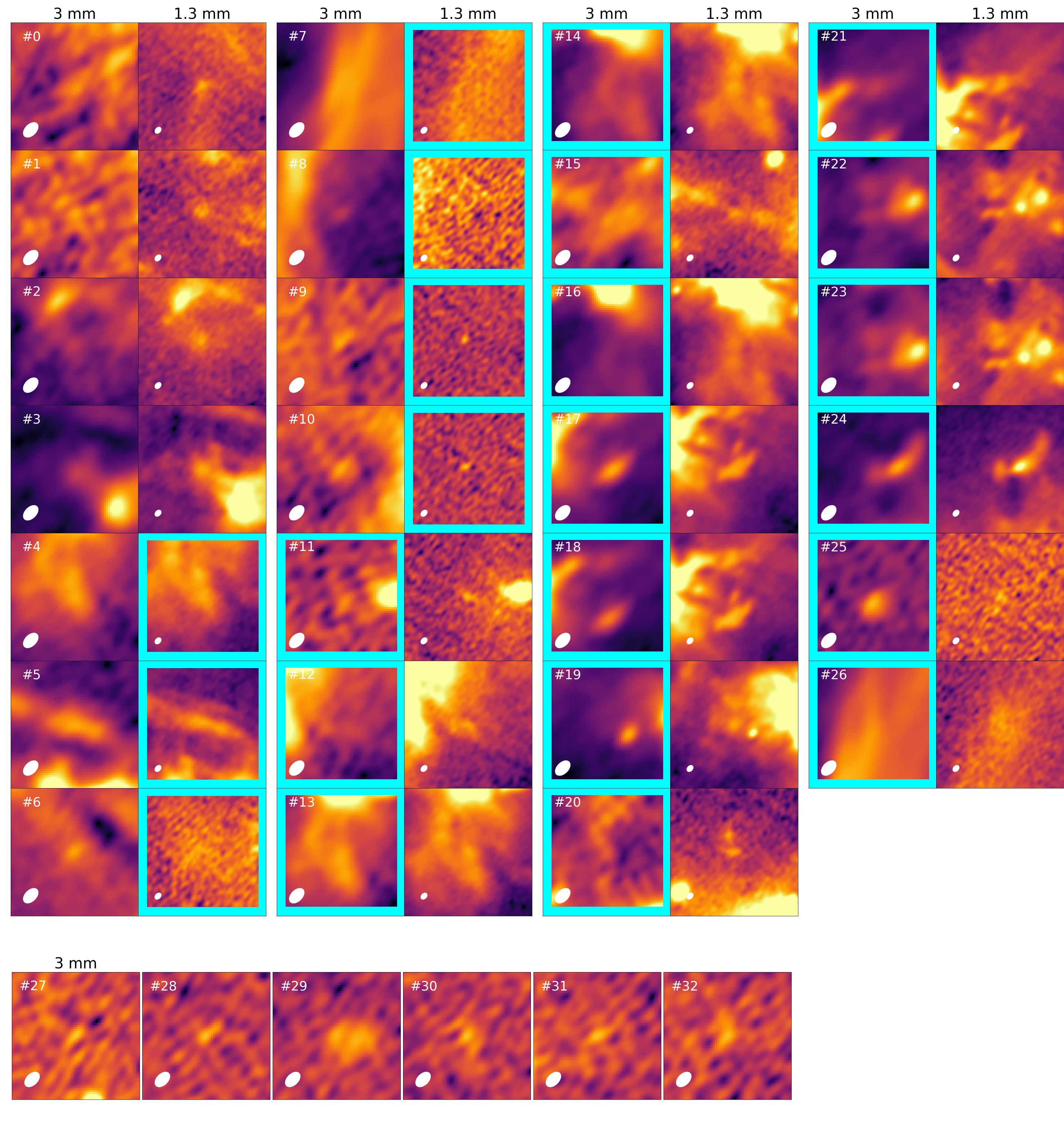}
    \caption{Snapshots of \textit{ambiguous} sources in W51-IRS2. The sources \#27--32 are outside of the field of view of the $1.3\,{\rm mm}$ continuum image. }
    \label{fig:snapshot1_ambiguous_w51n}
\end{figure*}

\begin{figure*}[h]
    \centering
    \includegraphics[scale=0.5]{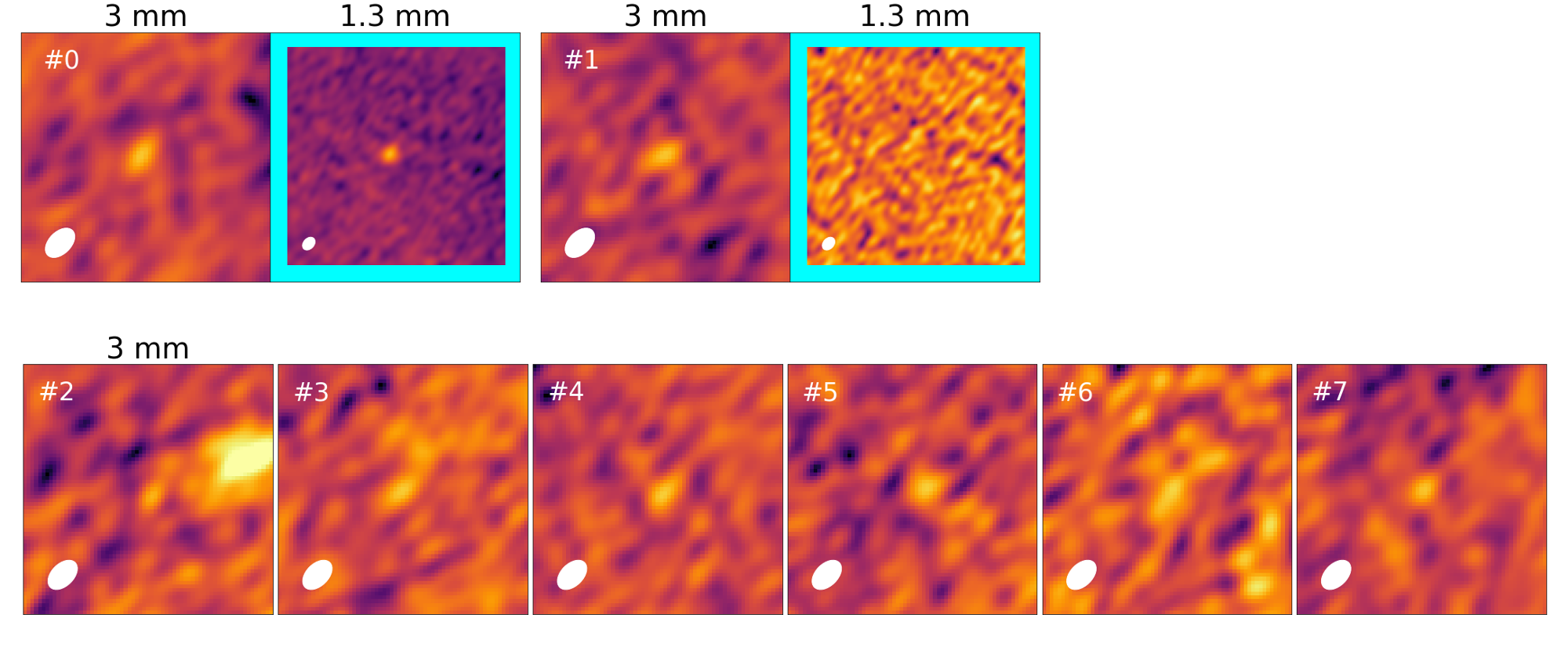}
    \caption{Snapshots of \textit{low S/N} sources in W51-E. The sources \#2--7 are outside of the field of view of the $1.3\,{\rm mm}$ continuum image.  }
    \label{fig:snapshot1_lowsn_w51e}
\end{figure*}

\begin{figure*}[h]
    \centering
    \includegraphics[scale=0.5]{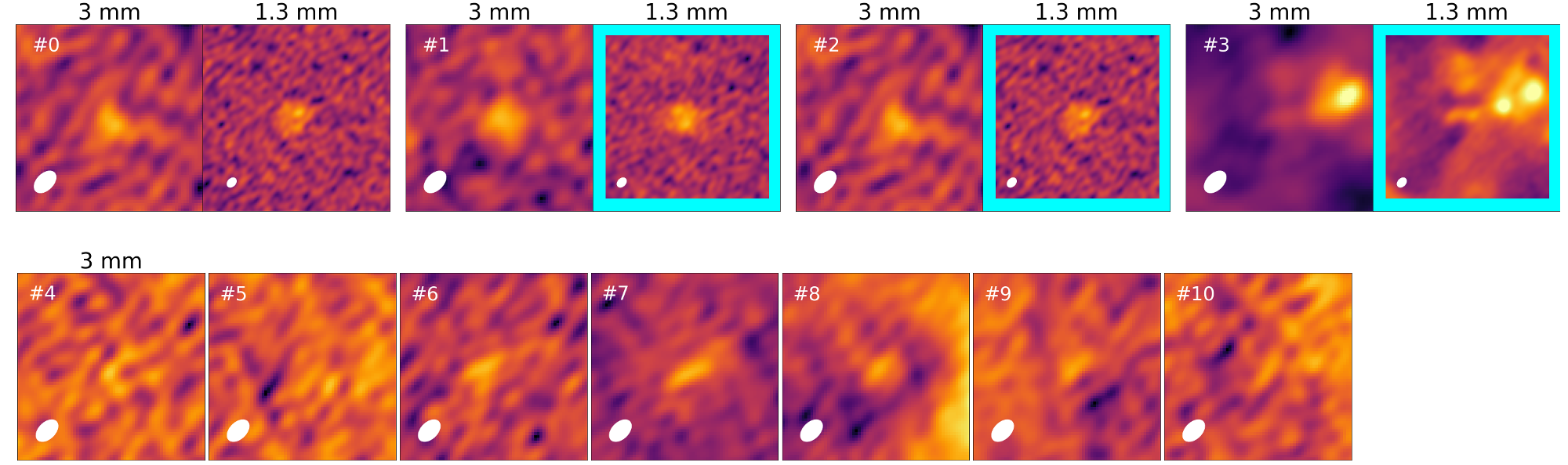}
    \caption{Fig.~\ref{fig:snapshot1_lowsn_w51e} continued. The sources \#4--10 are outside of the field of view of the $1.3\,{\rm mm}$ continuum image. }
    \label{fig:snapshot1_lowsn_w51n}
\end{figure*}

We provide snapshots of \textit{ambiguous} (Fig.~\ref{fig:snapshot1_ambiguous_w51e}--\ref{fig:snapshot1_ambiguous_w51n}) and \textit{low S/N} sources (Fig.~\ref{fig:snapshot1_lowsn_w51e} and \ref{fig:snapshot1_lowsn_w51n}) that were unselected in the final catalog of PPOs in Sec.~\ref{subsec:dendro}. Sources identified by only one of the three independent observers are classified as \textit{ambiguous}. In W51-E and W51-IRS2, 45 and 33 \textit{ambiguous} sources are identified. Among the observers, we found different assessments of compactness, significance, or independence from background for these sources. The \textit{ambiguous} sources may particularly include substructures of diffuse dust emission and hyper-compact HII regions (e.g, \#5 in W51-E; w51e2w \citealt{goddi16}). On the other hand, \textit{low S/N} sources are picked by more than two observers but are not identified in \texttt{dendrogram} due to their low S/N. In Fig.~\ref{fig:snapshot1_ambiguous_w51e}--\ref{fig:snapshot1_lowsn_w51n}, some sources have their counterparts not classified as \textit{ambiguous} or \textit{low S/N} catalogs; these counterparts are highlighted with cyan squares in their images. In some cases, these counterparts are real PPOs classified as single detection. For example, PPO \#25 in W51-IRS2 has single detection at $3\,{\rm mm}$ (Fig.~\ref{fig:snapshot1_w51n}). Its  $1.3\,{\rm mm}$ counterpart is classified as \#17 \textit{ambiguous} sources in Fig.~\ref{fig:snapshot1_ambiguous_w51n} as it is less centrally-peaked. The sky positions of these sources are provided by electronic form at \url{https://zenodo.org/records/16235156}. 




\end{document}